%% file: kintora-sc.tex
\pdfoutput=1
\documentclass[conference]{IEEEtran}
\newif\ifanon
\anonfalse
\ifdefined\buildanon \anontrue \fi

\usepackage{amsmath,amssymb}
\usepackage{graphicx}
\usepackage{booktabs}
\usepackage{array}
\usepackage{multirow}
\usepackage{xcolor}
\usepackage[hyphens]{url}
\usepackage[colorlinks=true,linkcolor=blue,citecolor=blue,urlcolor=blue]{hyperref}

\newcommand{\vpar}{v_\parallel}
\newcommand{\dd}{\mathrm{d}}

\begin{document}

\title{Scalable Solvers for Implicit Eulerian Models of Magnetized Plasmas}

\ifanon
\author{\IEEEauthorblockN{Anonymous submission}}
\else
\author{\IEEEauthorblockN{Mark F.~Adams}
\IEEEauthorblockA{Applied Mathematics and Computational Research Division\\
Lawrence Berkeley National Laboratory\\
Berkeley, CA, USA\\
mfadams@lbl.gov}
}
\fi

\maketitle

\begin{abstract}
Implicit Eulerian models of magnetized plasmas need GPU solvers for their stiff terms: the advection along the magnetic field, the
field equation and the collisions.  This paper describes such solvers in a discontinuous Galerkin code for gyrokinetic ions and
drift-kinetic electrons in tokamaks, written in Kokkos.  Its mesh extrudes flux-aligned triangles into field-aligned wedges, joined by
a mortar that conserves mass exactly; energy is conserved to the time integrator's $O(\Delta t^4)$ per step.  GMRES solves each implicit stage of the advection, preconditioned by an upwind sweep,
block Gauss--Seidel in the drift's upwind order.  With kinetic electrons at the physical mass ratio, Newton's method solves each stage
together with the potential.  Its linear solves are preconditioned by the sweep, by a fluid-electron block whose toroidal modes are
independent plane problems for a geometric multigrid, and by an approximate transport step that we propose, the inflow transfer:
inexpensive passes, each reusing the sweep's block inverses, that carry the fast electrons' inflow across the toroidal subdomains of
the parallel decomposition.  The solvers are PETSc objects and a custom batched GPU solver for the Landau
collisions, published as verified examples, with porting prompts, for an LLM code assistant to implement in a user's code.  The
Cyclone base case's linear ion-temperature-gradient growth rate with gyrokinetic ions at $\rho^* = 1/100$ is 2 to 10\% above GYSELA's,
and on a shaped plane the geodesic acoustic mode's frequency is within 2\% of the shaped-plasma relation, which is itself accurate to a few percent.
Weak scaling to 256 NVIDIA A100 and AMD MI250X devices and strong scaling to 192 measure the four-dimensional decomposition of the
five-dimensional phase space.
\end{abstract}

\section{Introduction}
\label{sec:intro}
Gyrokinetic and drift-kinetic codes for tokamak turbulence have moved to GPUs, continuum, semi-Lagrangian and particle codes
alike~\cite{Germaschewski2021,Mandell2024GX,Malaboeuf2025,XGCgpu2021}.  This paper describes an Eulerian discontinuous Galerkin (DG)~\cite{CockburnShu2001}
code, KINTORA, for gyrokinetic ions and drift-kinetic electrons that conserves mass exactly and energy asymptotically in the time step, includes
nonlinear Landau collisions~\cite{Landau1936} between every pair of species, runs on shaped flux-aligned meshes, and runs on
the NVIDIA A100~\cite{Choquette2021} and the AMD MI250X~\cite{Atchley2023} from one Kokkos~\cite{TrottKokkos2022} source.

\begin{table*}[t]
\centering
\caption{The solvers: what each is in PETSc~\cite{PETSc2024}, the structure the problem must supply, the section that
describes it, and its label in Fig.~\ref{fig:stack} (a letter, or the panel).}
\label{tab:solvers}
\footnotesize
\begin{tabular}{@{}>{\raggedright}p{0.12\textwidth}>{\raggedright}p{0.31\textwidth}>{\raggedright}p{0.36\textwidth}>{\raggedright}p{0.06\textwidth}>{\raggedright\arraybackslash}p{0.05\textwidth}@{}}
\toprule
solver & in PETSc & structure & Sec. & Fig.~\ref{fig:stack} \\
\midrule
K1, advection sweep & the preconditioner type \texttt{PCKINSWEEP} (\texttt{-pc\_type kinsweep}), under GMRES or as FieldSplit's kinetic block &
an upwind order: field-aligned wedges along $\varphi$, a topological order of the drift's upwind graph in the plane, the mirror force's
direction in $\vpar$ & \ref{sec:advection} & C \\
K2, plane multigrid & \texttt{PCMG} with Galerkin coarse operators, smoothed by a multiplicative, coloured, overlapping domain
decomposition over flux-surface annuli & a hierarchy of plane meshes, not nested, and the flux-surface
annuli of each & \ref{sec:fluid}, \ref{sec:fieldmg} & A, F \\
K3, batched block Thomas & a custom batched direct solver, called inside each Newton step of the collision stage, one GPU team per system & many
systems, one per species and point of space, block tridiagonal in $\vpar$, each block one $\vpar$ segment's nodes by every $\mu$ row & \ref{sec:collisions} & B \\
K4, low-storage ARKIMEX3 & the time integrator type \texttt{TSKINARKIMEX3} & an implicit-explicit split, $F(t, U, \dot U) = G(t, U)$, with
an identity or diagonal mass & \ref{sec:step} & (a) \\
K5, fluid-electron block & FieldSplit's Schur-complement preconditioner, specialized to this model; its plane systems solved by a sparse LU or by \texttt{PCMG} (K2) & toroidal symmetry: one plane system per toroidal Fourier mode & \ref{sec:fluid} & F \\
\bottomrule
\end{tabular}
\end{table*}

Kinetic electrons make the time step stiff in two ways.  Their thermal speed at equal temperature is $(m_D/m_e)^{1/2} \approx 61$
times the deuterium ions', so at the time step that accuracy sets, their toroidal Courant number per wedge is about 30, and implicit
gyrokinetic codes treat the parallel streaming implicitly for that reason~\cite{Kotschenreuther1995,Barnes2019}.  Their streaming also
couples to the potential in the $\omega_H$ mode, the electrostatic shear-Alfv\'en wave, which makes an explicit field unstable at any useful step~\cite{Lee1987,BelliHammett2005}.
The code's step therefore treats the streaming implicitly and, with kinetic electrons, solves each stage's potential together with
the distribution; for the ions alone the implicit streaming is not a stability requirement, since their Courant number per wedge is
below one (Section~\ref{sec:solver}).  Each species runs on its own velocity grid in its own thermal units, so the electrons run at the
physical mass ratio, on a grid graded toward $\vpar = 0$ (Section~\ref{sec:background}).  The weak-scaling series of Section~\ref{sec:scaling},
the Landau damping test, the collision tests and the shaped collisional model of Section~\ref{sec:turb} run at deuterium's ratio, 3670.5, and the comparison with the GENE code~\cite{Jenko2000} at hydrogen's, 1836.

The paper's subject is the solver algorithms that make the implicit step affordable on a GPU.  Implicit kinetic models of magnetized
plasmas share a structure: the advection is stiff along the magnetic field, the field equation is anisotropic along the flux surfaces,
and collisions couple only velocity space at each point of space.  Here the phase space is five-dimensional (Figure~\ref{fig:decomp}):
the poloidal plane $(R, Z)$, the toroidal angle $\varphi$, the parallel velocity $\vpar$ and the magnetic moment $\mu$, which carries
the perpendicular velocity, $\mu \propto v_\perp^2/B$.  Five solvers exploit this structure (Table~\ref{tab:solvers}): K1, a
sweep along the advection's upwind structure that preconditions GMRES~\cite{SaadSchultz1986}; K2, a geometric multigrid on the plane, smoothed over flux-surface annuli, which solves the kinetic electrons' fluid-electron
block's modes and, as an option, the field equations; K3, a batched direct solver for the collisions' linear systems, one over each species'
velocity grid at each point of space, block tridiagonal in $\vpar$, each block one $\vpar$ segment's nodes by every $\mu$ row; K4, a
low-storage implicit-explicit time integrator; and K5, the kinetic electrons' fluid-electron block, a solve specialized to the toroidal
symmetry of this model's field operator.  K1, K2 and K4 are PETSc objects configured by options, a registered preconditioner, a
configured multigrid and a registered time integrator~\cite{BrownComposable2012}; K3 is a custom batched direct solver.  All five run
on the GPU, except where the host is faster: the plane multigrid's two coarsest levels, the field solves' multigrid cycle and K5's
Fourier transform.

These solvers are specialized for strongly anisotropic systems.  This paper publishes them to bring the algorithms to other
applications through packaged expert workflows for a large language model (LLM) code assistant: each solver is a verified example of
its algorithm, which the assistant can apply flexibly in the context of a specific application code, and the application supplies the
structure the solver relies on.  The solvers' source and tests are public\ifanon\else~\cite{KintoraCode2026}\fi, with prompts for the
assistant to port each of K1 to K4 into another application or into PETSc.  K4 has been ported this way into a clone of PETSc, as a
low-storage option of \texttt{TSARKIMEX} (Section~\ref{sec:discipline}); K3 is a candidate, as a direct solver for PETSc's batched
solvers (\texttt{PCBJKOKKOS}).  The prompts direct the
assistant to report, before it changes any code, how the solver's structure maps onto the target application, what the application must supply, and what
it proposes to leave out.  Porting one solver this way differs from translating a whole code (Section~\ref{sec:related}): the solver
moves, and the target keeps its own discretization and data.  The code itself was written
with such an assistant (Acknowledgments).

This paper makes the following contributions:
\begin{itemize}
\item Five solvers published as verified examples, four with porting prompts for an LLM code assistant; one, K4, ported into a clone
of PETSc's main branch (Table~\ref{tab:solvers}, Section~\ref{sec:discipline}).
\item The advection solve (K1): GMRES preconditioned by one upwind sweep, on the GPU as dependence-ordered launches (Section~\ref{sec:advection}).
\item The kinetic-electron preconditioner (K1, K5): the sweep, an inflow transfer across the toroidal blocks, and a fluid-electron
block solved by toroidal modes (Sections~\ref{sec:advection}, \ref{sec:kinetic}).
\item The plane multigrid (K2): smoothed over flux-surface annuli, for the electrons' fluid-electron block and, as an option, the
field solves with the Pad\'e polarization (Sections~\ref{sec:fluid}, \ref{sec:field}).
\item The collision solve (K3): a batched block-tridiagonal direct solver for the Newton steps of Landau collisions (Section~\ref{sec:collisions}).
\item A conservative Eulerian DG discretization: exact mass; energy to the integrator's $O(\Delta t^4)$ on the circular meshes, and on
the shaped ones with kinetic electrons, where a term at the mortar between wedges, the seam kick, cancels the error of the potential's
jump (Sections~\ref{sec:hard}, \ref{sec:conservation}).
\item A gyroaverage on the unstructured P2 plane: the Galerkin form of the ring average, integrated exactly by polygon clipping, with
a charge deposit that is its exact adjoint (Section~\ref{sec:gyro}).
\item A 4D decomposition of the 5D phase space, measured in weak scaling to 256 devices and strong scaling to 192
(Sections~\ref{sec:parallel}, \ref{sec:scaling}).
\item The hardware utilization on the A100 of a shaped model with gyrokinetic ions, kinetic electrons at deuterium's mass ratio and
Landau collisions (Section~\ref{sec:utilturb}).
\end{itemize}

\section{The model, the grid and the decomposition}
\label{sec:background}
\textbf{The model.}  The code advances each species' full distribution $f_s(R, Z, \varphi, \vpar, \mu)$, $R$ the major radius, in the
guiding-centre (Littlejohn) form~\cite{Littlejohn1983,CaryBrizard2009}, with the magnetic moment $\mu$ an exact invariant and time in
units of $a/v_{ti}$, $a$ the minor radius.  Lengths are in units of $a$ and the field in units in which the ions' thermal Larmor radius
is $\rho_i = 1/B_0$, $B_0$ the field on the axis, so $\rho^* = \rho_i/a = 1/B_0$.  Each species runs in its own thermal units, $\vpar$
in $v_{ts}$ and $\mu$ in $T_s/B_0$, with the species numbers
$\sigma_s = v_{ts}/v_{ti}$ and $\tau_s = T_s/T_i$, and its Hamiltonian in units of $T_s$ is $H_s = \vpar^2/2 + \mu B + (Z_s/\tau_s)\,\phi_s$.
The ions are gyrokinetic: $\phi_s$ is the potential averaged over the gyro-ring of each magnetic moment $\mu$, $\langle\phi\rangle_\mu$
(Section~\ref{sec:gyro})~\cite{Dubin1983,BrizardHahm2007}.  The electrons are drift-kinetic, $\phi_s = \phi$, since $k_\perp \rho_e
\le 0.02$ in the runs here.  The potential solves the gyrokinetic quasineutrality condition: the gyroaveraged ion density and the ion
polarization density balance the electron density, the electrons adiabatic or kinetic~\cite{Lee1983,BrizardHahm2007}.  Collisions are
nonlinear Landau collisions between every pair of species (Section~\ref{sec:collisions}), and a Boltzmann inflow condition at the wall
(Section~\ref{sec:hard}) closes the domain.  This section describes the grid in the poloidal plane, the toroidal
direction and velocity space (Table~\ref{tab:grid}), and then its decomposition over the devices.

\begin{table}[t]
\centering
\caption{The discretization of the Cyclone base case (CBC)~\cite{Dimits2000} at point B of the weak series (points A, B and C are
levels 2, 3 and 4 of the plane, 256, 1,024 and 4,096 cells): $3.4 \times 10^8$ unknowns for the ions and $2.3 \times 10^8$ for the
electrons, and point C 16 times as many: each point halves the plane's cells in size and doubles the wedges and the $\vpar$ segments (Table~\ref{tab:weak}).}
\label{tab:grid}
\footnotesize
\begin{tabular}{@{}lll@{}}
\toprule
direction & discretization & nodes \\
\midrule
poloidal plane & DG $p = 2$, flux-aligned triangles & 6 per cell \\
toroidal $\varphi$ & 16 wedges, DG $p = 2$ & 3 per wedge \\
$\vpar$, ions & DG $p = 3$, 12 uniform segments to $6\,v_{ti}$ & 48 \\
$\vpar$, electrons & DG $p = 3$, 8 graded segments to $6\,v_{te}$ & 32 \\
$\mu$, each & 8 elements of $p = 2$ in $v_\perp$ & 24 rows \\
\bottomrule
\end{tabular}
\end{table}

\textbf{The plane.}  The poloidal cross-section is meshed by triangles aligned with the flux surfaces, either concentric circles or a
shaped Solov'ev equilibrium~\cite{Solovev1968,CerfonFreidberg2010}; each refinement \emph{level} splits every cell into four.  The cells lie in rings, one between each pair of adjacent
flux surfaces of the mesh; an annulus is one ring's cells.  A second family of circles, the isotropic-triangle meshes (L3a to L6a), sets each ring's cell count by the cells' aspect ratio;
Table~\ref{tab:mg} and the benchmark of Section~\ref{sec:turb} use it, and the other circular runs the first.  The cells are straight-sided, so the wall is a polygon inscribed in the last flux surface, and $f$ is
DG of degree 2 in the plane: six nodes per cell.

\textbf{The toroidal direction.}  The torus is cut into $N_w$ wedges.  Each wedge is a prism in
coordinates that follow the magnetic field line from the wedge's upstream plane, so the parallel streaming is one-dimensional, along $\varphi$
at a fixed point of the plane, and the magnetic field enters the guiding-centre flux only through $B_\varphi = F/R$, with $F$ the equilibrium's poloidal current function, a flux function, which is
evaluated exactly.  In these
coordinates the guiding-centre flux, built from the continuous interpolant of the Hamiltonian and from the unit vector
$\mathbf{b}$ pulled back through the wedge's map (Section~\ref{sec:hard}), is divergence-free and conserves energy pointwise~\cite{CaryBrizard2009}.  The distribution is DG of degree 2 along $\varphi$, collocated at three
Gauss--Lobatto stations per wedge.  Figure~\ref{fig:geometry}a draws two wedges with the map of a large-aspect-ratio circle, $\theta \to \theta + \Delta\varphi / q(r)$, with $\theta$
the poloidal angle and $q$ the safety factor.  Inside a wedge the mesh is conforming; where a wedge's downstream face
meets the next wedge, the field-line image of a triangle overlaps several triangles of the next plane.  A static supermesh of these overlaps, the mortar between the two wedges~\cite{FarrellMaddison2011}, carries one upwinded flux per quadrature point; that flux leaves the upstream element exactly as it enters the downstream one,
so mass is conserved for any numerical flux.  The image of a triangle edge is curved and is represented by a quadratic curve through the mapped end and mid points
(Section~\ref{sec:hard}).

\textbf{The velocity grids.}  Each species has its own velocity grid in its own units.  $\vpar$ is DG of degree 3, collocated at four
Gauss--Lobatto nodes per segment, on uniform segments for the ions and on segments graded geometrically toward $\vpar = 0$ for the
electrons, whose resonance with an ion-scale mode lies at $\vpar \sim v_{ti} \approx v_{te}/61$ (Table~\ref{tab:grid}).  The magnetic
moment is sampled at the 24 nodes of the collision operator's $v_\perp$ grid, eight elements of degree 2 (Table~\ref{tab:grid}), the
$\mu$ rows; each $\mu$ row is a separate 4D problem for the advection: the streaming along $\varphi$, the magnetic
drift in the plane and the mirror force.

\textbf{The field and the collisions.}  The potential is a C0 P2 function on two field stations a wedge, each a copy of the plane,
and is solved plane by plane (Section~\ref{sec:field}).  Collisions act at each spatial point on the velocity grids of every species
there, a batch (Section~\ref{sec:collisions}).

\subsection{The distributed-memory decomposition}
\label{sec:parallel}
\begin{figure*}[t]
\centering
\includegraphics[width=\textwidth]{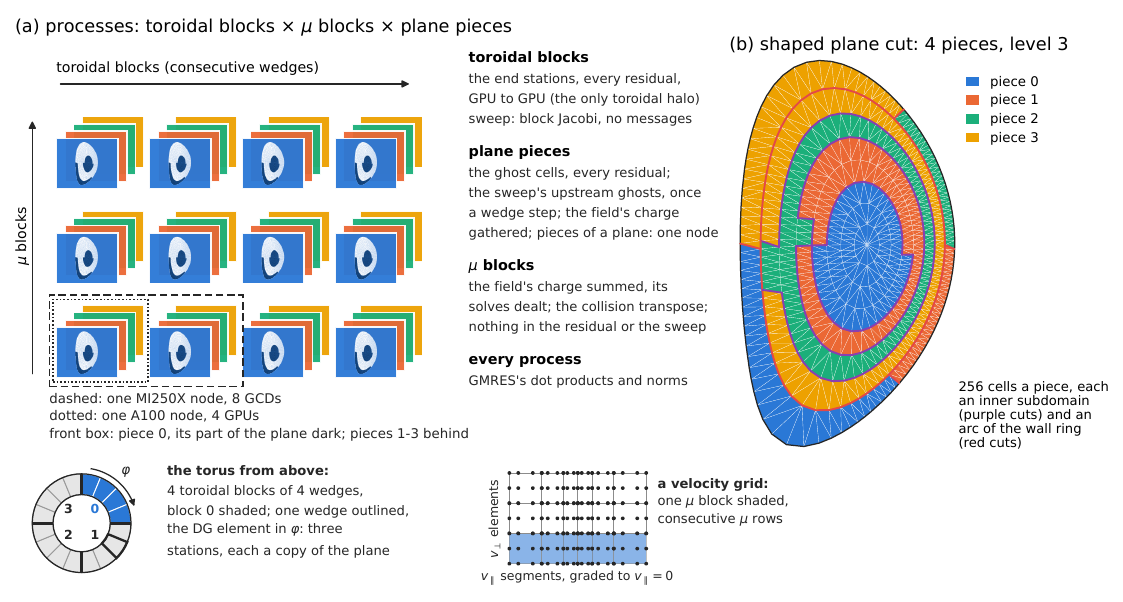}
\caption{The distributed-memory decomposition.  (a)~The processes as toroidal blocks by $\mu$ blocks by plane pieces, with the exchanges each
axis carries; below, what each axis splits.  (b)~The shaped (Solov'ev) plane at level~3 cut into four pieces by the code's rule.}
\label{fig:decomp}
\end{figure*}

The processes form a grid of toroidal blocks by $\mu$ blocks by plane pieces, one process per GPU or MI250X graphics compute die (GCD)
(Figure~\ref{fig:decomp}a).  A toroidal block owns consecutive wedges, a $\mu$ block owns consecutive $\mu$ rows, and a plane piece owns
part of the poloidal plane, the same part at every station of the process's wedges.

\textbf{The decomposition order.}  Four of the five dimensions are split, in the order of the coupling each adds.  $\vpar$ is not
split: the sweep crosses a cell's $\vpar$ segments in the mirror force's direction, and a collision batch needs the whole velocity grid.
\begin{enumerate}
\item $\mu$, by $\mu$ blocks, first.  Each $\mu$ row is a separate problem for the advection, so a $\mu$ block adds no halo and no
iteration, and it divides the memory of the block inverses that the sweep stores (Section~\ref{sec:advection}).  The $\mu$ blocks sum the charge for the field, share out the field solves for the potential's fluctuations about its toroidal mean (Section~\ref{sec:field}) and gather the result, and transpose the collision batches, each of which needs every $\mu$ row of every
species at a point of space (Section~\ref{sec:collisions}).
\item $\varphi$, by toroidal blocks of consecutive wedges.  The blocks exchange the first and last stations of every $\mu$ row in each
residual evaluation, and the sweep is block Jacobi across them, which the inflow transfer compensates (Section~\ref{sec:stage}).
\item The plane, $(R, Z)$, by plane pieces, last.  The pieces exchange their ghost cells, every station of every $\mu$ row, in each
residual evaluation, the largest exchange, and the sweep is block Jacobi between them, which adds one or two GMRES iterations to an advection solve (Section~\ref{sec:advection}).  At level 5 on 256 GCDs the step is 1.5$\times$ shorter with the $\mu$ rows split first than with the plane split first.
\end{enumerate}
GMRES's inner products and norms reduce over every process.  The halo exchanges are non-blocking sends and receives from device
buffers by GPU-aware MPI, completed before the kernel that reads them; the field's reductions and the collision transpose are staged
through host memory.  After gathering the charge over the pieces, every plane piece solves the whole plane's field by a direct
factorization, unlike the fluid-electron block, whose plane multigrid is distributed over the pieces (Sections~\ref{sec:fluid}
and~\ref{sec:limitations}).  The processes are numbered in the reverse of the splitting order: the plane piece varies fastest, then the toroidal block, then the
$\mu$ block, so a plane's pieces fill a node, up to four on a Perlmutter node of four A100s and up to eight on a Frontier node of
eight GCDs, and the plane halo moves over the node's links while the toroidal halo crosses nodes.  At level 5 on 16 pieces, two
Frontier nodes a plane, this order cut the step from 6.16 to 4.67\,s against numbering the toroidal blocks fastest.

\textbf{Several velocity grids.}  The $\mu$ blocks are placed on the species' grids in contiguous groups, and each grid divides its $\mu$
rows over its own group, so a process holds one grid's rows and every kernel reads one grid's tables.  The exchanges across the $\mu$
blocks are sums of plane-sized arrays, and the toroidal blocks and plane pieces join processes of the same $\mu$ rows, so several grids
add no exchange to those of one grid.  Since a process holds one grid, two grids need at least two $\mu$ blocks, and a run of two
species on one device runs two processes on it (point A of Table~\ref{tab:weak}).  The load balance between the grids is the choice
of block counts; by default the split minimizes the busiest process's $\mu$ rows times segments.  The collision transpose maps each process's rows to their
species and $\mu$ row through its grid.

\textbf{The cut.}  The plane is cut in its ring-walk order, ring by ring and around each ring.  The cells inside the wall ring, the outermost ring of cells, are divided into $P$ ranges of equal count, so the pieces are bands of rings whose boundaries may fall inside a ring, and the wall ring into $P$
arcs, one per piece (Figure~\ref{fig:decomp}b).  The wall ring carries the work of the wall slivers, where a wedge's field-line image overhangs the wall
(Section~\ref{sec:hard}); left to the outermost piece, that
work would load the piece to 2.5$\times$ the mean at four pieces.

\section{The implicit step and the advection solve}
\label{sec:solver}

\subsection{The time step}
\label{sec:step}
\textbf{The semi-discrete system.}  The state $f$ holds every species' distribution: per $\mu$ row, the DG coefficients at the three
stations of every wedge.  The potential $\phi$ lives on the field stations, two per wedge (the wedge's first plane and its middle), each a
C0 P2 function on the plane's $n_p$ nodes, so $\phi_w = (\phi_{w,0}, \phi_{w,1})$ for wedge $w = 0, \ldots, N_w - 1$.  The kinetic
equation and quasineutrality read
\begin{equation}
M \dot f + R(f, \phi) = M S, \qquad \hat L \phi = D f,
\label{eq:semi}
\end{equation}
where $M$ is the DG mass matrix, $R$ the DG residual of the drift-kinetic flux with the mortar and the wall, $S$ the collision rate from the last collision stage, held fixed until the next (Section~\ref{sec:collisions}), and $\hat L$ the
field operator, bordered by one row that fixes the potential's free constant (Section~\ref{sec:field}).  $D$ is the charge deposit: each species'
$Z_s \int f_s N_i$ on the field nodes, $N_i$ the plane's C0 P2 basis functions, through the station maps below, less the equilibrium density; an ion $\mu$ row deposits through
its gyroaverage (Section~\ref{sec:gyro}).  The residual splits as $R(f, \phi) = A f + N(f, \phi)$: $A$ is the advection at zero potential
(streaming along $\varphi$, the magnetic drift in the plane, the mirror force) and $N$ every term that reads $\phi$ (the $E \times B$
drift of the state, the parallel electric force, the wall inflow condition's response).  The kinetic stations read the field through
station maps: $E_0 = I$ at a wedge's first plane, $E_1$ at its middle, and $E_2$, the next plane's field carried back along the field
lines through the supermesh.

\textbf{The integrator.}  The step is the additive Runge--Kutta IMEX scheme ARKIMEX3~\cite{KennedyCarpenter2003}, with the tableau of
PETSc's ARKIMEX3~\cite{PETSc2024}.  The electron model sets how the field is integrated.  With adiabatic electrons the field is explicit, solved
from each stage's value; with kinetic electrons it is implicit, part of the state, and Newton's method solves each stage together with
quasineutrality (Section~\ref{sec:kinetic}).  The scheme integrates $\dot y = F_E(y) + F_I(y)$, the state $y$ being the distribution $f$, with the potential
added for kinetic electrons, in four stages, the first explicit and the next three implicit with the same diagonal coefficient
$\gamma = 0.4359$:
\begin{equation}
\begin{aligned}
Y_i &= Z_i + \gamma \Delta t\, F_I(Y_i), \\
Z_i &= y_n + \Delta t \textstyle\sum_{j < i} \bigl(a^E_{ij} F_E(Y_j) + a^I_{ij} F_I(Y_j)\bigr),
\end{aligned}
\label{eq:stage}
\end{equation}
$i = 1, \ldots, 4$.  $Y_i$ is the stage value, the state at stage $i$, with $Y_1 = y_n$, and $Z_i$ is what the earlier stages contribute
to it; the step ends with $y_{n+1} = y_n + \Delta t \sum_j b_j (F_E(Y_j) + F_I(Y_j))$, $b$ the implicit tableau's last row.  The
implicit stage is a linear or nonlinear equation in $Y_i$ at the shift $s = 1/(\gamma \Delta t)$, the same in every stage.  In the runs here
the time step is set by accuracy; the explicit terms' two limits, the wall's (Section~\ref{sec:hard}) and the explicit field's
(Section~\ref{sec:fieldop}), do not bind in these runs: $\Delta t$ is 0.1 to 0.5\,$a/v_{ti}$ in the ITG runs here, 0.5
to 2 in the weak series, and 0.25 and 1 in the nonlinear runs.

\textbf{A low-storage integrator.}  The scheme runs as \texttt{TSKINARKIMEX3}, a time integrator registered as a PETSc TS type.  A diagonal
mass, one on the distribution's rows and zero on the field's, gives the kinetic-electron stages their semi-explicit
form~(\ref{eq:dae}).  As each stage completes, its two slopes, $F_E(Y_i)$ and $F_I(Y_i)$, are added at once into the right side $Z_j$ of every later stage and into
the solution, so no stage's slopes are stored; a stage's own $Z_i$ then holds its implicit slope until the next stage needs it.  The
integrator keeps six state-sized work vectors, against sixteen for PETSc's general ARKIMEX integrator with the same tableau, the
nonlinear solver's residual counted in both, and a seventh for the embedded pair's local error estimate.  At points B and C of the weak
series a state-sized vector holds 0.34\,GB on an ion process, so the ten vectors the integrator does not store would add 3.4\,GB to each
ion process's device, where the busiest A100 holds 29.3\,GB at B.  A nonlinear stage starts from the last stage's value or from an extrapolation of the earlier stages' slopes on the distribution's
rows, whichever has the smaller residual, at two residual evaluations a stage (the predictor guard,
\texttt{-ts\_kinarkimex3\_predictor\_guard}); Newton's tolerance is relative to the residual at the last stage's value, whichever start is taken.  The weak series
uses the guard; the shaped collisional model starts from the last stage's value.

\textbf{Adiabatic electrons: the field explicit.}  The implicit part is the whole advection at zero potential, $F_I(f) = -M^{-1}
A f$, and every term that reads the potential is explicit, as is the collision source: the explicit part is $G(f) = F_E(f) = -M^{-1}
N(f, \phi(f)) + S$, with $\phi(f) = \hat L^{-1} D f$.  Each implicit stage solves one linear system, the advection solve,
\begin{equation}
(s M + A)\, Y_i = s M Z_i,
\label{eq:pathI}
\end{equation}
by GMRES preconditioned with one sweep (Section~\ref{sec:advection}), then the field solve of its value, $\phi_i = \hat L^{-1} D Y_i$
(Section~\ref{sec:field}; every run of this paper factors the field directly), then $G(Y_i)$; the field is also solved once at the end of the step (Figure~\ref{fig:stack}a).

\subsection{The advection solve}
\label{sec:advection}
With adiabatic electrons each implicit stage solves~(\ref{eq:pathI}).  The
operator is applied matrix-free, and PETSc's GMRES~\cite{SaadSchultz1986} solves the system to a relative residual of $10^{-10}$,
restarted every ten iterations and right-preconditioned by one sweep.  The sweep is \texttt{PCKINSWEEP}, a custom preconditioner registered as a PETSc PC type (\texttt{-pc\_type kinsweep}), so the
solvers that use it, this solve and the kinetic-electron stages' solves, compose it through PETSc's options~\cite{BrownComposable2012}.  With adiabatic electrons
the sweep's blocks depend only on the shift $s$, which is the same in every stage, so the preconditioner is one fixed linear operator for
the whole run.
With kinetic electrons the blocks carry the potential's $E \times B$ drift and are rebuilt at the potential every step.  The streaming along $\varphi$ is the part of $A$ that is stiff
for the electrons: at the largest $\vpar$ the toroidal Courant number per wedge is 0.55 for the CBC's ions at $\Delta t = 0.1$, and 61
times that for the electrons, whose grid reaches $6\,v_{te}$.

\begin{figure*}[t]
\centering
\includegraphics[width=\textwidth]{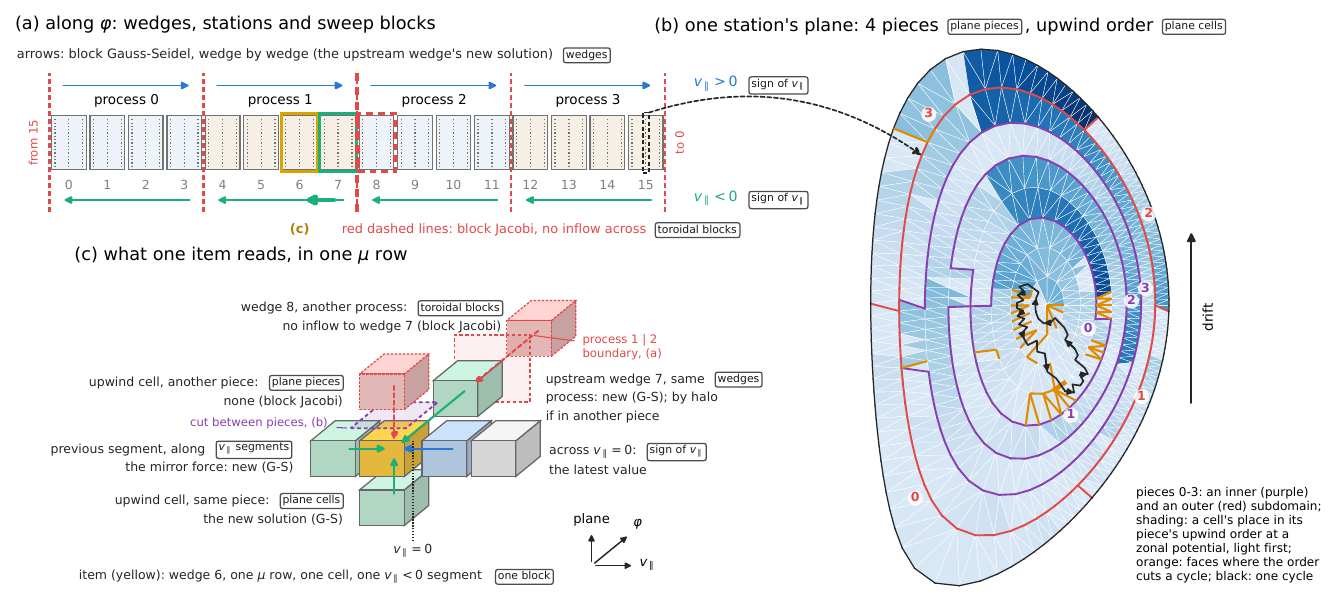}
\caption{The sweep's block Gauss--Seidel and block Jacobi structure; the boxed labels name the levels of Figure~\ref{fig:levels}a.  (a)~Along $\varphi$: 16 wedges on 4 processes, one sweep block of four each, as on the torus of
Figure~\ref{fig:decomp}a.  (b)~The plane of Figure~\ref{fig:decomp}b at one
station: the cut lines and circled numbers are the four pieces, and the shading is the sweep's order in each piece for a potential constant on each flux surface, from light (upwind) to dark (downwind), restarting at each cut; orange faces cut the upwind graph's cycles.  (c)~One item of the device sweep (Section~\ref{sec:advection}), yellow, and what it reads; red marks what
it does not read.}
\label{fig:sweep}
\end{figure*}

\textbf{The sweep.}  The sweep is a block Gauss--Seidel pass along the operator's upwind structure, with block Jacobi wherever that
structure crosses a process boundary (Figure~\ref{fig:sweep}):
\begin{enumerate}
\item no implicit term couples two $\mu$ rows, so each $\mu$ row is swept on its own, with its own blocks and order;
\item a process's wedges form one sweep block with kinetic electrons, and blocks of four consecutive wedges with adiabatic
electrons, eight once the toroidal Courant number reaches 3; the blocks of different processes are swept at once: block Jacobi, which omits the mortar inflow across a block boundary and around the torus;
\item inside a block, the segments with $\vpar > 0$ are swept forward along $\varphi$ and those with $\vpar < 0$ backward, interleaved,
and each wedge takes its mortar inflow from the upstream wedge's new solution;
\item inside a wedge, the cells of a plane piece are visited in a topological order of the in-plane flux's upwind graph, and the faces to
another piece carry no inflow: block Jacobi between the pieces;
\item inside a cell, the $\vpar$ segments are visited in the direction of the mirror force;
\item each block of one wedge, cell, $\vpar$ segment and $\mu$ row, $72 \times 72$ for three $\varphi$ stations $\times$ six plane nodes
$\times$ four $\vpar$ nodes, is solved with a stored inverse.
\end{enumerate}
On the GPUs the inverses are kept in single precision~\cite{Anzt2019}, which halves their memory and leaves the iteration counts
unchanged.  With kinetic electrons the runs drop the ions' coupling between $\vpar$ nodes, the mirror force's, which splits each ion
block into four $18 \times 18$ blocks, a quarter of a $72 \times 72$ block's memory, at the same iteration count; the electrons keep
the $72 \times 72$ blocks.
The sweep sends no messages between toroidal blocks.  Between plane pieces it exchanges, before each wedge after a sweep
block's first, the upstream station of the ghost cells that the wedge reads, so the mortar inflow from another piece is exact.  At zero potential the in-plane flux is the magnetic drift alone, because the wedge's coordinates follow the field line and the parallel
streaming has no component in the plane.  The drift's upwind graph is acyclic on the concentric meshes, and on the shaped ones no flux runs against the order on 23 of the 24 $\mu$ rows, and on the last the couplings against it amount to a
Courant number of $3 \times 10^{-7}$, so item 4 is exact, or nearly,
inside a piece.  With kinetic electrons the order is rebuilt at the potential every step, and the $E \times B$ drift's closed
streamlines, around the eddies and between counter-flowing zonal layers, give the graph cycles.  The order breaks each at the unplaced
cell farthest upwind along the mean in-plane flux, and across the face it cuts the sweep is block Jacobi (Figure~\ref{fig:sweep}b);
Hackbusch and Probst gather a cycle's cells into one block instead~\cite{HackbuschProbst1997}.
Gauss--Seidel with the unknowns numbered along the flow is the robust smoother for
convection-dominated problems~\cite{BeyWittum1997,HackbuschProbst1997}, and point-block Jacobi carries the advection's error only one
cell downstream an iteration.  In a test of our check suite (Section~\ref{sec:discipline}) on the circle at levels 1 to 3, $B_0 = 50$ and 200, the sweep takes four to five GMRES iterations in every case,
where point-block Jacobi in the plane takes six to nineteen, more at each level and as $B_0$ falls.  At $\Delta t = 0.1$ the CBC's advection solve takes 5 and 6 GMRES iterations at levels 4 and 5 on one plane piece and 7 at level 6 on
two; splitting the plane into up to 16 pieces adds at most two.

\textbf{The sweep on the device.}  The sweep is one kernel driven by a list: each item (direction, wedge, $\mu$ row, cell) is a team that solves the
cell's segment blocks, $72 \times 72$, with the stored inverses.  An item joins the earliest launch that its reads allow~\cite{AndersonSaad1989}: after the
cell's upwind neighbours in the plane, after the upstream wedge's mortar inflow and, where the sequential sweep's order requires, after the other
direction's pass over the same cell.  Every read sees the value the sequential sweep's order produces, so the result is bitwise the sequential sweep's.  Against one launch per wavefront of the sequential order, the schedule cut the launches from 484 to 175 and the sweep from 58.7 to 33.2\,ms on one GCD (level 3, sweep blocks of four wedges).
  Issuing the block inverses' loads ahead of their
multiply-adds, tabulating the face coefficients at set-up and caching each team's mortar pieces in scratch memory gave a further
1.74$\times$.  The sweep's critical path is the in-plane wavefront, one dependent launch per wavefront of the upwind order, independent of
how many wedges a process owns.

\textbf{Segment items.}  When a process holds few $\mu$ rows or a small plane piece, its launches hold few items, and a launch with
few items is bound by its fixed cost, not by its work.  The sweep then schedules each item as one segment of a cell rather than all of the cell's segments in
series: an item (direction, wedge, $\mu$ row, cell, segment) waits for the cell's previous segment and for the same segment of the cell's upwind
neighbours, so a launch's dependent chain is one segment long.  The reads and their order are unchanged, so the result is bitwise the
cell-item sweep's.  At one $\mu$ row per process (level 5, 192 GCDs) segment items cut the sweep from 1.69 to 1.00\,s per step; at
12 $\mu$ rows per process, where the launches are full, segment items make the sweep 21\% slower.  The sweep therefore uses segment items whenever the plane is cut into pieces or a process holds three $\mu$ rows or fewer.

\section{The wedge discretization}
\label{sec:hard}
Three results of this section enter the solvers.  The mortar's fluxes are integrated on a static supermesh of each cell's field-line
image and the next plane's cells, so the flux that leaves one wedge enters the next and mass is conserved for any numerical flux.  A
kick at the mortar, the parallel electric field's delta function where the potential jumps, cancels the seam term, the energy error from the potential's jump at the mortar; it reads the potential, so with kinetic electrons it is a term of the stage's Jacobian.  At the wall the Boltzmann inflow condition adds its response to the operator that solves for the potential's toroidal mean, which
makes the relaxation of the wall's zonal potential implicit.

Four parts of the discretization are hard: carrying a constant state exactly through these curved images, the magnetic drift at the
shaped mortar, the wall, and the mortar's seam.

\begin{figure*}[t]
\centering
\includegraphics[width=\textwidth]{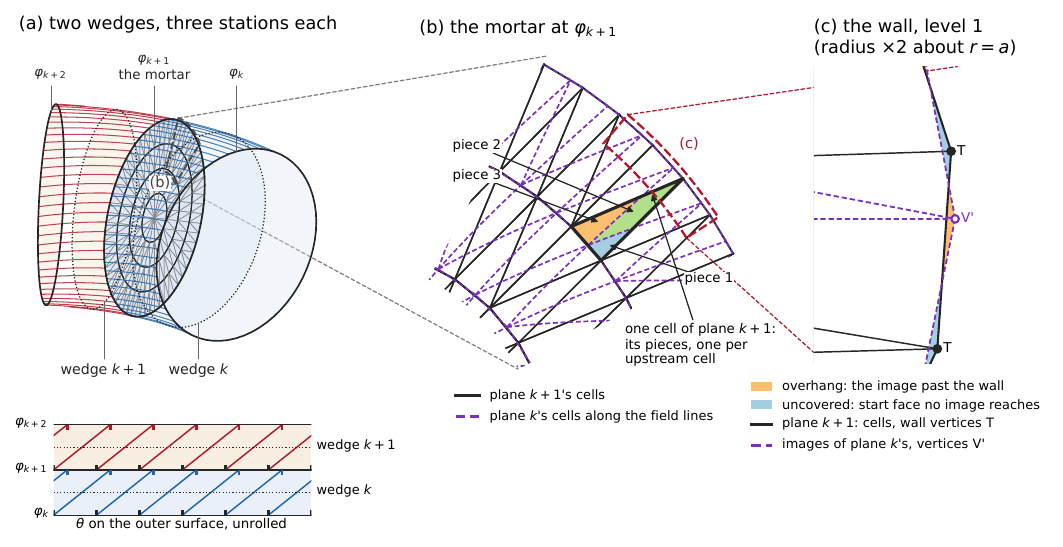}
\caption{The wedge geometry.  (a)~Two wedges of the level-2 circle, each with its three stations drawn as planes (dotted, a middle
station), and the mortar at $\varphi_{k+1}$; blue and red, the wedges' grids on the outer flux surface, along the field lines; the dashed
outline marks panel (b); below, that surface unrolled, where wedge $k$'s lines miss wedge $k+1$'s vertices (black ticks) at the mortar.  (b)~The mortar at $\varphi_{k+1}$, the supermesh pieces of one cell shaded; the
red outline marks panel (c).  (c)~The wall of the level-1 plane, turned so the wall runs up the page: the slivers between plane $k+1$'s
wall polygon and the image of plane $k$'s.}
\label{fig:geometry}
\end{figure*}

\textbf{Constants through curved images.}  A field-line image of a straight cell edge is curved.  Straight chords for the images would
miss 6--8\% of a cell at level 4, a first-order defect.  Quadratic image curves alone shrink the missing area to $10^{-5}$ of a cell's area but still
leave a constant's rate of change at $10^{-4}$.  The discretization therefore satisfies a geometric conservation law~\cite{ThomasLombard1979}: the toroidal component of the field enters as $\tilde B^\varphi = B_\varphi(X_h) \det(D_\xi X_h)$, the exact
closed 2-form pulled back by the discrete map $X_h$, and the supermesh pieces tile the mapped cells exactly, so the missing area is zero by
construction.  $\tilde B^\varphi$ carries $1/R$ through $X_h$ and is not a polynomial, so the streaming preserves constants to
rounding only with a plane rule of degree 14.  The runs use degree 7, where a constant's rate on the circle is $7 \times 10^{-6}$ at
level 3 and $3 \times 10^{-7}$ at level 4, against $10^{-12}$ at degree 14 and a streaming term of order $10^{-2}$; the well-balanced
subtraction, of the discrete operator applied to the equilibrium, removes the equilibrium's share.
The supermesh itself is built by clipping curved edges, with every root bisected to the last bit and every decision about a
shared edge made from data that both cells sharing the edge see, so the tiling closes to rounding.  The step's sums are
compensated~\cite{Kahan1965}, so the mass holds to rounding in floating point.

\textbf{The drift at the shaped mortar.}  On the Solov'ev geometry the flux carries the magnetic drift through the mortar, and the two sides
of a mortar must evaluate the drift consistently.  Interpolating $\tilde b$, the unit vector $\mathbf{b}$ in the wedge's coordinates, at
the nodes would leave a first-order defect in a constant
state, because the interpolant's curl is $O(h)$; the code pulls $\tilde b$ and its curl back pointwise, as for $\tilde B^\varphi$, which leaves
a constant state's rate of change at the end stations at $2.3 \times 10^{-4}$, against $1.4 \times 10^{-2}$ with nodal interpolation.  The remainder is the gradient-$B$ drift: each side has its own
quadratic interpolant of $|B|$, and the end stations' defect converges with the mesh.  $|B|$ cannot be pulled back, because the energy identity requires $|B|$ in the solution's
space, so the remedy is resolution: the shaped geodesic acoustic mode's (GAM's)~\cite{Winsor1968} damping rate differs from the axisymmetric solver's (the code's DG
discretization in $(R, Z, \vpar, \mu)$ for states independent of $\varphi$, with no wedges and no mortar) by $-11.5\%$ at level 2 and
by $+1.2\%$ at level 3, and the well-balanced subtraction (Well-balancing, below) confines the defect to the perturbation.

\textbf{The wall.}  The plane's outer boundary is a polygon inscribed in the last flux surface.  The field-line map rotates the polygon along the
surface, so the polygon's image does not match the next plane's polygon (Figure~\ref{fig:geometry}c): slivers of the image overhang the wall and
slivers of the next plane's start face receive no image.  Flux that leaves the plasma goes to the wall; where flux enters, the inflow
condition prescribes the state outside as $f_0(H_0 + \phi)$, with $H_0 = \vpar^2/2 + \mu B$ the Hamiltonian at zero potential: the
equilibrium at the total energy, a canonical Maxwellian times the
Boltzmann factor of the wall potential; an ion row's state reads the row's own gyroaveraged potential, $f_0(H_0 + (Z_s/\tau_s)
\langle\phi\rangle_\mu)$.  We use this Boltzmann inflow condition because it gives the wall an exact nonlinear Lyapunov
functional (Appendix~\ref{app:wall}).  Its upwind term at the lateral wall sets the wall's explicit time-step limit (Appendix~\ref{app:wall}),
which, among the runs here, binds only the Landau damping test of Appendix~\ref{app:landau}; at the wall potentials of the edge the term must be treated implicitly, for which we
have a Newton design and no implementation.  The inflow condition relaxes the wall potential through the zonal
polarization response, at a rate that grows as $B_0^2$, too fast for an explicit field; the field operator of the toroidal mean, the
potential averaged over the field stations, carries the inflow's linear response to the wall potential, which makes that relaxation
implicit (Section~\ref{sec:field}).  Each sliver is
integrated as its own region (Appendix~\ref{app:wall}).

\textbf{The mortar's seam.}  The potential is continuous inside a wedge but not across a mortar: a wedge's end face reads the next
plane's potential carried back along the field lines, and the next wedge's start face reads that plane's potential itself, so the
discrete Hamiltonian jumps by $[H_h]$ across the mortar.  The flux through the mortar times that jump is the seam term of the energy
identity (Section~\ref{sec:conservation}).  On smooth states the jump is a projection error and the seam converges as $h^4$, but with kinetic electrons it drives a growing grid
mode.  The remedy is the term the jump implies: the parallel electric
field has a delta function at the mortar, which the volume term, built from each wedge's own potential, omits.  The kick is that term,
the shift in $\vpar$ that a particle crossing the jump receives, which keeps its energy; the distribution jumps at the same mortar, so
the term is a nonconservative product~\cite{DalMasoLeFlochMurat1995}.  At each mortar quadrature point the residual adds the kick, the $\vpar$ advection of the upwind trace by the impulse $\sigma_s \tilde B^\varphi [H_h]$ of the row's species, tested on the upwind
side, with central fluxes between $\vpar$ segments.  The kick conserves mass and cancels the seam term in the energy and in the linear
free energy exactly, on any velocity grid; an ion row's kick is built from the jump of its own gyroaveraged potential, which cancels
the row's seam term the same way (Section~\ref{sec:gyro}).  The kick carries no flux through $\vpar = 0$, where a flux would drive a grid mode of its
own.  On the shaped meshes the mortar's flux also carries the
in-plane drift, whose share of the seam the kick does not cancel.  On the shaped meshes with adiabatic electrons the kick is off: it would cancel only the streaming's share of the seam, and without it
a seeded grid mode decays.  With kinetic electrons the kick is on: without it the electrons'
streaming seam drives a mode near the axis, and the two species' remaining drift shares, of opposite sign, cancel to $7 \times 10^{-5}$ of either in a test at a fixed state at the mass ratio 100.
On the circle, where the kick is on with either electron model, it adds 9\% to the adiabatic CBC step at level 3 on the A100.

\textbf{Well-balancing.}  The code evolves the full distribution, not a perturbation about an equilibrium.  The canonical
equilibrium $f_0(P_\varphi, \mathcal{E}, \mu)$~\cite{Angelino2006} is a function of the invariants, the canonical toroidal momentum
$P_\varphi$, the energy $\mathcal{E}$ and $\mu$, so the continuous operator
annihilates $f_0$, but the discrete operator does not annihilate the interpolant $\Pi f_0$: its truncation error would move the
equilibrium itself, a grid-scale relaxation that no physics drives.  The residual therefore subtracts the fixed vector
$R_h(\Pi f_0)$~\cite{Berberich2021}, so that $\Pi f_0$ is an exact discrete steady state and what changes in a run is the perturbation.  Every run of this paper uses the
subtraction; the null test of Appendix~\ref{app:null} holds the unseeded equilibrium to a relative change of $1.5 \times 10^{-15}$ of the state's maximum over 1000 steps.

\section{Kinetic electrons: the stage and the fluid-electron block}
\label{sec:kinetic}
\begin{figure*}[t]
\centering
\includegraphics[width=\textwidth]{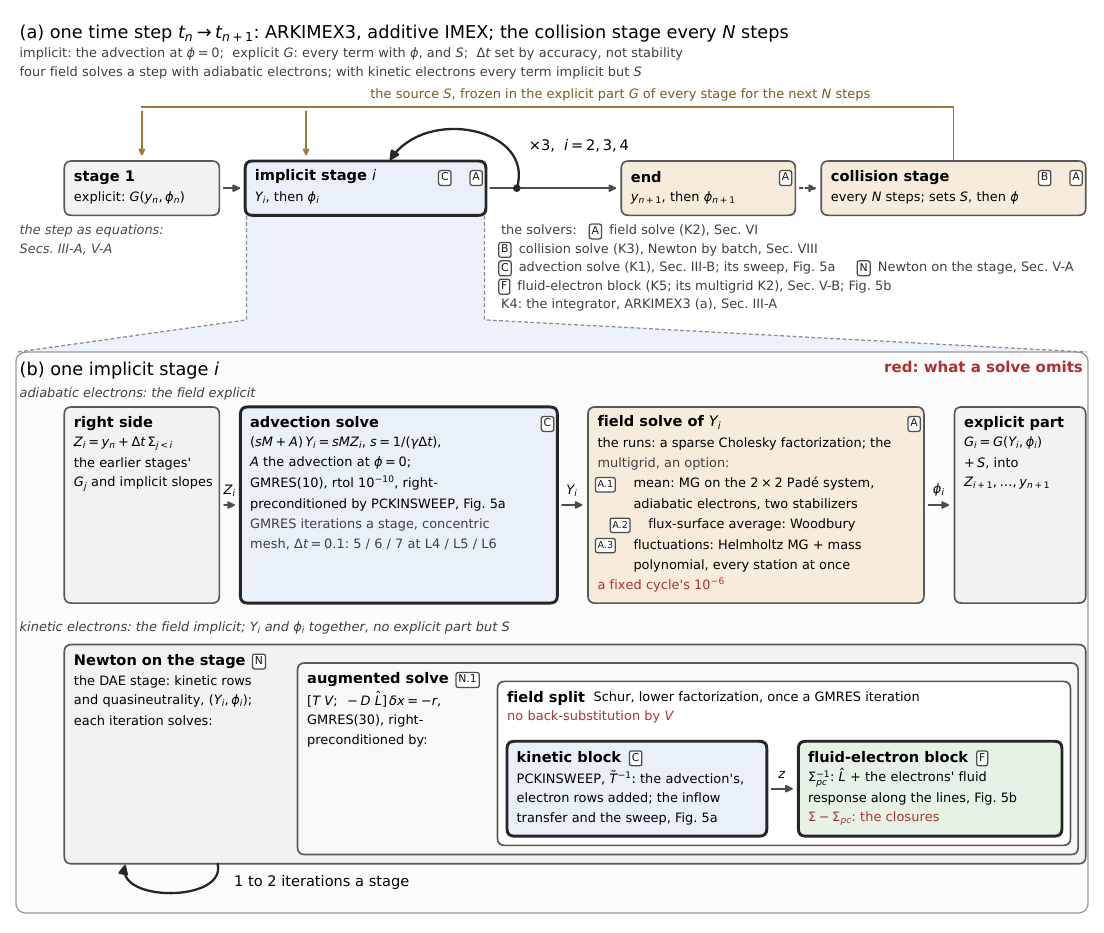}
\caption{One time step and its solvers; (b) zooms into the bold box of (a).  (a)~One step of ARKIMEX3~\cite{KennedyCarpenter2003}.
(b)~One implicit stage, with each electron model.}
\label{fig:stack}
\end{figure*}

\begin{figure*}[t]
\centering
\includegraphics[width=\textwidth]{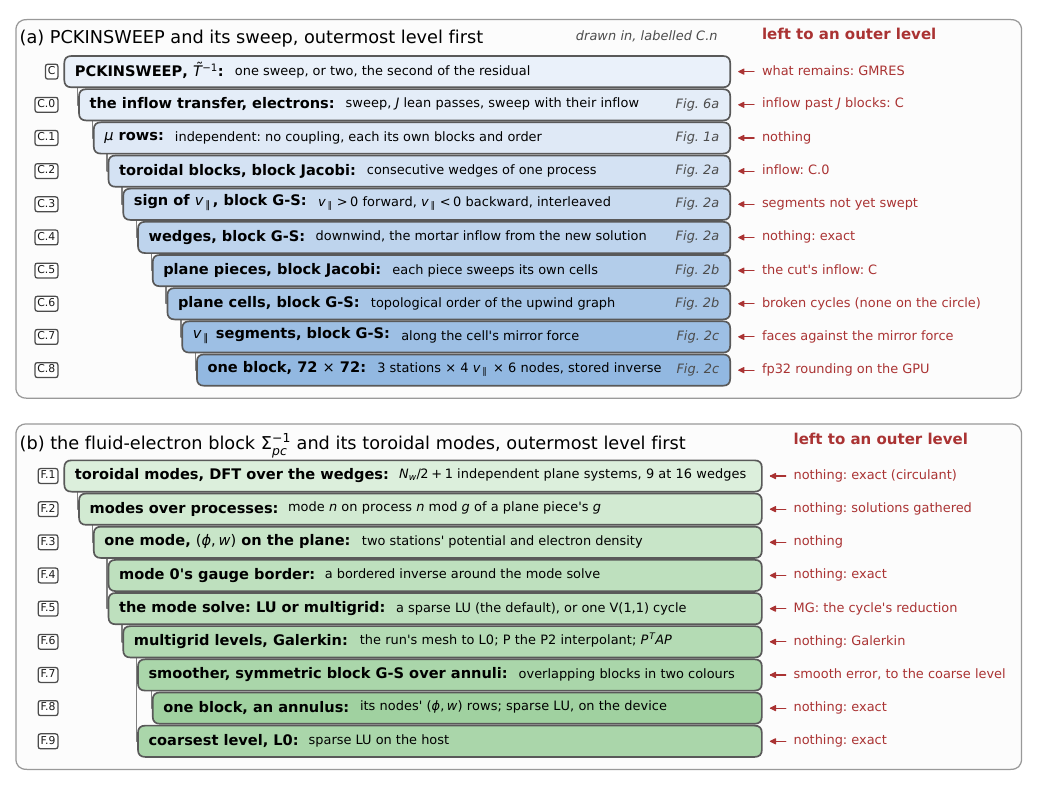}
\caption{The levels of the solvers in Figure~\ref{fig:stack}b, outermost first.  (a)~\texttt{PCKINSWEEP}, the advection solve's
preconditioner and the kinetic block, and its sweep.  (b)~The fluid-electron block.}
\label{fig:levels}
\end{figure*}

Figure~\ref{fig:stack} zooms from one step (a) to one implicit stage with each electron model (b), and Figure~\ref{fig:levels}
gives the levels of the sweep (a) and of the fluid-electron block (b); their letters are those of Table~\ref{tab:solvers}.  The plane multigrids of the fluid-electron block and of the field's two solves, for the toroidal mean and the fluctuations
(Section~\ref{sec:field}), share one hierarchy of plane meshes and one smoother, block Gauss--Seidel over the plane's annuli
(Figure~\ref{fig:annuli}).

\subsection{The stage}
\label{sec:stage}
\textbf{Kinetic electrons: the field implicit.}  The electrons are a second species of $\mu$ rows on their own velocity grid.
Their streaming couples to the potential in a mode, the $\omega_H$ mode~\cite{Lee1987}, whose frequency at the physical mass ratio limits
an explicit field to a step about 180 times shorter than $\Delta t = 0.1$.  So every term of $R$ is implicit and only the collision
source stays explicit.  The field becomes part of the state, $x = (f, \phi, \alpha)$ with $\alpha$ the border's multiplier, and
quasineutrality its algebraic rows~\cite{HairerWanner1996}; the stage value is $X_i = (Y_i, \phi_i, \alpha_i)$, and with $\mathbb{M} =
\mathrm{diag}(I, 0, 0)$ each implicit stage solves
\begin{equation}
\mathcal{F}(X_i) = s\,\mathbb{M}(X_i - Z_i) + \begin{bmatrix} M^{-1} R(Y_i, \phi_i) \\ \hat L \phi_i - D Y_i \end{bmatrix} = 0
\label{eq:dae}
\end{equation}
by Newton's method~\cite{KnollKeyes2004}.  Each Newton step solves the augmented system
\begin{equation}
\begin{bmatrix} T & V \\ -D & \hat L \end{bmatrix}
\begin{bmatrix} \delta f \\ \delta\phi \end{bmatrix} = -\mathcal{F}, \qquad
\begin{aligned} T &= s I + M^{-1} \partial_f R, \\ V &= M^{-1} \partial_\phi R, \end{aligned}
\label{eq:aug}
\end{equation}
both blocks at the Newton iterate: $T$ is the advection with the iterate's $E \times B$ drift and $V$ how a potential moves the state.
Eliminating $\delta f$ leaves the field's Schur complement
\begin{equation}
\Sigma = \hat L + D T^{-1} V = K_{\mathrm{pol}} + Q_i + Q_e,
\label{eq:schur}
\end{equation}
where $K_{\mathrm{pol}}$ is the polarization stiffness (the Pad\'e operator with gyrokinetic ions, Section~\ref{sec:fieldop}) and
$Q_s$ is species $s$'s charge response, within one stage, to a potential.  GMRES(30) solves~(\ref{eq:aug}), right-preconditioned
by a lower block factorization~\cite{MurphyGolubWathen2000,BenziGolubLiesen2005}, PETSc's FieldSplit preconditioner
(\texttt{PCFIELDSPLIT}) with a Schur complement,
\begin{equation}
P^{-1} \begin{bmatrix} r_f \\ r_\phi \end{bmatrix} = \begin{bmatrix} z \\ \Sigma_{\mathrm{pc}}^{-1} (r_\phi + D z) \end{bmatrix},
\qquad z = \tilde T^{-1} r_f,
\label{eq:pc}
\end{equation}
with $\tilde T^{-1}$ the sweep of the advection solve with the electron rows added, applied twice or combined with the inflow transfer
(below), and
$\Sigma_{\mathrm{pc}}$ an approximation of
$\Sigma$ that can be factored, the fluid-electron block of Section~\ref{sec:fluid}.  $Q_e$ is the expensive part: within a stage at
$\Delta t = 0.1$ a thermal electron crosses a tenth of a poloidal turn along the line, so $Q_e$ couples a node to a band of its whole flux surface, and its columns cannot be estimated by graph-coloured probing~\cite{CurtisPowellReid1974} at a cost independent of $h$.  GMRES supplies what the preconditioner
omits: the lower factorization's back-substitution by $V$, the sweep's block Jacobi across processes, and the fluid
closure in place of the kinetic response.

\textbf{Two sweeps.}  $\tilde T^{-1}$ applies the sweep twice, the second time to the residual the
first leaves, which comes from the couplings the first sweep omits, the inflow across the block Jacobi boundaries along $\varphi$ and
between plane pieces among them.  \texttt{PCKINSWEEP} forms that residual without an application of the operator: during the sweep it
stores each block's right side less the couplings the sweep kept, and one pass over the final solution's couplings gives the residual as
the difference (\texttt{-pc\_kinsweep\_sweeps 2}).  Richardson with two iterations on FieldSplit's kinetic block, the sweep its
preconditioner, gives the same GMRES count with one application of $T$.  On the shaped plane, where the residual pass costs as much as an application
of $T$, Richardson's step is 3--6\% shorter at every process layout we tested, and the runs of Section~\ref{sec:utilturb} use it.  On the shaped plane of Section~\ref{sec:turb}, at the mass ratio 100, a second sweep cuts the GMRES iterations per Newton iteration
from 26 to 18 and the step by 12\%; a third cuts them to 16 at the same step time.

\begin{figure}[t]
\centering
\includegraphics[width=\columnwidth]{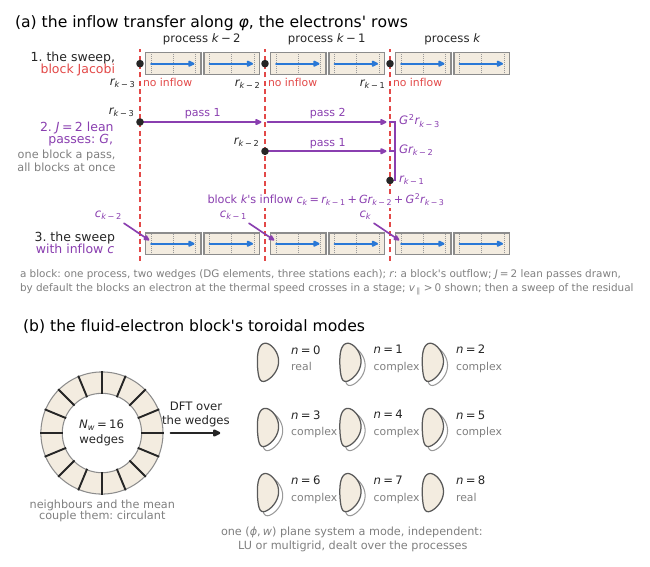}
\caption{The kinetic-electron stage's couplings along $\varphi$.  (a)~The inflow transfer: the sweep, two lean passes, drawn for block $k$'s inflow, and the sweep with the block boundaries' inflow.  (b)~The fluid-electron block's transform over 16 wedges into nine independent plane
systems.}
\label{fig:elecpc}
\end{figure}

\textbf{The inflow transfer.}  The sweep's block Jacobi along $\varphi$ moves information one toroidal block per application of the
preconditioner, but in a stage a fast electron streams through several blocks, and there the GMRES iterations grow: the parallel block
Jacobi of discrete-ordinates transport degrades as the subdomains thin~\cite{YavuzLarsen1992,AnistratovAzmy2015}.  We propose an
approximate transport step for the inflows $c$ at the block boundaries (Figure~\ref{fig:elecpc}a).  They satisfy $c = r + Gc$, where
$r$ is the outflow that the sweep delivers to each boundary and $G$ carries a block's inflow all the way through the block, with no
source, to the next block's inflow; the system couples every block around the torus, $(I - G)\,c = r$.  The transfer truncates its
Neumann series with $J$ iterations of $c \leftarrow r + Gc$ from $c = r$.  By default $J$ is the number of blocks an electron at the
thermal speed crosses in a stage, rounded up, a heuristic: in a stage, information travels about as far as the electrons do.
An application of $G$, a \emph{lean pass}, reuses the $72 \times 72$ block inverses that the sweep stores
(Section~\ref{sec:advection}) and carries only the inflow along $\varphi$, at a fifth to a quarter of a sweep's time.
The preconditioner then applies three sweeps: the sweep, the $J$ lean passes, the sweep with the inflow $c$, and a sweep of the
residual (Appendix~\ref{app:transfer}).
At point B of the weak series, at $\Delta t = 1$, four times the step of the series' base, the transfer cuts the GMRES iterations per
Newton iteration from 23 with two sweeps to 15.  With the transfer, and with collisions every step instead of every fourth, the series
takes that step at 1.9 to 2.6 times the base's cost a step, so it runs 1.5 to 2.0 times as fast per unit of simulated time
(Section~\ref{sec:scaling}).
As the toroidal blocks thin to one wedge, the GMRES iterations of two sweeps triple and those with the transfer stay flat
(Table~\ref{tab:transfer}): the lean passes restore, at a fraction of a sweep's cost, the coupling that the cuts remove, so the code
can cut the torus into the thin blocks that parallelism needs.

\textbf{The kinetic-electron counts.}  Table~\ref{tab:weak} gives the weak series' Newton and GMRES counts at the physical mass ratio,
with collisions every step and the fluid block's modes by the multigrid: 13 GMRES iterations per Newton iteration on one device and
on 16, and 17 on 256, the first point split in the plane.  Newton takes one or two iterations per implicit stage (Table~\ref{tab:weak}).
In the weak series an implicit stage's first linear solve stops at a relative residual of the smaller of $10^{-2}$ and a tenth of
the residual reduction that the previous stage's first Newton iteration reached, so GMRES does not over-solve it.  In the shaped
collisional model the stage's later linear solves stop when their residual reaches 0.3 of Newton's tolerance, which in a test on the
model's plane cut the GMRES iterations by a quarter at the same Newton count and step error.

\subsection{The fluid-electron block}
\label{sec:fluid}
\textbf{The fluid closure.}  $\Sigma_{\mathrm{pc}}$ replaces $Q_e$ by the electrons' fluid response along the field lines at the stage
shift~\cite{ChenChacon2014}, the isothermal two-moment closure of the drift-kinetic equation~\cite{HammettPerkins1990},
\begin{equation}
\Sigma_{\mathrm{pc}} = \hat L + M' X (I + X)^{-1}, \qquad X = (\sigma_e/s)^2\, M'^{-1} D_\parallel,
\label{eq:fluid}
\end{equation}
with $M' = (n_e/\tau_e) M_p$ on every station, $M_p$ the plane's mass matrix, $\sigma_e = v_{te}/v_{ti}$, and $D_\parallel$ the along-line
stiffness of the stations' P2 field in $\varphi$ (each wedge's three stations, through $E_0$, $E_1$, $E_2$).  Under gyrokinetic ions
$\hat L$ keeps the long-wavelength polarization $K_{\mathrm{pol}}$ in place of the Pad\'e form.  The closure is exact in both limits:
the adiabatic response $M'$ when the electrons cross the mode's parallel wavelength within a stage ($X \gg 1$), and the inertial
response $(\sigma_e/s)^2 D_\parallel$ when they do not ($X \ll 1$); between, it differs from the kinetic response $1 + \zeta
Z(\zeta)$~\cite{FriedConte1961} by up to 30\%.  It is a preconditioner only: Newton converges the kinetic equations, and the fluid block
changes the GMRES count, never the solution.  The code does not form $(I + X)^{-1}$; it adds the electrons' density $w$ as an unknown
and solves the sparse system
\begin{equation}
\begin{bmatrix} \hat L + M' & -M' \\ -M' & M' + (\sigma_e/s)^2 D_\parallel \end{bmatrix}
\begin{bmatrix} \phi \\ w \end{bmatrix} = \begin{bmatrix} r \\ 0 \end{bmatrix},
\label{eq:fluidsys}
\end{equation}
whose elimination of $w$ is~(\ref{eq:fluid}).  Each ion species adds a term of the same form at its own thermal speed, $X_s = X (\sigma_s/\sigma_e)^2$ on $M'$ weighted by $Z_s^2 (n_s/\tau_s)/(n_e/\tau_e)$, so that $\Sigma_{\mathrm{pc}}$
also approximates $Q_i$.

\textbf{The block's further terms.}  Three more terms of the kinetic operator are in $\Sigma_{\mathrm{pc}}$: the electrons' response to
the Boltzmann inflow condition at the wall, a diagonal $W_e/s$ on the density rows, with $W_e$ the electrons' charge response through
the wall's inflow term to a uniform unit potential; the station maps $E_1$ and $E_2$ in $D_\parallel$ by $L^2$ projection in place of interpolation, when cuDSS's sparse LU factors the modes; and the whole of $\tilde{\mathbf b} \cdot
\tilde\nabla$ in $D_\parallel$, including its in-plane component, which is nonzero where the wedge's quadratic map departs from the field
line.  With the kinetic block solved exactly, the three terms together cut the GMRES count by 44\% on the shaped plane, most of it
from the inflow condition's response and the projection, and by 15\% on the circle (both at the mass ratio 100); the ions' term cuts
it a further 14\% on the shaped plane.

\textbf{The toroidal modes.}  Every operator in~(\ref{eq:fluidsys}) is the same in every wedge: the equilibrium is axisymmetric, every
plane carries the same mesh, and the field-line map from a plane to the next is the same rotation.  $\hat L$ couples a station only to
the stations of its own wedge and, through the mean, to all of them; $D_\parallel$ couples a wedge's stations to the next wedge's first.
So~(\ref{eq:fluidsys}) is block circulant in the wedge index~\cite{Davis1979}, and the discrete Fourier transform over the wedges,
\begin{equation}
\hat x_n = \sum_{w=0}^{N_w-1} x_w\, e^{-2\pi i n w / N_w}, \qquad \hat\Sigma_n = \sum_{d} \Sigma_d\, e^{-2\pi i n d / N_w},
\label{eq:dft}
\end{equation}
makes it block diagonal (Figure~\ref{fig:elecpc}b): one system per toroidal mode $n$, with $\Sigma_d$ the coupling between wedges $d$ apart ($d = -1, 0, 1$).  The
data are real, so $\hat x_{N_w - n} = \overline{\hat x_n}$ and the modes $n = 0, \ldots, N_w/2$ suffice: nine at $N_w = 16$.  Modes
$0$ and $N_w/2$ are real; the other seven are complex and stored as real systems of twice the size~\cite{DayHeroux2001}, $[\mathrm{Re}\,{-\mathrm{Im}};
\mathrm{Im}\ \mathrm{Re}]$.  Mode 0 alone carries the mean's operator $\bar\Lambda$ of~(\ref{eq:Lop}) and the gauge border.  A mode's unknowns are the two stations'
potential and density on the plane: $4 n_p$ for a real mode (plus the border at $n = 0$) and $8 n_p$ for a complex one.  The shift and $\sigma_e$ do not change, so each mode is set up once per run.

The nine systems are independent, and they are distributed round-robin over the processes that hold the same plane piece, its toroidal and $\mu$ blocks: of
those $g$ processes, the $r$-th owns the modes $n$ with $n \bmod g = r$, sets them up and solves them, and the solutions are gathered back to
all $g$.  One application of $\Sigma_{\mathrm{pc}}^{-1}$, once per GMRES iteration, gathers the field, forms the owned modes' right
sides, solves them, gathers the solutions, and transforms back.

\begin{figure*}[t]
\centering
\includegraphics[width=\textwidth]{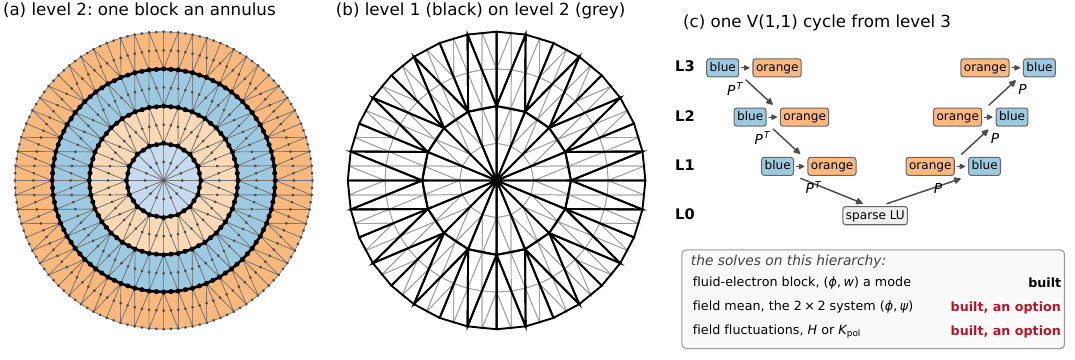}
\caption{The plane multigrid's hierarchy and smoother.  (a)~The level-2 plane cut into the smoother's blocks, one annulus
each: heavy lines and large dots are the nodes two blocks share, blue and orange the two colours.  (b)~The level-1 mesh (black) on the
level-2 mesh (grey), not nested.  (c)~One V(1,1) cycle from level 3, and the solves that use the hierarchy.}
\label{fig:annuli}
\end{figure*}

\textbf{The mode solve.}  Each mode is solved by a sparse LU, the default (cuDSS~\cite{cuDSS} on the A100, PETSc's Kokkos LU with a supernodal
triangular solve on the MI250X), or by one V(1,1) cycle of geometric multigrid~\cite{Brandt1977}, whose set-up and memory grow in
proportion to the mesh while the LU's fill grows faster~\cite{George1973} (Table~\ref{tab:mg}); the weak series uses the multigrid,
the shaped collisional model the LU.  On an earlier version of the code (Table~\ref{tab:mg}b) the worst mode's convergence factor per cycle is 0.79 at the physical mass ratio, against 0.17 at the ratio 100, yet one cycle
takes as many GMRES iterations as the sparse LU at both ratios.  With either mode solve the GMRES iterations rise from 12 at the ratio
100 to 20 at the physical ratio, so the rise comes from the fluid closure, not from the multigrid.  The multigrid's levels are the run's plane mesh and its
coarser refinement levels down to level 0 (Figure~\ref{fig:annuli}), the prolongation the coarse P2 interpolant at the fine nodes (the
identity over a mode's components), and the coarse operators Galerkin, $P^T A P$.  On every level above level 0 the smoother is a
symmetric block Gauss--Seidel over blocks of whole annuli, a block's nodes those of its annulus' cells, so adjacent blocks share the
nodes of one flux surface (Figure~\ref{fig:annuli}a).  The blocks alternate in two colours, which the pre-smoother runs in order and the
post-smoother in reverse (Figure~\ref{fig:annuli}c); each block is solved by a sparse LU, and level 0 is a sparse LU.  Two blocks of one colour share no cell, so a colour's blocks are solved at once against the same residual, which for the field
operators is the same as solving them in turn; the fluid block's station maps reach across an annulus and couple blocks of one colour,
and that coupling enters as block Jacobi.  Levels 1 and 0 run on the host, where they are faster, and the finer levels on the GPU.
Mode 0's gauge border is inside its LU, or around the cycle as a bordered inverse.  Under the multigrid the mode solve is one cycle,
and GMRES also supplies the error that cycle leaves.

\textbf{The fluid-electron block over plane pieces.}  When the plane is cut into pieces (Section~\ref{sec:parallel}), the sweep
treats the pieces as block Jacobi, but the fluid-electron block does not inherit the cut.  Its along-line stiffness $D_\parallel$ is
assembled over the whole plane, and each mode's multigrid deals whole annulus blocks to the processes of the pieces, so its smoother
is the same at every piece count: the cycle's convergence factor does not change to the printed digits.  The cut lets one complex
mode at level 5, which a single 40\,GB A100 cannot hold, run over four pieces.

\section{The field solves}
\label{sec:field}
Quasineutrality is a continuous Galerkin solve, C0 P2 on the plane, at each field station, where the DG wedges' end stations share the
field: one system for the toroidal mean, the potential's average over $\varphi$, its axisymmetric part, and one for the fluctuations
about it.  With gyrokinetic ions each is the Pad\'e block system in the potential
and an auxiliary potential, sparse, two unknowns a node.  Every run of this paper factors these systems by a sparse Cholesky method, replicated on each plane piece, and the replicated
factorization does not scale: at level 6 it no longer fits a GCD (Table~\ref{tab:mg}c).  The field's path to scale is a geometric
multigrid over flux-surface annuli, distributed over the pieces and built as an option (Section~\ref{sec:fieldmg}).  In two nonlinear runs it reproduces the factorization's Newton and GMRES counts and its state (Section~\ref{sec:fieldmg}).

\subsection{The operators}
\label{sec:fieldop}

\textbf{The field as run.}  The field operator acts station by
station about the toroidal mean $\bar\phi = \sum_F W_F \phi_F / \sum_F W_F$, with weights $W_F = \Delta\varphi/3$ at a plane and
$2\Delta\varphi/3$ at a middle (the C0 P2 quadrature along $\varphi$):
\begin{equation}
(L\phi)_F = W_F \bigl[\Lambda_0 (\phi_F - \bar\phi) + \bar\Lambda \bar\phi\bigr].
\label{eq:Lop}
\end{equation}
Summing~(\ref{eq:Lop}) over the stations leaves the mean's equation, $\bar\Lambda \bar\phi = \sum_F l_F / \sum_F W_F$ with $l_F$ station $F$'s load, and
subtracting it leaves one plane operator $\Lambda_0$ for every station's fluctuation, one right side a station.
With adiabatic electrons, $\Lambda_0 = K_{\mathrm{pol}} + M_{n_0}$, the polarization stiffness $K_{\mathrm{pol}} = \tau \int n_0
\nabla_\perp N_i \cdot \nabla_\perp N_j / B^2$ plus the adiabatic response on the density-weighted mass matrix, the field equation being
quasineutrality multiplied by $\tau = T_e/T_i$; and the mean's $\bar\Lambda$ adds two stabilizers of the explicit coupling: a parallel-polarization term (below), which damps the grid-scale ion-acoustic mode, and the wall inflow condition's response.  The mean can also carry
the flux-surface-averaged electron response~\cite{Hammett1993,DorlandHammett1993,AdamsHPC2026}, which subtracts the adiabatic response's
flux-surface average, a rank-$K$ term for $K$ surfaces.  The drift-kinetic ITG verification runs use plain adiabatic electrons, as
their semi-Lagrangian reference does; the other adiabatic runs, the gyrokinetic CBC rows among them, carry the response.  The CBC comparisons can mix the two because the seeded $n = 5$ mode has no flux-surface average: at a small seed its growth rate
and frequency do not change with the response.
With kinetic electrons the electrons supply the response themselves, $\Lambda_0 = K_{\mathrm{pol}}$, and a border row fixes the one
constant the operator does not see.  With gyrokinetic ions $K_{\mathrm{pol}}$ is replaced by the Pad\'e operator $P$, solved through the
block system~(\ref{eq:pade}), and $R$ reads each ion row's potential through its gyroaverage $J_r$ (Section~\ref{sec:gyro}).  Each field operator, the mean's and the
fluctuations', is factored once by a sparse Cholesky method: cuDSS~\cite{cuDSS} on the A100, and on the MI250X PETSc's Kokkos Cholesky,
whose triangular solves we replace by a supernodal solve~\cite{Yamazaki2020}: two launches per height of the supernodal elimination tree, 46 at mesh level 6, against one launch per level set in Kokkos
Kernels' triangular solve~\cite{KokkosKernels2021}, 1782.  Every plane piece factors and solves
the whole plane's field (Section~\ref{sec:parallel}), so the factorization's cost per process grows with the plane at fixed work per
device.

\textbf{The polarization.}  The polarization operator is the Pad\'e form~\cite{Jolliet2007,Grandgirard2016},
$P = \tau\,[M_{n_0} - M_{n_0}(M_{n_0} + K_{\mathrm{pol}}/\tau)^{-1}M_{n_0}]$, with $M_{n_0}$ the density-weighted mass matrix and
$K_{\mathrm{pol}}$ the polarization stiffness.  Its symbol is
$\tau n_0 b/(1 + b)$, $b = k_\perp^2\rho_i^2$, against the exact $1 - \Gamma_0(b)$, 1.7\% under it at the CBC's $k_\perp\rho_i = 0.28$, and
its limit as $\rho_i \to 0$ is $K_{\mathrm{pol}}$, the drift-kinetic field's long-wavelength operator.  $P$ is symmetric positive
semidefinite and $P\mathbf{1} = 0$ by construction.  The field solves it as one sparse block system of $K_{\mathrm{pol}}$'s sparsity,
\begin{equation}
\begin{bmatrix} \tau M_{n_0} + \Theta & -\tau M_{n_0} \\ -\tau M_{n_0} & \tau M_{n_0} + K_{\mathrm{pol}} \end{bmatrix}
\begin{bmatrix} \phi \\ \psi \end{bmatrix} = \begin{bmatrix} l \\ 0 \end{bmatrix},
\label{eq:pade}
\end{equation}
with $\psi$ an auxiliary potential and $l$ the charge load, whose Schur complement in $\phi$ is $P + \Theta$, with $\Theta$ the adiabatic, flux-surface-averaged and wall terms above;
the system is factored once, as the drift-kinetic operator is.  The fluid-electron block, which preconditions the kinetic-electron stages, keeps $K_{\mathrm{pol}}$ in its
approximate field operator (Section~\ref{sec:fluid}).

\textbf{The explicit field's stiff modes.}  With the parallel streaming implicit and the force from the potential explicit, the pair forms
a grid-scale ion-acoustic loop, the ions' counterpart of the electrons' $\omega_H$ mode~\cite{Lee1987}, unstable at $\Delta t = 0.1$
for $B_0 = 200$.  A parallel-polarization term $\kappa\,\Delta t^2 \tau \int n_0
(\mathbf{b} \cdot \nabla N_i)(\mathbf{b} \cdot \nabla N_j)$ added to the mean's field operator bounds the loop's $\omega \Delta t \le \kappa^{-1/2}$;
the runs use $\kappa = 0.19$.

The rest of this section is the multigrid that replaces the factorizations, built as an option.

\subsection{The multigrid}
\label{sec:fieldmg}
\begin{table}[t]
\centering
\caption{The plane multigrid as built.  (a)~The field solves' convergence factor per V(1,1) cycle, the residual's mean reduction per cycle over the last 10 of 20 cycles,
at levels 3 to 5 of the circle's isotropic-triangle meshes (L3a--L5a), $B_0 = 70$, $\Delta t = 0.25$: $H$ and $K_{\mathrm{pol}}$ the
fluctuations' operators, the mean the toroidal mean's block system.  (b)~The fluid-electron block's mode
solve, the CBC at level 4 with kinetic electrons, $\Delta t = 0.1$, on 64 GCDs.  (c)~The field solve at scale, kinetic electrons on the
isotropic-triangle meshes with 48 wedges, on 48 GCDs (L5a) and 192 (L6a).}
\label{tab:mg}
\footnotesize
\setlength{\tabcolsep}{4pt}
\begin{tabular}{@{}lrrrrrr@{}}
\toprule
(a) & \multicolumn{3}{c}{annulus} & \multicolumn{3}{c}{Jacobi--Chebyshev} \\
\cmidrule(lr){2-4}\cmidrule(l){5-7}
operator & L3a & L4a & L5a & L3a & L4a & L5a \\
\midrule
$H$, adiabatic & 0.006 & 0.032 & 0.106 & 0.084 & 0.087 & 0.089 \\
mean, adiabatic & 0.031 & 0.031 & 0.105 & 0.087 & 0.086 & 0.114 \\
$K_{\mathrm{pol}}$, kinetic e$^-$ & 0.108 & 0.127 & 0.135 & 0.106 & 0.118 & 0.145 \\
mean, kinetic e$^-$ & 0.111 & 0.165 & 0.284 & 0.148 & 0.185 & 0.355 \\
\quad two-grid, L4a exact & & & 0.170 & & & \\
\quad W-cycle & & & 0.177 & & & \\
\bottomrule
\end{tabular}

\smallskip
\begin{tabular}{@{}lrr@{}}
\toprule
(b) & sparse LU & multigrid \\
\midrule
factors' nonzeros over the mode matrix's & 26.6$\times$ & 5.8$\times$ \\
set-up, the slowest process (s) & 707.5 & 21.7 \\
a mode solve on the GCD (ms) & 10.8 & 4.8 \\
the block's share of the step & & 3\% \\
worst mode's factor, $m_i/m_e$ 100; 3670.5 & & 0.17; 0.79 \\
GMRES a Newton iteration, 100; 3670.5 & 12.0; 20.2 & 11.9; 19.9 \\
\bottomrule
\end{tabular}

\smallskip
\begin{tabular}{@{}lrrrr@{}}
\toprule
(c) & \multicolumn{2}{c}{a field solve (ms)} & \multicolumn{2}{c}{the field a step (s)} \\
\cmidrule(lr){2-3}\cmidrule(l){4-5}
field solve & L5a & L6a & L5a & L6a \\
\midrule
replicated, sparse Cholesky & 79 & & 2.77 & out of memory \\
replicated, multigrid & 265 & 693 & 9.22 & 40.47 \\
distributed, multigrid & 204 & 232 & 7.09 & 13.55 \\
\bottomrule
\end{tabular}
\end{table}

\textbf{The fluctuations: a Helmholtz solve and a mass solve.}  For the fluctuations $\Theta$ is the adiabatic response alone, $\Theta = c\,M_{n_0}$,
a multiple of the density-weighted mass matrix that $P$ is built from, with $c = 0$ for kinetic electrons.  Then the Schur complement of
the block system~(\ref{eq:pade}), $S = \Theta + P$, factors as $S = \tau M_{n_0} X H$ with $X = (\tau M_{n_0} + K_{\mathrm{pol}})^{-1}$, and
\begin{equation}
H = c\,M_{n_0} + (1 + c/\tau)\,K_{\mathrm{pol}}, \qquad S^{-1} = \frac{\tau H^{-1} + M_{n_0}^{-1}}{\tau + c}.
\label{eq:identity}
\end{equation}
$H$ is a Helmholtz operator, a mass plus a stiffness, and the identity is exact: it follows from $P = \tau M_{n_0} X K_{\mathrm{pol}}$
and the partial fraction $K_{\mathrm{pol}} = \tau(H - c M_{n_0})/(\tau + c)$.  The equivalent product form, $S^{-1} = H^{-1}(I +
K_{\mathrm{pol}} M_{n_0}^{-1}/\tau)$, is the discrete counterpart of multiplying the Pad\'e quasineutrality condition by $1 -
\rho_i^2\nabla_\perp^2$, which turns it into a Helmholtz equation.  The multigrid solves the sum form: with $H^{-1}$ and $M_{n_0}^{-1}$
replaced by symmetric polynomials of fixed degree, the sum stays symmetric, which the energy identity requires
(Section~\ref{sec:gyro}), and the product does not.  It matches the block system's Cholesky solve to rounding
(Appendix~\ref{app:fieldmg}).  A fluctuation solve is therefore one multigrid solve with $H$ and one
mass polynomial (Section~\ref{sec:apply}), on all the field stations at once.  With kinetic electrons $H = K_{\mathrm{pol}}$ is singular:
its multigrid carries the constant null space, which the prolongation reproduces exactly, and the gauge is fixed after the solve.

\textbf{The mean: a $2 \times 2$ block multigrid.}  The mean's $\Theta$ also carries the stabilizers, $\Theta = c\,M_{n_0} + S_\parallel + W$ with
$S_\parallel$ the parallel-polarization term and $W$ the wall term ($\Theta = S_\parallel + W$ with kinetic electrons), and the identity~(\ref{eq:identity}) has
no sparse form for it.  The multigrid therefore solves the mean's block system~(\ref{eq:pade})
itself.  The system is symmetric positive definite, sparse, with four blocks of $K_{\mathrm{pol}}$'s sparsity, and exact for any $\Theta$; it has the structure of the fluid
block's~(\ref{eq:fluidsys}), two unknowns a node, so it runs on the fluid block's multigrid with $(\phi, \psi)$ in place of $(\phi, w)$.
The mean is one right side per field solve, against one per field station for the fluctuations.

\textbf{The flux-surface average.}  The flux-surface-averaged response stays the rank-$K$ Woodbury correction outside the solve,
$\phi = S^{-1} l + G C^{-1} G^T l$, with $G = S^{-1}U$, the capacitance $C = D_E - U^T G$, $U = M_{n_0}E$, $D_E = E^T M_{n_0}E$ and $E$
the indicator functions of the $K$ surfaces.  Formed from an inexact solve, $G$ and $C$ would amplify the
solve's error in the zonal response (Appendix~\ref{app:fieldmg}), so the multigrid's path forms them once, at set-up, with the cycle
iterated to $10^{-12}$, and applies $G C^{-1} G^T l$ per solve.

\textbf{A fixed linear cycle.}  Each multigrid solve is a fixed number $k$ of Chebyshev iterations preconditioned by a symmetric V(1,1) cycle $B$,
started from zero, which is the polynomial $p(BS)B$ in the operator: linear in the load, and symmetric for symmetric $B$ and $S$.  A
tolerance-based stop makes the solve nonlinear in the load, and a nonzero initial guess makes it affine; either breaks the symmetry.
With the fixed cycle the field operator is symmetric and static, so the energy identity of Section~\ref{sec:conservation} holds for it
exactly.  The solve takes five Chebyshev iterations, one cycle each: a relative error of $7 \times 10^{-8}$ for a cycle's convergence factor
of 0.12, about Table~\ref{tab:mg}a's largest but the kinetic-electron mean's, and $7 \times 10^{-6}$ at that mean's 0.28 at level 5.

\textbf{The smoother.}  The parallel-polarization term is anisotropic, strong along the flux surfaces, and grows against
$K_{\mathrm{pol}}$ as $B_0^2$.  The default smoother, the one-annulus block Gauss--Seidel of Figure~\ref{fig:annuli}, holds against
that anisotropy in the design's test (Appendix~\ref{app:fieldmg}).  As built, its V(1,1) factors on the isotropic-triangle meshes stay at or below 0.135 from level 3 to level 5 for the Helmholtz
operator, the adiabatic mean and $K_{\mathrm{pol}}$, and the kinetic-electron mean's grows from 0.11 at
level 3 to 0.28 at level 5, set by the strength of its wall term $W$; a W-cycle holds it at 0.18 at level 5 (Table~\ref{tab:mg}).
Point Jacobi with Chebyshev acceleration~\cite{Adams2003}, which needs only matrix--vector products on the device, is an option per
solve, and each cycle's factor is measured and printed at set-up.
On the device the cycle is bound by its launch count, 368 launches a V-cycle, 224 of them the annulus blocks' supernodal triangular solves, and takes six times the host's cycle at
level 4, so the cycle runs on the host.  A batched banded solve of the annulus blocks, one launch per colour per level, would remove that
count; it is not built, since at level 6 a faster field solve would shorten the step by at most 5\%.

\textbf{Hierarchy and distribution.}  The field multigrids use the fluid block's hierarchy (Section~\ref{sec:fluid}).  The field gets its own partition into whole-ring annuli over the plane pieces, reached by a scatter from the plane pieces of the kinetic equation (Section~\ref{sec:parallel}), so no
piece holds the whole plane's field; the gyroaverage's operator $B_r$ is then held by piece, its rows for the owned nodes with a halo of
the largest Larmor radius, which shrinks with $\rho^*$.  With kinetic electrons on the isotropic-triangle meshes, from level 5 to level 6,
four times the plane over twice the pieces, the distributed solve's time grows 14\%, the replicated multigrid's 2.6 times, and the
replicated factorization no longer fits a GCD (Table~\ref{tab:mg}).  At level 5a the distributed multigrid's field solve takes 204\,ms against the factorization's 79\,ms (Table~\ref{tab:mg}c).
\textbf{As built.}  Over the first 40 and 20 steps of two nonlinear runs at level 3, with adiabatic electrons and with kinetic electrons, the multigrid,
as the preconditioner of a fixed number of Chebyshev iterations with its cycle on the host, reproduces the factorization's Newton and GMRES counts and the state to $4 \times 10^{-11}$ with the annulus smoother, and lengthens the step by 4\%
and 0.5\%.  Over four and eight plane pieces it reproduces one piece's load to rounding.

\subsection{The mass solve and the Pad\'e operator's application}
\label{sec:apply}
\textbf{The mass solve.}  The plane's mass inverses, in the gyroaverage's push $J_r = M_p^{-1}B_r$ and deposit and in the fluctuations'
identity~(\ref{eq:identity}), become a Chebyshev polynomial $p(M_p)$ on the Jacobi-scaled P2 mass matrix, whose condition number, 5.3 at
level 2, is bounded independently of $h$~\cite{Wathen1987}, so degree 26 reaches a relative error of $10^{-10}$.  $p$ is symmetric, so the
push and the deposit stay an exact transpose pair, and $J_r\mathbf{1} = \mathbf{1}$ holds to the polynomial's error.

\textbf{The Pad\'e operator's application.}  With kinetic electrons the field rows of the augmented system~(\ref{eq:aug}) apply $\hat L$, and with
gyrokinetic ions its Pad\'e part contains $(\tau M_{n_0} + K_{\mathrm{pol}})^{-1}$: the runs here apply it by a solve with that block's
Cholesky factors, one per field station per GMRES iteration, which under multigrid would become a cycle per iteration.  With the
multigrid the stages instead carry the auxiliary potential $\psi$ of~(\ref{eq:pade}) as an unknown of the stage system, so that the field rows are the two
sparse block rows of~(\ref{eq:pade}), $(\tau M_{n_0} + \Theta)\phi - \tau M_{n_0}\psi - D f$ and $-\tau M_{n_0}\phi + (\tau M_{n_0} +
K_{\mathrm{pol}})\psi$, with no inner solve.  Under the multigrid the fluid-electron block's modes carry $\psi$ as well, three unknowns a node; on the level-1 circle at $B_0 = 20$
with kinetic electrons that takes 35\% fewer GMRES iterations per Newton iteration than keeping $K_{\mathrm{pol}}$ in the block, as
the runs' block does, and leaving $\psi$ to GMRES.

\section{The gyroaverage}
\label{sec:gyro}
The gyroaverage of each ion $\mu$ row is the Galerkin form of the ring average on the C0 P2 plane, integrated exactly by clipping each
cell against the translated cells, and its charge deposit is the exact transpose of its push.  Mass, the energy identity and the seam
kick's cancellation therefore hold with it as without it.  The gyroaverage adds two mass solves a row and station to each residual and,
with kinetic electrons, 2 to 12\% more GMRES iterations on the weak series (Table~\ref{tab:gyrocost}).

For an ion $\mu$ row $r$ the gyroaverage of the potential at a gyrocentre $x$ is the ring average
$\langle\phi\rangle_r(x) = (2\pi)^{-1}\oint \phi(x + \rho_r \mathbf{e}(\alpha))\,\dd\alpha$, with $\rho_r$ the row's Larmor radius and
$\mathbf{e}(\alpha)$ the unit vector at the angle $\alpha$ in the poloidal plane.  The ring is sampled at $N_r = \max(4,
2\lceil \pi\rho_r/h\rceil)$ points, with $h$ the poloidal cell size, one in each cell the ring crosses, as particle codes grow the count
with the Larmor radius~\cite{Hatzky2002}.  The average enters the ions twice: the row's
Hamiltonian reads $\langle\phi\rangle_r$ (the push), and the row's density enters quasineutrality averaged over the same ring (the
deposit).  The energy identity needs the deposit to be the exact adjoint of the push~\cite{ScottSmirnov2010,Gorler2011}.  This section
builds that pair for the C0 P2 potential on the unstructured plane; the polarization operator that goes with it is the Pad\'e form
of Section~\ref{sec:fieldop}.

\textbf{The Galerkin ring form.}  Sampling the P2 interpolant at each node's ring points gives an accurate push, but its transpose, the
deposit of Lee's four-point average~\cite{Lee1987}, scatters point charges into the vertex entries of the load, whose P2 basis functions
have zero integral: the deposited load of a smooth density is 5--15\% off its Galerkin load, and the double average it implies is
indefinite.  On the linear and spline bases of particle codes~\cite{Hatzky2002,Ku2009} that defect does not arise.  The code instead discretizes the ring average's bilinear form,
\begin{equation}
B_{r,ij} = \frac{1}{N_r}\sum_{q=1}^{N_r} \int_\Omega N_i(x)\,N_j(x + \rho_r \mathbf{e}_q)\,\dd A,
\label{eq:ringform}
\end{equation}
with $N_i$ the P2 basis, and takes the push as the L2 projection of the ring-averaged potential, $J_r = M_p^{-1}B_r$, and the deposit as
its exact adjoint, $J_r^T = B_r^T M_p^{-1}$, with $M_p$ the plane's P2 mass matrix.  The integral is exact: the overlap of a cell with a
translated cell is a convex polygon, found by Sutherland--Hodgman clipping~\cite{SutherlandHodgman1974}, and a degree-4 rule on the
polygon's fan triangles integrates the product of two P2 functions exactly.  The Larmor radius varies over the plane as $B^{-1/2}$, so
the clipping takes $\rho_r$ at each cell's centroid, and the translated cells stay straight triangles.  Each row of $B_r$ is the ring
average of the partition of unity, so $B_r \mathbf{1} = M_p\mathbf{1}$ at every node, the wall's included: the push of a constant is that
constant, and the deposit conserves the ions' charge.  We measured $B_r\mathbf{1} = M_p\mathbf{1}$ to $5 \times 10^{-15}$ and the pair's
adjointness to $3 \times 10^{-16}$; away from the wall $M_p^{-1}B_r$ is an L2 contraction.  The cells' own
quadrature rule is no substitute for the clipping, since the translated basis functions have kinks inside a cell: that assembly is 74--80\%
asymmetric and not a contraction.

\textbf{Where it acts.}  The potential lives on the field stations, two a wedge, each a copy of the poloidal plane, so the ring's
translations are rigid and one $B_r$ per ion $\mu$ row serves every station of every wedge.  The row's potential enters every site its
Hamiltonian enters: the drifts, the parallel force, the seam kick's jump and the wall's inflow state.

\textbf{What stays exact.}  The energy identity of Section~\ref{sec:conservation} holds with the gyroaverage for any linear $J_r$ whose
exact transpose is the deposit, provided the same $J_r$ acts at every site the row reads the potential and the field operator is
symmetric and static: the volume term vanishes pointwise, the lateral faces telescope, since $J_r\phi$ is a C0 P2 function on the station,
and the kick built from the row's own jump cancels the mortar's seam.  A positive semidefinite field operator, as the Pad\'e form is,
makes the free energy decrease at flat profiles; the nodal sampler's indefinite double average is excluded by this, not only by its
accuracy.  The Boltzmann inflow condition keeps its Lyapunov functional with the row's own inflow state, and mass, the well-balanced
equilibrium and constants do not involve the potential.  The approximations are in the operator, not in the identities: the ring's
quadrature, the P2 projection of the ring average, the Pad\'e polarization, and the ring taken in the poloidal plane rather than normal to $\mathbf b$ (Section~\ref{sec:limitations}).

\section{Collisions and the batched block-tridiagonal solve}
\label{sec:collisions}
The collision stage applies nonlinear Landau collisions between every pair of species at each spatial quadrature point, on the velocity
grids of every species there, a batch.  It steps backward Euler by a Newton iteration with the Picard Jacobian, the operator with its
coefficients frozen at the iterate, and each iteration solves one
block-tridiagonal system along $\vpar$ a species and batch, by direct elimination with one GPU thread team a system.  The stage runs
every $N$ steps, $N$ from 1 to 5 in the runs here, and leaves its rate as a source frozen in the explicit part until the next stage, so the
collision operator enters no implicit stage's system.

Collisions are the nonlinear Landau operator~\cite{Landau1936} between every pair of species, each species on its own velocity grid and
in its own units.  With $u = v/v_{ts}$ and $f_s$ a density per $\dd^3u$, species $b$ changes species $a$ at the rate
\begin{multline}
\Gamma Z_a^2 Z_b^2 \frac{\sigma_a}{\tau_a}\, \nabla_u \cdot \int U(\sigma_a u - \sigma_b u') \Bigl[\frac{\sigma_a}{\tau_a} f_b'\,
\nabla_u f_a \\ - \frac{\sigma_b}{\tau_b} f_a\, \nabla_{u'} f_b'\Bigr] \dd^3u',
\label{eq:landau}
\end{multline}
with $U$ the Landau tensor at the physical velocity difference, in units of $v_{ti}$, $f_b' = f_b(u')$, and $\Gamma$ the ions' collision
coefficient.  One Coulomb logarithm serves every pair.  The collision frequency is derived from a reference
plasma, its density, temperature and minor radius, with the ion--ion Coulomb logarithm of~\cite{Huba2019}, 16.9 at $10^{20}\,$m$^{-3}$ and
1\,keV.

\textbf{The discretization.}  The operator is discretized by DG~\cite{HirvijokiAdams2017,AdamsLandau2025} on the species' own grids at
every spatial quadrature point.  A batch is the velocity grids of every species at one point, its channels, the velocity points (one $\vpar$ node and one $\mu$ row
each), ordered species by species,
and each pair has its own table of velocity differences, exactly antisymmetric between the two orders of the pair.  The weak form
conserves each species' mass and each pair's momentum and energy: we measured them at $10^{-16}$ of their terms on one batch.  The stage
takes two backward Euler substeps, each solved by Newton's method to a relative residual of $10^{-10}$, or $10^{-6}$ in the weak series
of Section~\ref{sec:scaling} and the shaped collisional model of Sections~\ref{sec:turb} and~\ref{sec:utilturb}.  At $10^{-6}$ a stage
changes the mass by at most $1.2 \times 10^{-8}$ and the energy by $8.1 \times 10^{-8}$ of their totals on the series, and by $6.1
\times 10^{-11}$ and $4.5 \times 10^{-10}$ in the model.  The Jacobian's fixed-coefficient part, which Newton's linear solves use, is
block diagonal by species, so each Newton iteration solves one system per species, and the residual couples them.
  The stage's discrete entropy is not monotone by
construction: the weak form differentiates the polynomial $f$, not $\ln f$, which a metriplectic discretization
avoids~\cite{KrausHirvijoki2017}.  Measured per batch as $S = -\sum_j \omega_j g_j \ln g_j$, with $g_j$ the distribution at channel $j$ and $\omega_j$ its quadrature
weight, over the channels with $g_j > 0$, the entropy's largest decrease over the 100 collision stages of the collisional zonal-flow
decay test (Section~\ref{sec:verif}) is $2.8 \times 10^{-7}$ of $|S|$; on the same problem in the axisymmetric solver it is below
$10^{-9}$ on velocity grids of $72 \times 36$ nodes and finer.

\textbf{The coupling.}  The stage runs every $N$ steps, $N = 1$ in the weak series and in the shaped collisional
model, and leaves its rate as a source frozen in the
explicit part until the next stage, a balanced coupling~\cite{Speth2013} whose fixed points are the steady states of the coupled
equations.  Mass holds to the Newton tolerance, and the coupling's error in the energy $K + U$ of Section~\ref{sec:conservation} grows in proportion to the interval $N\Delta t$.  A batch couples every
$\mu$ row of every species, so the batches are transposed across the $\mu$ blocks before the stage and back after it.  With kinetic electrons the potential is part of the integrator's state.  With gyrokinetic ions the ions' charge is the
deposit of every $\mu$ row, each through its own ring, and the stage moves density between $\mu$ rows, so it changes the charge; the field
is solved again after each stage, which restores quasineutrality to $10^{-8}$ of the load.

\textbf{On the device.}  The coefficient sum, the integral of~(\ref{eq:landau}) over every pair of a batch's channels, is one kernel
that evaluates $U$ as it sums rather than reading stored tables, which grow with the pairs to 0.9\,GB a batch for 64 graded electron
segments.  $U$ depends on a batch only through $\sqrt{B/B_0}$ in the perpendicular coordinate, and the wedges of a process repeat $B/B_0$ at
each point of the plane, so one evaluation of $U$ serves a group of batches of equal $B/B_0$, each summed in its own order, which keeps
the result bitwise.

\textbf{The batched block-tridiagonal solve.}  Each Newton step's linear systems, one per species and batch, are block tridiagonal in $\vpar$ with the whole $\mu$ column in each block.  A block is one $\vpar$ segment's four nodes by the
species' 24 $\mu$ rows, 96 unknowns, a system has one block a segment, and a segment's unknowns couple to the neighbouring segments'
only through the rows of its end nodes.  The solve is a
block Thomas elimination, one GPU thread team per system, which eliminates each segment's block with its right side and its coupling to
the next segment as extra columns.  Gauss--Jordan elimination with partial pivoting factors the block in registers: each thread holds 24 entries of one column of the $96 \times 96$ block, each pivot step passes the pivot's column and row through the team's shared memory, and no block is kept there, so
a compute unit holds several teams; a blocked LU in shared memory is an option.  The factorization runs in single precision while the
residual and Newton's test stay in double, and Newton's iteration counts do not change.  The direct solve replaces Jacobi-preconditioned
TFQMR~\cite{Freund1993}, whose iterations grow with the grading of the electron grids (Table~\ref{tab:collsolve}).

\begin{table}[t]
\centering
\caption{The collision stage's block-tridiagonal solve against its alternatives, in seconds.  The proxy is the relaxation
of~\cite{AdamsLandau2025} on one device to 20 of its time units, on point B's two velocity grids (B) or on six $\vpar$ segments by six elements in
$v_\perp$ a species (S).  TFQMR $n$: its iterations a Newton iteration.  Point B's rows: its plane and wedges at $m_i/m_e$ 1836, with the electron segments given.  The collisional step is three steps of kinetic electrons at the mass ratio 100 on the shaped level-1 plane,
four wedges, one device, collisions every step.}
\label{tab:collsolve}
\footnotesize
\setlength{\tabcolsep}{4pt}
\begin{tabular}{@{}llrr@{}}
\toprule
configuration & device & before & after \\
\midrule
\multicolumn{4}{@{}l}{\emph{Jacobi TFQMR $\to$ the direct solve: a stage's solve, point B, $m_i/m_e$ 1836}} \\
16 graded e$^-$ segments, TFQMR 48 & A100 40\,GB & 26.2 & 21.1 \\
32 graded e$^-$ segments, TFQMR 215 & A100 80\,GB & 249.4 & 34.9 \\
\multicolumn{4}{@{}l}{\emph{the factorization in double $\to$ single precision: the solve}} \\
proxy B, 1024 batches & A100 & 30.4 & 30.1 \\
\multicolumn{4}{@{}l}{\emph{the blocked LU $\to$ Gauss--Jordan in registers}} \\
proxy B, 256 batches, the solve & GCD & 18.8 & 8.2 \\
proxy S, 256 batches, the solve & GCD & 6.04 & 2.45 \\
the collisional step & GCD & 128.0 & 83.0 \\
the collisional step & A100 & 78.3 & 61.3 \\
\bottomrule
\end{tabular}
\end{table}

\textbf{The kernels' costs.}  Sharing one evaluation of $U$ among four batches of equal $B/B_0$ makes the coefficient sum
1.6$\times$ faster, with the kernel held to 128 registers a thread by launch bounds.  The direct solve is 1.24$\times$ faster than
Jacobi TFQMR on 16 graded electron segments and 7.1$\times$ on 32 (Table~\ref{tab:collsolve}); Table~\ref{tab:utilturb} gives the
kernels' rates in the shaped collisional model.

\textbf{Checks against physics.}  Two checks test the operator beyond its conservation (Table~\ref{tab:verification}): the
ion--electron temperature exchange and friction between two Maxwellians at the physical mass ratio~\cite{Huba2019}, and the anisotropic
relaxation of an electron--deuterium plasma of~\cite{AdamsLandau2025} on our grids, where the ions deviate the most, since our one
Coulomb logarithm replaces that paper's ion--ion value.

\section{Conservation}
\label{sec:conservation}
Mass is exact for any numerical flux.  It holds to rounding where the stages are solved to rounding, and to the solves' tolerances
otherwise: $1.2 \times 10^{-7}$ of the total a step in the shaped collisional model, and $2 \times 10^{-11}$ with its solves 100 times
tighter (Section~\ref{sec:turb}).  The semi-discrete energy $K + U$ changes only through the
wall and the mortar's seam, and the seam kick cancels the seam where it is on.  What remains in a step is the time integrator's
$O(\Delta t^4)$, the seam on the shaped meshes with adiabatic electrons, the collision stage's balanced coupling, the solves' tolerances
and rounding.  The identity needs a field operator that is symmetric and static, which is why the field multigrid of
Section~\ref{sec:fieldmg} is a fixed linear cycle and not a solve to a tolerance.

The energy is $K + U$, the kinetic energy $K = (f, H_0)$ plus the field
energy $U$, and the semi-discrete scheme satisfies
\begin{equation}
\frac{d}{dt}(K + U) = -F_{\mathrm{wall}} - S_{\mathrm{seam}},
\label{eq:energy}
\end{equation}
where $F_{\mathrm{wall}}$ is the energy flux through the wall and its inflow condition, an exchange with the boundary, and $S_{\mathrm{seam}} = \sum G\,[H_h]$
is the mortar's seam term, summed over the mortar's quadrature points: $[H_h]$ the jump of the discrete Hamiltonian across the mortar
and $G$ the flux through it.  On the circular
meshes, and on the shaped ones with kinetic electrons, the seam kick cancels $S_{\mathrm{seam}}$ term by term (Section~\ref{sec:hard});
with adiabatic electrons on the shaped meshes it is off and $S_{\mathrm{seam}}$ remains.  With gyrokinetic ions the identity holds
with each ion row's own Hamiltonian (Section~\ref{sec:gyro}).  A gate, a case of our check suite that asserts a stated bound, checks Equation~(\ref{eq:energy}) to $3 \times 10^{-15}$ of
the flux scale or better.

Besides the wall's flux, $K + U$ changes through four terms (Table~\ref{tab:energy}): the seam, where the kick
is off; the time integrator, at $O(\Delta t^4)$ per step, since ARKIMEX3 does not preserve the quadratic field energy~\cite{Cooper1987} (halving $\Delta t$
divides the integrator's defect by 16.0); the solves, through their tolerances, the advection solve's relative residual of $10^{-10}$ (the
field's factorization is direct), which Table~\ref{tab:energy} does not list apart; and rounding.  On the CBC configuration at level 3 the kick takes the seam from $1.2 \times 10^{-7}$ of the
energy residual's terms to rounding.  Without the kick the seam falls to $8.5 \times 10^{-9}$ at level 4, order 3.9 in $h$ against the expected 4.  The linear ITG run, seeded
at $10^{-4}$, holds $K + U$ to $5 \times 10^{-15}$ of the initial kinetic energy per step with and without the kick (10 steps measured at
level 3, 4 at level 4), since only the seed's energy changes.

The collision stage conserves energy~\cite{HirvijokiAdams2017} to its Newton tolerance, and the step's $K + U$ defect in the collisional decay test is the same with and without collisions to three
digits.  With ions and kinetic electrons colliding, the balanced coupling's defect in $K + U$ grows with the stage's interval, from
$1.8 \times 10^{-9}$ per step at $N = 1$ to $3.7 \times 10^{-8}$ at $N = 5$ (Table~\ref{tab:energy}).

\begin{table}[t]
\centering
\caption{The energy budget's terms besides the wall flux, per step.  The kick is on for the circle by default, and on the shaped meshes
with kinetic electrons only.  The shaped collisional model's rows, at $\Delta t = 1$, are at the run's solve tolerances and, after the
semicolon, with the solves 100 times tighter (Section~\ref{sec:turb}).}
\label{tab:energy}
\footnotesize
\begin{tabular}{@{}lll@{}}
\toprule
term & configuration & per step \\
\midrule
\multicolumn{3}{@{}l}{\emph{relative to the flux scale}} \\
identity~(\ref{eq:energy}) & check suite & $\le 3 \times 10^{-15}$ \\
                          & gyroaverage on, L1 & $1.4 \times 10^{-15}$ \\
\multicolumn{3}{@{}l}{\emph{relative to the summed magnitudes of the energy residual's terms}} \\
seam, kick & CBC, L3 & $1.1 \times 10^{-16}$ \\
seam, no kick & check suite, circle & $10^{-12}$--$10^{-11}$ \\
              & \quad shaped, with $\phi$ & $10^{-9}$--$10^{-7}$ \\
     & CBC, L3 / L4 & $1.2 \times 10^{-7}$ / $8.5 \times 10^{-9}$ \\
\multicolumn{3}{@{}l}{\emph{relative to the initial kinetic energy}} \\
ARKIMEX3 & circle L1, $\Delta t$ 0.1, 0.05 & $1.8 \times 10^{-11}$, $1.1 \times 10^{-12}$ \\
         & GAM, circle L2, $\Delta t$ 0.2 & $3.6 \times 10^{-9}$ \\
rounding & equilibrium, L2 & $1.9 \times 10^{-16}$ \\
coupling & CBC L3, $N = 1$ & $1.8 \times 10^{-9}$ \\
         & CBC L3, $N = 5$ & $3.7 \times 10^{-8}$ \\
all, $K + U$ & ITG, seed $10^{-4}$, L3 / L4 & $5 \times 10^{-15}$ \\
\multicolumn{3}{@{}l}{\emph{relative to the energy before the stage}} \\
collision stage & zonal-flow decay, L3 & $4.7 \times 10^{-14}$ \\
\multicolumn{3}{@{}l}{\emph{relative to the totals}} \\
all, $K + U$ & shaped collisional model & $7 \times 10^{-8}$; $1.2 \times 10^{-8}$ \\
all, mass & shaped collisional model & $1.2 \times 10^{-7}$; $2 \times 10^{-11}$ \\
\bottomrule
\end{tabular}
\end{table}

\section{Verification discipline and a port}
\label{sec:discipline}
Three checks hold each solver to a reference.  Every GPU kernel matches its host counterpart to $10^{-12}$ relative or better, the
sweep with double-precision inverses; the device sweep, a level schedule of the sequential sweep's reads, reproduces the sequential sweep
bitwise; and a run over several processes reaches one process's state within the solves' tolerance, $10^{-8}$ relative or better.
Each check needs only the solver and its reference, so a port can carry all three: the host version against
the device, the sequential order against a scheduled sweep, and one process against a decomposition.  The port below is checked the same way, against PETSc's own integrator.

The code is written in Kokkos and runs on the NVIDIA A100 (CUDA) and the AMD MI250X (HIP) from one
source.  Single precision, the GPUs' default for the sweep's inverses, changes only the preconditioner, not the converged state: our check
suite's kinetic-electron run reaches the double-precision run's state to Newton's tolerance with no more iterations.  We hold every
performance change to a stricter rule where we can: a reordering of a kernel's loads or threads must reproduce the previous state
bitwise on both GPUs before the reordering is kept.

Our check suite runs on the host build for each change to the code, and on the code of this paper every case passes with no compiler warnings.

The gyroaverage's
gates assert $B_r\mathbf{1} = M_p\mathbf{1}$, the pair's adjointness, the contraction away from the wall, the push's symbol against $J_0$
to $10^{-4}$ for $k h \le 1$, and the Pad\'e operator's symmetry, null space and positivity; with it on, the energy identity holds per row
to $1.4 \times 10^{-15}$ of its terms.  The collision stage's gates with several species assert each species' mass and each pair's
momentum and energy on one batch to $10^{-16}$ of their terms, that two identical ion species split 0.3 and 0.7 are one species to
$9 \times 10^{-15}$, and that the device's stage matches the host's to $3 \times 10^{-14}$.  Gates of this kind test what the code computes, not what it should
compute: conservation and the Maxwellian fixed point do not depend on the collision rate, so a separate gate checks the relaxation rate
against the NRL formula~\cite{Huba2019}.

\textbf{A port.}  We gave an LLM code assistant (Claude Opus~5.5) the public snapshot and the low-storage integrator's porting prompt
in a clone of PETSc's main branch.  In 50 minutes the assistant read the material, proposed a plan, which one
person approved unchanged, and wrote the integrator as a low-storage option of \texttt{TSARKIMEX} for any additive tableau with an
explicit first stage, in two commits.  Its new test compares the option with \texttt{TSARKIMEX} on a nonlinear advection--diffusion
problem in nine variants, among them a DAE with a mass, adaptive steps and two processes, and the two agree to $7.2 \times 10^{-14}$
relative or better; PETSc's whole \texttt{TS} test suite passes.  The new option adds what PETSc's step requires and
KINTORA's type lacks: a rejected step rolls back.  Three gaps in the prompt left the assistant to decide: what the porting notes' two counts of work
vectors include, which differ and state no basis; the names, since an option of \texttt{TSARKIMEX} cannot keep KINTORA's type name as
the prompt asked; and which of KINTORA's options are experiments to leave out.  The public prompts now state all three.  The
branch is not yet submitted to PETSc.

\section{Evaluation}
\label{sec:evaluation}

\subsection{Setup}
\label{sec:setup}
\begin{table}[t]
\centering
\caption{Machines and software~\cite{Atchley2023,Choquette2021}.}
\label{tab:setup}
\footnotesize
\begin{tabular}{@{}lll@{}}
\toprule
 & Frontier & Perlmutter \\
\midrule
device & MI250X GCD, 64\,GB & A100, 40 and 80\,GB \\
devices per node & 8 GCDs & 4 GPUs \\
compiler & GCC 14.2 & GCC 14.3 \\
GPU stack & ROCm 6.4.2 & CUDA 13.2 \\
MPI & Cray MPICH 9.0.1 & Cray MPICH 9.1.0 \\
PETSc fork (64-bit) & eda0b4c370a & eda0b4c370a \\
Kokkos & 5.2.2 & 5.2.2 \\
field factorization & Kokkos Cholesky & cuDSS 0.8.0 \\
\bottomrule
\end{tabular}
\end{table}
Table~\ref{tab:setup} lists the machines and software, and the artifact description\ifanon\else{} (Appendix~\ref{app:ad})\fi{} gives the commit each result ran on.  Point C's
$5.4 \times 10^9$ ion unknowns are past 32-bit indices, so every run of this paper uses PETSc with 64-bit indices.  Every run
but the weak series' point A places one process on each GPU or GCD, with GPU-aware MPI, the weak series on 40\,GB A100s and the model of
Section~\ref{sec:utilturb} on 80\,GB ones, and times its steps after a warm-up step, which carries the set-up's lazy allocations.  The verification rows come from the runs
the artifact description maps, several on code before the seam kick; the code with the kick reproduces the level-3 CBC growth rates to 0.06\%, the
shaped GAM's history to $9 \times 10^{-12}$ and the collisional decay's slowdown factor to three digits.  The optimization steps and the
layout comparisons of Sections~\ref{sec:parallel}, \ref{sec:advection} and~\ref{sec:hard} and Appendix~\ref{app:util} and the weak
series' base of Table~\ref{tab:weakparts} are ablations on the code of their time, and we have not re-timed them or the strong scaling.  On the MI250X each point of the
level-3 strong scaling is run five times, each in its own job and allocation, with a range of 0.2--1.5\% of the median step, which the
figure gives; every other point is one run.

\subsection{Verification}
\label{sec:verif}
Table~\ref{tab:verification} compares each test with its reference.  At level 4 the drift-kinetic CBC growth rate is 4.4\% above the
semi-Lagrangian solver's, and the gyrokinetic rate 1.02--1.10 times GYSELA's.  Kinetic electrons raise the rate 1.63--1.68 times,
against GENE's 2.00.  Toroidal Landau damping with kinetic electrons is 2.5\% from the root of the dispersion relation.  The shaped GAM
agrees with the axisymmetric solver to 2.1\% in frequency, damping and residual, and with the shaped-plasma frequency to 2\%.  The
collisional zonal-flow decay agrees with the axisymmetric solver to under 1\% rms.  Landau collisions conserve the moments to
$1.2 \times 10^{-16}$, match the exact exchange and friction to 2.2\%, and reproduce DMPlexLandau's temperatures to 1.8\,eV.

\begin{table*}[t]
\centering
\caption{The verification tests.  DK and GK: drift-kinetic and gyrokinetic ions; the ion tests without a label ran with drift-kinetic
ions.  SL: the semi-Lagrangian solver of~\cite{AdamsMethods2026} on the level-4 mesh.  GYSELA: the gyrokinetic rate of GYSELA's Table~10,
interpolated to $k_\theta\rho = 0.28$~\cite{Grandgirard2016}.  GENE~\cite{Jenko2000}: the kinetic-electron rate and frequency over the adiabatic ones at the
mass ratio 1836~\cite{Gorler2016}.  Root: the two-species dispersion relation's.  Exact: the exchange and friction between two
Maxwellians~\cite{Huba2019}.  DMPlexLandau: the temperatures of~\cite{AdamsLandau2025}, $t_0$ its time unit.  The
shaped-GAM and decay rows compare the 5D wedge with the axisymmetric solver (Section~\ref{sec:hard}) on the same mesh and velocity
grid.
Gao: the shaped-plasma GAM frequency of~\cite{Gao2010} at the seeded surface.  HR: Hinton and Rosenbluth's short-time law~\cite{HintonRosenbluth1999}; Xiao: the closed form of~\cite{Xiao2007} as printed
in~\cite{Frei2021}; decay time: the factor on the collision time that fits each law to $R(t)$; $\nu^* = qR_0/(\epsilon^{3/2} v_{ti}\tau_{ii})$: the ion
collisionality at mid-radius, $\epsilon = r/R_0$, $\tau_{ii}$ Braginskii's ion collision time~\cite{HelanderSigmar2002}.}
\label{tab:verification}
\footnotesize
\begin{tabular}{@{}llll@{}}
\toprule
test & configuration & measured & reference \\
\midrule
toroidal ITG (CBC), DK & circle L3, 16 wedges & $\gamma = 0.140 \pm 0.002$, $\omega = -0.220$ & \\
                   & circle L4, 16 wedges & $\gamma = 0.142 \pm 0.003$, $\omega = -0.222$ & SL, L4: $0.136 \pm 0.003$, $-0.222$ \\
toroidal ITG (CBC), GK & circle L3, 16 wedges, $B_0 = 50$ & $\gamma = 0.105$, $\omega = -0.237$ & GYSELA: $0.087$, $-0.206$ \\
                   & circle L4, 32 wedges, $B_0 = 100$ & $\gamma = 0.096 \pm 0.004$, $\omega = -0.242$ & GYSELA: $0.087$, $-0.206$ \\
CBC, GK, kinetic / adiabatic & circle L3, $B_0 = 50$, 1836 & $\gamma$ $1.63\times$, $\omega$ $1.27\times$ & GENE: $2.00\times$, $1.04\times$ \\
                   & circle L4, $B_0 = 100$, 1836 & $\gamma$ $1.68\times$, $\omega$ $1.17\times$ & GENE: $2.00\times$, $1.04\times$ \\
Landau damping, kinetic e$^-$ & slab, L2, 16 wedges, 3670.5 & $\gamma = -0.0893$ & root: $-0.0871$ \\
shaped GAM         & Solov'ev L3, 16 wedges & $\omega$ $+0.19\%$, $\gamma$ $+1.2\%$, residual $-2.1\%$ & axisymmetric, same mesh \\
                   & Solov'ev L4, 16 wedges & $\omega$ $+0.3\%$, $\gamma$ $+0.1\%$ & axisymmetric, same mesh \\
                   & & $\omega = 0.68$--$0.70$ & Gao: $0.688$ \\
collisional zonal-flow decay & circle L3, 4 wedges, $\nu^* = 0.25$ & $R(t)$ within 0.8\% rms & axisymmetric, same mesh \\
                     & & decay time $1.8\times$ / $0.90\times$ & HR / Xiao \\
                     & circle L3, 4 wedges, $\nu^* = 0.063$ & $R(t)$ within 0.9\% rms & axisymmetric, same mesh \\
                     & & decay time $1.6\times$ / $0.79\times$ & HR / Xiao \\
Landau collisions, 3670.5 & one batch, e$^-$ uniform, graded & moments $\le 1.2 \times 10^{-16}$ & 0 \\
                     & & exchange $1.022\times$, friction $1.006\times$ & exact \\
e--D relaxation, 3670.5 & one A100 & $T$ to 0.7\,eV by 20\,$t_0$, 1.8\,eV by $1.2 \times 10^4$\,$t_0$ & DMPlexLandau \\
\bottomrule
\end{tabular}
\end{table*}
\begin{figure}[t]
\centering
\includegraphics[width=\columnwidth]{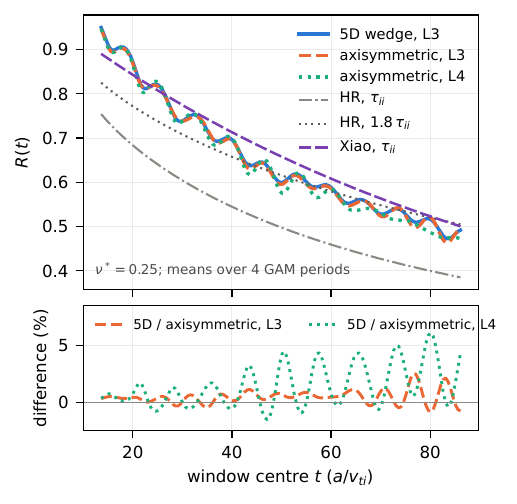}
\caption{The collisional decay of the zonal-flow residual~\cite{RosenbluthHinton1998}: $R(t)$, the collisional run's zonal
potential over its collisionless twin's.  Top: the 5D wedge and the axisymmetric solver against Hinton and Rosenbluth's short-time
law~\cite{HintonRosenbluth1999} and Xiao's closed form~\cite{Xiao2007,Frei2021}.  Bottom: the 5D result relative to each axisymmetric
one.}
\label{fig:hr}
\end{figure}

Appendices~\ref{app:landau}--\ref{app:null} give
three more: toroidal Landau damping, and a slab ITG mode with and without the gyroaverage, against analytic references, and the
conservation and null tests;
Appendix~\ref{app:wall} gives the wall's analysis.

\textbf{The toroidal ITG mode.}  The CBC~\cite{Dimits2000} runs a linear ITG mode on the circle at $B_0 = 50$: the $(m, n) = (7, 5)$
mode seeded on a canonical-Maxwellian equilibrium whose profile parameters give effective gradients $(R/L_n, R/L_T) = (2.2, 6.9)$~\cite{AdamsMethods2026}, with
$\Delta t = 0.1$ and no filter.  We fit the $(7, 5)$ projection of the potential in radial bins at mid-radius, over the same window as
the semi-Lagrangian reference on the level-4 mesh~\cite{AdamsMethods2026}.  At level 4 the frequency agrees, and the growth rate is 4.4\%
higher, 1.4 standard deviations of the two fits combined (Table~\ref{tab:verification}); the level-3 and level-4 rates agree within their
fits' errors.  The A100 reproduces the MI250X's level-3 rates to 0.09\% or better.
No filter is needed: apart from the seeded mode, its poloidal sidebands and the zonal component (6\% of the seeded mode), every
resolved mode stays below 0.1\% of the seeded mode.

With gyrokinetic ions (Section~\ref{sec:gyro}) the same case runs on the weak series' point B grid, $\Delta t = 0.25$, against the
gyrokinetic rate of GYSELA for the same configuration at $\rho^* = 1/180$, interpolated to $k_\theta\rho = 0.28$ and converted to
$c_s/a$~\cite{Grandgirard2016}, which matches the Cyclone benchmark's codes~\cite{Dimits2000}.  The gyroaverage lowers the
drift-kinetic rate by a quarter, as finite Larmor radius should~\cite{Romanelli1989,DongHortonKim1992}: at $B_0 = 50$, $\rho^* =
1/50$, to 1.15--1.21$\times$ GYSELA's over the fit windows, with the frequency 15\% higher (Table~\ref{tab:verification}).  There the
canonical equilibrium spreads its profiles over an ion banana width, the radial excursion of a trapped ion's orbit, so the ion temperature at the mode is 1.31 times the electrons' where GYSELA's
local Maxwellian has them equal.  At $B_0 = 100$, with the $(14, 10)$ mode at the same $k_\theta\rho$ on level 4 and 32 wedges, that
ratio is 1.05, and the rate is 1.02--1.10$\times$ GYSELA's, still falling slowly at the run's end.  Both cases have 18 plane nodes a
poloidal wavelength; with 9, $B_0 = 100$ on level 3, every rate falls by a factor of 1.2 to 2.2, so we do not compare at fewer.

\textbf{The ITG mode with kinetic electrons.}  GENE gives the CBC's kinetic-electron mode at the mass ratio 1836 as 2.00$\times$
the adiabatic case's growth rate and 1.04$\times$ its frequency~\cite{Gorler2016}.  With gyrokinetic ions and the electrons on 16 graded segments, collisionless, the kinetic electrons raise the growth rate 1.63 and
1.68 times and the frequency 1.27 and 1.17 times at $B_0 = 50$ and 100 (Table~\ref{tab:verification}); drift-kinetic ions give 1.67 and
1.62, and 1.11 and 1.04.  The electron grid does not set the gap: 16, 32 and 64 segments move
the rate and frequency by 1\%.  Nor does the step: at the weak series' step at point B, $\Delta t = 1$, with the inflow transfer, the rate
moves 0.3\% from $\Delta t = 0.25$ and its ratio to the adiabatic rate 0.2\%.
Against drift-kinetic ions at the same $B_0$, the gyroaverage raises the kinetic-electron mode's frequency by 22\% at $B_0 = 50$
and 18\% at $B_0 = 100$, and the adiabatic mode's by 7\% and 6\%,
and both ratios move toward GENE's from $B_0 = 50$ to 100.  The canonical equilibrium also leaves the two species' densities
unequal, since the ions' profiles spread over their banana width and the electrons' do not: at the mode the ion density is 1.06 times
the electrons' at $B_0 = 50$ and 1.01 times at $B_0 = 100$, where GENE's are equal.  At the series' own mass ratio, 3670.5, on 8 segments, the collisionless
mode is the 1836 mode to 2\%.  With Landau collisions every four steps the rate falls through the run, from 0.099 over $t \in [15,
24]$ to 0.064 over $[30, 48]$, at the collisionless frequency, which suggests that the collisions damp the electrons' response; the run has no settled
rate to compare.

\textbf{The shaped GAM.}  On the Solov'ev mesh a zonal seed rings the geodesic acoustic mode.  The 5D wedge and the axisymmetric
solver on the same level-3 mesh and velocity grid agree in frequency, damping and residual to 2.1\% or better
(Table~\ref{tab:verification}), and the damping's offset shrinks with the mesh (Section~\ref{sec:hard}).  The comparison tests the 5D operator's axisymmetric limit on shaped geometry,
including the mortar.  At level 4 the two agree in frequency and damping to 0.3\%, and the absolute frequency, 0.68 to 0.70 by fit
window, lies within 2\% of the shaped-plasma relation~\cite{Gao2010} at the seeded surface, 0.688; the relation's truncation at
$O(1/q^2)$, its omitted triangularity term and its choice of the major radius place the exact frequency 3--6\% above it.  The run is
too short for the residual, for which shaped-plasma theory also gives a reference~\cite{XiaoCatto2006shaping}.

\textbf{The collisional zonal-flow decay.}  A zonal potential on the circle at $B_0 = 500$ ($k_r \rho_p = 0.196$, with $\rho_p$ the poloidal Larmor radius) runs with and
without collisions, and the ratio $R(t)$ of the two divides out the discretization's collisionless residual~\cite{RosenbluthHinton1998}; both runs are averaged over
four GAM periods, whose amplitude still exceeds the residual's, before the ratio is formed.  At both
collisionalities the 5D wedge agrees with the axisymmetric solver on the same mesh to 0.8\% and 0.9\% rms (Table~\ref{tab:verification}), closer
than the axisymmetric result's own change from level 3 to level 4, 2.2\% rms at $\nu^* = 0.25$ (Figure~\ref{fig:hr}).  In both solvers and at both collisionalities the residual decays
1.6--1.8$\times$ slower than Hinton and Rosenbluth's short-time law~\cite{HintonRosenbluth1999}, which rests on a variational
estimate of the neoclassical polarization.  An eigenfunction expansion of the pitch-angle scattering operator with momentum restoration
gives the decay in closed form~\cite{Xiao2007}; in the form and collision frequency of~\cite{Frei2021}, where GENE matches it at
$\nu^* = 3.1$, our decay is 1.1--1.3$\times$ faster than it at the same collision time.  The closed form omits energy scattering.

\input{sec-turbulence}

\subsection{Scaling}
\label{sec:scaling}
\begin{figure*}[t]
\centering
\includegraphics[width=\textwidth]{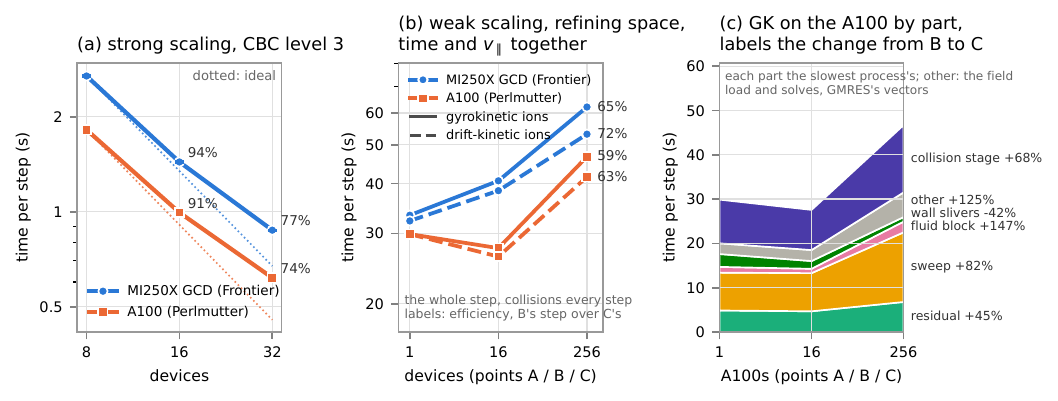}
\caption{Scaling of the CBC configuration, time per step after one warm-up step.  (a)~Strong scaling at level~3 from 8 to 32 devices: 8 toroidal blocks by 1, 2 and 4 $\mu$ blocks, drift-kinetic ions and adiabatic electrons; on the MI250X the median of five runs, the bars their range; the labels give the
efficiency, the 8-device step times 8 over the step times the devices.  (b)~The weak series of Table~\ref{tab:weak} at the physical
mass ratio.  (c)~The gyrokinetic series on the A100 by part.}
\label{fig:scaling}
\end{figure*}

The strong-scaling runs use the CBC configuration of Section~\ref{sec:verif}, with drift-kinetic ions and adiabatic electrons,
and time four steps after one warm-up step
(Figure~\ref{fig:scaling}; the repetitions are those of Section~\ref{sec:setup}).  Strong scaling at level 3 splits the $\mu$ rows: from 8 to 32 devices the step is 77\% efficient on the MI250X GCDs and 74\% on the
A100s, with the advection solve's count unchanged, and the GCD's step over the A100's is 1.42--1.48 at equal device
counts.  The same split carries level 4 from 16 to 192 GCDs at 61\%; beyond 192 GCDs the plane pieces add a halo and an iteration.

\begin{table*}[t]
\centering
\caption{The weak series at the physical mass ratio, 3670.5, with kinetic electrons and Landau collisions every step: each point
doubles the resolution in space, in time and in $\vpar$ for both species, so the unknowns and the devices grow 16$\times$ a point;
the $\mu$ grid stays at 24 rows a species.  Layout: toroidal blocks $\times$ $\mu$ blocks $\times$ plane pieces, the $\mu$ blocks split
evenly between the two grids; A's two processes share one device.  GK and DK: gyrokinetic and drift-kinetic ions.  Newton: iterations
per implicit stage; GMRES: iterations per Newton iteration; the counts are the same on the two machines.  Seconds per step, the whole
step with its collision stage, over seven steps after one warm-up, one run per point.}
\label{tab:weak}
\footnotesize
\begin{tabular}{@{}llllllllrrrr@{}}
\toprule
 & & & & & & \multicolumn{2}{c}{GK / DK} & \multicolumn{2}{c}{A100} & \multicolumn{2}{c}{MI250X GCD} \\
\cmidrule(lr){7-8}\cmidrule(lr){9-10}\cmidrule(l){11-12}
point & plane, wedges & $\Delta t$ & $\vpar$ nodes, ions / electrons & layout & devices & Newton & GMRES & GK & DK & GK & DK \\
\midrule
A & level 2, 8  & 2   & 24 / 16 & $1 \times 2 \times 1$  & 1   & 1.88 / 1.88 & 12.7 / 12.4 & 29.9 & 29.9 & 33.3 & 32.2 \\
B & level 3, 16 & 1   & 48 / 32 & $2 \times 8 \times 1$  & 16  & 1.88 / 1.88 & 13.2 / 12.6 & 27.6 & 26.3 & 40.6 & 38.4 \\
C & level 4, 32 & 0.5 & 96 / 64 & $4 \times 16 \times 4$ & 256 & 1.75 / 1.75 & 17.4 / 15.6 & 46.7 & 41.6 & 62.1 & 53.1 \\
\bottomrule
\end{tabular}
\end{table*}

\begin{table*}[t]
\centering
\caption{The weak series of Table~\ref{tab:weak} by part, gyrokinetic ions, seconds per step.  Step: the whole step; collisionless:
the step without its collision stage.  Sweep: the advection solves' sweeps, in parentheses their share of the collisionless step.
Field load: assembling the field solves' charge.  Collision stage: the step less the collisionless step, in parentheses its share of
the step; Newton: its iterations per batch, summed over its two substeps.  Memory: the busiest device's peak, of the A100's 40\,GB and
the GCD's 64.  Gain: the base series' time per unit of simulated time over this series', the base at a quarter of the time step, with
collisions every fourth step, two sweeps in place of the inflow transfer and no predictor guard, gyrokinetic / drift-kinetic ions.
Efficiency from B to C: B's seconds per step over C's, gyrokinetic / drift-kinetic ions.}
\label{tab:weakparts}
\footnotesize
\setlength{\tabcolsep}{4pt}
\begin{tabular}{@{}llrrrrrrrccc@{}}
\toprule
 & & & & & & & & & & \multicolumn{2}{c}{efficiency B to C, GK / DK} \\
\cmidrule(l){11-12}
device & point & step & collisionless & sweep & field load & collision stage & Newton & memory, GB & gain, GK / DK & step & collisionless \\
\midrule
A100 & A & 29.88 & 20.02 & 8.54 (43\%)  & 0.81 & 9.86 (33\%)  & 7.1 & 27.3 & 2.01 / 1.97 & & \\
     & B & 27.56 & 18.50 & 8.61 (47\%)  & 1.58 & 9.06 (33\%)  & 5.2 & 29.3 & 1.80 / 1.77 & & \\
     & C & 46.68 & 31.50 & 15.71 (50\%) & 8.94 & 15.18 (33\%) & 4.1 & 33.4 & 1.52 / 1.47 & 59\% / 63\% & 59\% / 65\% \\
GCD  & A & 33.32 & 23.19 & 11.66 (50\%) & 1.36 & 10.14 (30\%) & 7.1 & 28.9 & 1.98 / 1.89 & & \\
     & B & 40.62 & 28.59 & 15.75 (55\%) & 2.67 & 12.03 (30\%) & 5.2 & 29.7 & 1.73 / 1.69 & & \\
     & C & 62.14 & 47.28 & 27.47 (58\%) & 8.93 & 14.87 (24\%) & 4.1 & 33.6 & 1.67 / 1.61 & 65\% / 72\% & 60\% / 69\% \\
\bottomrule
\end{tabular}
\end{table*}

The weak series refines space, time and $\vpar$ together, so the unknowns and the devices grow 16$\times$ a point; the $\mu$ grid
stays at 24 rows a species (Table~\ref{tab:weak}).  The time step halves with the mesh, so the in-plane and toroidal Courant numbers are the same at every point,
and so is the distance a thermal electron streams in a stage, three toroidal blocks, which the inflow transfer carries
(Section~\ref{sec:stage}); the collision stage runs every step.  Point B is the CBC grid of Table~\ref{tab:grid}.

Table~\ref{tab:weak} gives the series with gyrokinetic and with drift-kinetic ions on both machines; each configuration's Newton and
GMRES counts are the same on the two machines.  Point A is not a scaling baseline: its two processes, one for each species' velocity grid, share one device, so A's time
carries the sharing, and the efficiencies start at B.  Table~\ref{tab:weakparts} splits the gyrokinetic step by part.  From B to C the
whole step's efficiency is 59\% on the A100 and 65\% on the GCD with gyrokinetic ions, and 63\% and 72\% with drift-kinetic ions.  C is the first point split in the plane.  There the stage's GMRES count grows by four iterations per Newton iteration, and the collisionless
step's loss is in the sweep and in the field load, which exchanges the charge between the four plane pieces and grows 5.7 times from B on the A100 and 3.3 times on the GCD; the fluid block, 8\% of the collisionless step at C, grows 2.5$\times$.  On the A100 the collision stage grows by two thirds from B to C, in its slowest process's wait and in the transpose before it, and
on the GCD by a quarter.  The wall slivers raise these efficiencies: the slivers' integral (the wall slivers' part in Figure~\ref{fig:scaling}c), the costliest of the mortar's faces, acts
on the wall ring's cells, whose number doubles from one point to the next while the plane's cells quadruple, so its share of a
device's work halves at each point.  On the adiabatic CBC at level 4 it takes 5.1\% of the step (Appendix~\ref{app:wall}); we have not
measured its share on the series.

Per unit of simulated time the series is 1.5 to 2.0 times as fast as its base, the same points at a quarter of the step with
collisions every fourth step, two sweeps and no predictor guard (Table~\ref{tab:weakparts}).  The gain is largest at A and smallest at
C, where the count and the field load grow most, and the base is more efficient from B to C, at seven tenths on the A100.  At B and C,
on the same commit and counts, the GCD's collisionless step is about 1.5$\times$ the A100's and its sweep 1.8$\times$; its collision
stage is at most a third longer than the A100's, and at C, where the A100's stage spends its added time waiting, the two take the same time, so the stage's share of the step is
smaller on the GCD.

Table~\ref{tab:gyrocost} gives the gyroaverage's cost on the series, the gyrokinetic step against its drift-kinetic twin.  The
gyroaverage adds 2 to 12\% to the GMRES count, the most at C.  Priced at the drift-kinetic cost of an iteration, the extra iterations
are 43 to 50\% of the collisionless step's increase at B and C on both machines; the rest is the gyroaverage and the Pad\'e
field within each iteration.  At A on the A100 the two steps are equal to the printed digits.

\begin{table}[t]
\centering
\caption{The gyroaverage's cost on the weak series, the gyrokinetic step against its drift-kinetic twin (Table~\ref{tab:weak}).  GMRES:
the gyrokinetic count over the drift-kinetic one.  Collisionless and step: the increase of the collisionless step and of the whole
step.  Iterations: the extra iterations' share of the collisionless step's increase, priced at the drift-kinetic cost of an iteration.
Within: the rest, the gyroaverage and the Pad\'e field within each iteration, as a share of the gyrokinetic collisionless step, not of
the increase.  A dash:
no difference to the printed digits.}
\label{tab:gyrocost}
\footnotesize
\begin{tabular}{@{}llrrrrr@{}}
\toprule
point & device & GMRES & collisionless & step & iterations & within \\
\midrule
A & A100 & 1.02 & +0\%  & +0\%  & -- & -- \\
  & GCD  &      & +5\%  & +3\%  & 25\% & 4\% \\
B & A100 & 1.05 & +8\%  & +5\%  & 44\% & 4\% \\
  & GCD  &      & +9\%  & +6\%  & 43\% & 5\% \\
C & A100 & 1.12 & +19\% & +12\% & 50\% & 8\% \\
  & GCD  &      & +23\% & +17\% & 43\% & 11\% \\
\bottomrule
\end{tabular}
\end{table}

\subsection{The shaped collisional model on the A100}
\label{sec:utilturb}
The shaped collisional model of Section~\ref{sec:turb} is the regime the solvers are built for.  Its step is profiled on 8 A100s of
80\,GB, one $\mu$ block each, the ions' rows on four and the electrons' on four.  The run starts from the state at $t = 811$, the peak of the second burst, of a $\Delta t = 0.25$ run of the same model with
collisions every fourth step, and continues at $\Delta t = 1$.  This is one configuration and
one run.

\begin{table*}[t]
\centering
\caption{Per-kernel utilization of one A100 80\,GB in the shaped collisional model, process 4 of 8, which holds the electrons' lowest $\mu$ rows and
whose step is the slowest, over the eight timed steps after four warm-up steps; each part's heading also gives the same case's part on eight
MI250X GCDs.  The rows are the kernels that take 1\% of the step or
more, under the part of the step they belong to; each part's time is the slowest process's.  \% step and ms, the time per launch, come
from Nsight Compute's timing of every launch in the timed steps, the rates from its sampled launches.  GB/s is the DRAM throughput; the percents are Nsight Compute's DRAM, L2, L1 and SM
throughput~\cite{YangKurthWilliams2020}, DRAM of this device's 2.039\,TB/s; occ.\ is the active warps (\%); regs the registers per
thread; local is the local-memory traffic per thread (kB), spilled registers and per-thread arrays; lanes are the active lanes of 32;
FP64 and FP32 are percents of the 9.7 and 19.5\,TFLOP/s rates~\cite{Choquette2021}.  The Jacobian rows are the parts of the residual applied as the stage's linearized operator, which GMRES applies.  Bold marks the roof a kernel runs against, the largest of DRAM, L2, L1 and SM where
it reaches 60\%.}
\label{tab:utilturb}
\footnotesize
\setlength{\tabcolsep}{3pt}
\input{fig/tab_kernels_turb.tex}
\end{table*}

Table~\ref{tab:utilturb} gives the step by part and the kernels of the slowest process.  The implicit stages take two Newton iterations
each and 10.3 GMRES iterations per Newton iteration, 16 at most; the collision stage, at its tolerance of $10^{-6}$, takes 4.6 Newton
iterations per batch on average and 7 at most.  The counts are the same on both devices.  The step is 34.31\,s on the A100 and 40.85\,s
on the GCD, where the sweeps take 2.0$\times$ the A100's time and the rest of the step 1.10$\times$.  Per unit of simulated time the
A100 runs 1.84$\times$ as fast as at $\Delta t = 0.25$ with collisions every fourth step, which has the same interval between collision
stages: the collisionless work per unit of simulated time falls 2.8$\times$, and the collision stages cost the same.  The table shows that:
\begin{itemize}
\item the collision stage takes 53\% of the step, and its coefficient sum, two fifths of the step, runs at 80\% of the SM's peak
throughput and 63\% of the FP64 rate: it is bound by arithmetic, whose amount grows with the square of a batch's channels;
\item the Gauss--Jordan factorization of the collision solve's blocks takes 8\% of the step, in registers, in single precision, at no
roof;
\item the collisionless step's largest kernel, the sweep's launches, takes 10\% of the step: about 26,000 dependent launches a step at
0.14\,ms each and 15\% occupancy, bound by latency;
\item the build of the electrons' $72 \times 72$ sweep-block inverses takes 4\% of the step, at the L2 roof;
\item the Jacobian's wall-sliver, per-point and lateral-face sections reach the SM roof, at 51\%, 37\% and 37\% of the FP64 rate;
\item averaged over the step the device runs at 34\% of the A100's FP64 peak, 25 points of it in the coefficient sum, and at 40\% over
the 86\% of the step that the profiled kernels take.
\end{itemize}
The device is busy for 87\% of this process's step.  The norms' reductions take 1.6\,s of a step on the process that waits longest
and 0.26\,s on the one that waits least, which indicates that the ions' processes wait for the electrons': the electrons' sweep blocks
keep the couplings between $\vpar$ nodes that the ions' blocks drop, and each process holds one species' $\mu$ rows.  The coefficient
sum and the sweep's launches take half the step, the Jacobian's sections a fifth.  The table points to three changes, none made
here: a cheaper evaluation of the collision coefficients, shortening each sweep item's dependent memory accesses, and giving the electrons' rows more processes than the
ions'.

\section{Related work}
\label{sec:related}
\textbf{Continuum codes.}  Eulerian (continuum) gyrokinetic codes discretize the distribution on a fixed phase-space grid: GENE~\cite{Jenko2000,Gorler2011}, ported to GPUs~\cite{Germaschewski2021}; CGYRO, a spectral solver for collisional plasmas~\cite{Candy2016}, also on
GPUs~\cite{CGYROgpu2018}; GT5D, a conservative global full-$f$ code~\cite{Idomura2008}; COGENT, a continuum code for neoclassical
transport and the edge~\cite{Dorf2013,Dorf2020}; Gkeyll, which applies discontinuous Galerkin methods to full-$f$ electromagnetic
gyrokinetics in the tokamak edge~\cite{Mandell2020}; and GX, a GPU-native spectral code for tokamak and stellarator
design~\cite{Mandell2024GX}.  Gkeyll uses DG in phase space, as this code does.  The closest discretization is the explicit DG solver
of Jo et al.~\cite{Jo2025}: prisms extend unstructured poloidal triangles along the field and exchange fluxes across the nonconforming
planes on a sub-triangulation of the two meshes, without a field solve so far.  This code adds flux-aligned meshes in the plane, a
conservative mortar between the wedges, the quasineutral field with kinetic electrons, and an implicit treatment of the parallel streaming.

\textbf{Semi-Lagrangian and particle codes.}  GYSELA advances a global full-$f$ distribution semi-Lagrangianly~\cite{Grandgirard2006,
Grandgirard2016} and now runs on AMD GPUs through OpenMP and Kokkos~\cite{Malaboeuf2025,Bourne2025}.  \ifanon A semi-Lagrangian DG
solver on the same meshes is described in~\cite{AdamsHPC2026,AdamsMethods2026}\else The semi-Lagrangian DG solver that precedes this one
on the same meshes and in the same code is described in~\cite{AdamsHPC2026,AdamsMethods2026}\fi; that solver supplies the growth-rate
reference for the ITG test here.  Particle-in-cell codes, XGC~\cite{Ku2009,XGCgpu2017,XGCgpu2021} and ORB5~\cite{Lanti2020},
carry the distribution on markers.

\textbf{Implicit time integration.}  GS2 treats the linear gyrokinetic terms implicitly and solves the field through a response
matrix~\cite{Kotschenreuther1995}; stella keeps that implicit parallel streaming inside an operator split that takes the drifts and the
nonlinearity explicitly~\cite{Barnes2019}.  GT5D integrates the linear part of the motion implicitly and solves each stage by a Krylov
method, made communication-avoiding and mixed-precision on Fugaku and Summit~\cite{Idomura2020}.  COGENT's semi-implicit additive
Runge--Kutta scheme first took the collisions implicitly~\cite{Ghosh2018}; its implicit full-$F$ neoclassical runs take the Vlasov and
Fokker--Planck operators implicitly, with GMRES preconditioned by algebraic multigrid (AIR~\cite{Manteuffel2018}) on a first-order
upwind matrix of the Vlasov part~\cite{Dorf2025}.  The advection solve here differs in its preconditioner: one sweep of the high-order DG
blocks themselves in their upwind order.  The closest solver on the same hardware is a sweep-based implicit DG solver for the BGK
equation on unstructured grids, whose graph-ordered GPU sweeps in Kokkos ran on Frontier~\cite{Evans2026}.  Low-storage
implicit--explicit Runge--Kutta schemes are new tableaus constructed to run in few state-sized registers~\cite{HiguerasRoldan2016}, two
to four in~\cite{CavaglieriBewley2015}; K4 instead keeps the Kennedy--Carpenter tableau~\cite{KennedyCarpenter2003} unchanged and
stores its step in six vectors (Section~\ref{sec:step}).

\textbf{Sweeps.}  The preconditioner is a transport sweep in the sense of discrete-ordinates neutron transport.  Parallel sweeps on
unstructured meshes schedule the cells as a directed acyclic graph~\cite{Pautz2002,Plimpton2005}, pipeline it across processes
(KBA~\cite{BakerKoch1998}), and break the graph's cycles where it has them~\cite{Plimpton2005,Vermaak2021,Haut2019}; our sweep breaks
the cycles of the $E \times B$ drift at the unplaced cell farthest upwind along the mean in-plane flux and uses block Jacobi in place of the pipeline.  Its GPU version is a
level-scheduled block triangular solve~\cite{AndersonSaad1989}; the level schedule keeps every dependence of the sequential order, so
the GPU sweep reproduces the sequential sweep bitwise.  Kokkos sweeps for
discrete-ordinates DG on GPUs~\cite{Suau2025,Choi2026} build on batched small dense solves~\cite{Suau2025}, as the sweep items here do.
Krylov methods around sweeps~\cite{Warsa2004}, their acceleration by a low-order elliptic correction~\cite{AdamsLarsen2002}, and AIR
multigrid in place of sweeps~\cite{Hanophy2020} are standard in that field.  Downwind-ordered Gauss--Seidel and cellwise DG block
smoothers make multigrid robust for advection-dominated problems~\cite{BeyWittum1997,HackbuschProbst1997,Kanschat2008}; here the
same ordering is a single application of a Krylov preconditioner.  For anisotropic diffusion, overlapping Schwarz smoothers with blocks
of fixed size lose their smoothing as the anisotropy grows, and robust smoothing needs blocks that are global along the strong
direction, as line smoothers are~\cite{Krzysik2026}; the plane multigrid's annulus blocks are global along the flux surface, the
direction in which the field operator is strong.

\textbf{Field alignment and mesh coupling.}  Following the field line from one poloidal plane to the next, rather than meshing in flux
coordinates, is the flux-coordinate-independent (FCI) approach of fluid edge codes~\cite{Ottaviani2011,Hariri2013,Stegmeir2016}, also used by the
gyrokinetic code GENE-X~\cite{Michels2021} and, as field-aligned interpolation, by GYSELA~\cite{Latu2017}.  A field-aligned
DG mesh converges with fewer toroidal intervals than a toroidally extruded one~\cite{Jo2025}, and field-aligned DG with
nonconforming interfaces is more accurate than a non-aligned grid at equal degrees of freedom~\cite{Dingfelder2020}.  DG schemes for parallel
derivatives on non-aligned grids~\cite{Held2016} and a finite-volume FCI scheme that integrates fluxes over field-aligned volumes, which
improves mass conservation by orders of magnitude over interpolation~\cite{Wiesenberger2023}, are the closest conservative couplings.
The mortar here integrates its fluxes on the supermesh of a cell's field-line image and the next plane's cells.  That supermesh is
the intersection mesh of conservative Galerkin projection between unstructured meshes~\cite{FarrellMaddison2011}, here built once and
used to evaluate fluxes rather than to remap.  Mortars keep DG's local conservation at non-conforming interfaces~\cite{Kopriva1996,Laughton2021}, and Gkeyll couples sheared
field-aligned grids at flux-tube boundaries by a conservative DG projection~\cite{Francisquez2024}.  The well-balanced residual subtracts
the discrete operator applied to the equilibrium, as high-order well-balanced schemes for balance laws do~\cite{Berberich2021}.

\textbf{Gyroaverages.}  The ring average of the field and of the charge goes back to gyrokinetic particle simulation, a four-point average
of each~\cite{Lee1983,Lee1987}, with the number of points growing with the Larmor radius in later particle codes~\cite{Hatzky2002};
XGC precomputes gyroaverage matrices on its unstructured mesh~\cite{Ku2009}, ORB5 pairs gyroaveraged markers with a long-wavelength or
Pad\'e polarization~\cite{Jolliet2007}, and Lin and Lee solve the full double-gyroaverage Poisson equation by iteration~\cite{LinLee1995}.
Continuum and semi-Lagrangian codes interpolate the ring on tensor
grids~\cite{Grandgirard2016,Steiner2015,CrouseillesMehrenbergerSellama2010}, and global GENE applies it as a matrix with its adjoint
in the field equation~\cite{Gorler2011}, the adjoint pair that energetic consistency requires~\cite{ScottSmirnov2010,BrizardHahm2007}.  Gkeyll keeps
the long-wavelength limit~\cite{Mandell2020}.  The gyroaverage here is the Galerkin form of the ring average on a C0 P2 unstructured
plane, integrated exactly by clipping, with its exact adjoint carried through the field-aligned station maps.

\textbf{Collisions and software.}  The collision operator is the conservative Landau discretization of~\cite{HirvijokiAdams2017}, here on a
velocity grid per species, and K3 replaces its batched Krylov solvers~\cite{AdamsLandau2025} by a direct solve; XGC applies a nonlinear multi-species
Fokker--Planck--Landau operator of its own~\cite{Hager2016}.  Gyrokinetic codes more often use linearized
operators, such as the gyrokinetic Coulomb and Sugama operators~\cite{Frei2021}; the operator here is nonlinear and has no
finite-Larmor-radius terms.  The code is built on PETSc~\cite{PETSc2024} and Kokkos~\cite{TrottKokkos2022}.  Batched direct
solvers factor many small band matrices at once on GPUs~\cite{Abdelfattah2023,Nayak2025}, among them matrices from
XGC~\cite{Nayak2025}, and Kokkos Kernels batches sparse Krylov solves~\cite{Liegeois2023}; block-tridiagonal GPU solvers treat one large
system, with blocks up to $4 \times 4$~\cite{KleinStrzodka2023} or by a recursive Schur-complement reduction over batched dense
kernels~\cite{Jin2025}.  K3 eliminates many block-tridiagonal systems, one GPU team a system, with blocks of 96 unknowns factored in
registers.

\textbf{LLM-assisted ports.}  LLM code assistants have translated whole HPC codes into C++ and Kokkos: the global gyrokinetic particle
code ORB5, whose translation agrees with the Fortran original to near machine accuracy on its first tests~\cite{Sadr2026}, and an ocean model of
about 74,000 lines of Fortran, through C, each stage validated against an acceptance criterion of its own~\cite{Koldunov2026}.  An
agentic workflow translates Fortran kernels to Kokkos without a person~\cite{Gupta2026}.  LLMs have been evaluated on HPC
kernels~\cite{Godoy2024}, and differential testing of LLM-translated HPC applications finds compilation a poor proxy for
correctness~\cite{Ludwig2026}.  The port of Section~\ref{sec:discipline} moves one solver, not a whole code, between two codes
built on PETSc's interfaces, and tests it against PETSc's own integrator.

\section{Limitations}
\label{sec:limitations}
The sweep couples processes by block Jacobi along $\varphi$ and between plane pieces, so the iteration count grows with the split.  We do not
pipeline the sweep across processes, as KBA does for neutron transport~\cite{BakerKoch1998}, because the exact coupling serializes the
pieces' chains; we have not measured it against the iterations it saves.  The sweep's depth grows with the mesh at fixed work per
device, and the time per launch approaches the launch's fixed cost.  The field's reductions and the collision transpose are staged through
host memory, and every plane piece factors and solves the whole plane's field.  Of the runs on the current code only the weak series' point C
splits the plane, and there every sparse triangular solve, the replicated field's among them, together takes under 1\% of the step.
The field multigrids of
Section~\ref{sec:field} distribute the field over the pieces but are not the default: on a level-6
plane over 192 GCDs the replicated field does not fit in device memory, and the distributed one takes 13.6\,s of a 174\,s step, 7.8\% (Table~\ref{tab:mg}c).  A process holds one grid's $\mu$ rows, so the load balance between
the species is the split of the $\mu$ blocks between the grids: split evenly, an electron process at point B holds two thirds of an ion process's state, and the electron processes wait; in the shaped
collisional model, whose ion sweep blocks drop the couplings between $\vpar$ nodes, the ion processes wait instead
(Section~\ref{sec:utilturb}).  The
series stops at point C, where a process holds 3 of a species' 24 $\mu$ rows, so the next doubling would split the plane into 32
pieces, an irregular layout; the large-scale series is a ladder in $\rho^*$ on the shaped collisional
model, which refines the plane with the velocity grid and the time step fixed, and follows this paper.
The sweep's block inverses, $72 \times 72$ per $\mu$ row, cell and segment, or four $18 \times 18$ for the ions with kinetic
electrons, divide over the $\mu$ blocks and plane pieces.  With kinetic electrons the blocks carry the potential, so a process holds
one set per wedge, and the inverses divide over the toroidal blocks only down to one wedge a process.  They set a floor on the devices
a level needs: at level 6 one $\mu$ row's $72 \times 72$ inverses for one wedge take 15\,GiB in single precision, and the ions'
$18 \times 18$ a quarter of that.
Level 6 of the concentric family has not been run with kinetic electrons.  The collision stage's discrete entropy is not monotone by construction (Section~\ref{sec:collisions}).  The shaped wedges' mortar carries the
gradient-$B$ mismatch of Section~\ref{sec:hard}, which converges with the mesh, and with adiabatic electrons keeps the mortar's seam: the
seam kick cancels only the streaming's share of the seam there, and the drift's share needs an in-plane kick on the mortar face, a
two-dimensional DG advection with its own edge fluxes, which we have not built.

In nonlinear runs the seam kick piles density at $\vpar = 0$ on the mortar's end stations: the linearized kick reflects the
population that the potential's jump would reflect and leaves it at the node that does not stream.  In an ion run with adiabatic
electrons at $\rho^* = 1/70$ and 48 wedges, its field signature, an alternation between the field stations at the toroidal mode $n =
N_w$, holds 19\% of the potential's spectral energy over $0.3 < r/a < 0.7$ and carries 0.3--1.3\% of the ion heat flux.  Removing the
pile needs a nonperturbative treatment of the potential's step at the mortar, which we have not built.

The weak series and the linear tests with kinetic electrons are measured at the physical mass ratio in the linear phase, over the
first steps of a run.  In a developed state at the mass
ratio 1836 on 24 $\vpar$ segments the stage's Newton solve fails: at the failing step the Jacobian at Newton's iterate is near-singular,
and a full Newton step grows the residual; on 25 segments, which put no node at $\vpar = 0$, it does not.  At 1836 the linear growth rate of the shaped plane's $(2, 1)$ mode is not converged in $\vpar$: it falls 10\% from 12 to 18 segments
and 8\% from 18 to 24, and 25 segments, with no node at $\vpar = 0$, give a rate a quarter above 24's.  The shaped collisional
model of Section~\ref{sec:turb} runs at deuterium's ratio, 3670.5, on 25 segments, so by this measure its electrons are not resolved
in $\vpar$, and no saturated-turbulence result with kinetic electrons is compared with a published one.  The benchmark's ion heat
diffusivity is half the published value inside the benchmark's transient, and the zonal shear layer that goes with the gap is not
explained (Section~\ref{sec:turb}).

The physics is electrostatic, and the electrons are drift-kinetic, with $k_\perp\rho_e \le 0.02$ in the runs here.  The gyroaverage acts on
the field, not on the collision operator, which omits the finite-Larmor-radius correction of order $k_\perp^2\rho_i^2$; the ring lies in the
poloidal plane, 0.8\% off the plane normal to $\mathbf{b}$; and the polarization is the Pad\'e form, 1.7\% under $1 - \Gamma_0$ at the
CBC's $k_\perp\rho_i$.  We have not tested the gyroaverage on the zonal-flow residual at finite $k_r\rho_i$ against Xiao and
Catto~\cite{XiaoCatto2006}: at the $k_r$ the test needs, the level-3 mesh holds one or two cells a wavelength, and the test needs levels 4
and 5.  The CBC comparisons have 18 plane nodes a poloidal wavelength, and we have not refined them further.  One Coulomb logarithm
serves every pair of species.

\section{Conclusion}
\label{sec:conclusion}
This paper has described five solvers for an implicit Eulerian gyrokinetic code: the upwind sweep, the plane multigrid smoothed by a
multiplicative, overlapping decomposition over flux-surface annuli, the batched block-tridiagonal collision solve, the low-storage
integrator and the fluid-electron block.  They are PETSc objects and a custom batched GPU solver, published as verified examples, with
porting prompts, that an LLM code assistant can apply flexibly in another application code\ifanon\else{}, listed with the prompts
in PETSc issue \#2015~\cite{PETScIssue2015}\fi; an assistant ported the integrator into
a clone of PETSc's main branch in 50 minutes.  The sweep preconditioner holds the advection solve to five to seven GMRES
iterations on one plane piece of the concentric meshes, from level 4 to level 6, and the sweep's dependence-ordered launches reproduce the sequential sweep
bitwise while cutting the launches from 484 to 175 (Section~\ref{sec:advection}).  With kinetic electrons at the physical mass ratio the stage's preconditioner, with the inflow transfer
across the toroidal blocks, holds GMRES to 13 to 17 iterations per Newton iteration over the weak series, from one device to 256.

The code they serve
carries gyrokinetic ions, drift-kinetic electrons at the physical mass ratio and nonlinear Landau collisions between every pair of
species, and runs on both vendors' GPUs from one source.  The decomposition splits the magnetic-moment rows before the plane, because the rows add no iterations: at
level 5 on 256 GCDs that choice is 1.5$\times$ faster than splitting the plane first.  With gyrokinetic ions the Cyclone base case's
growth rate at $\rho^* = 1/100$ is 2 to 10\% above GYSELA's.  In the shaped collisional model on eight A100s at $\Delta t = 1$ the
collision stage takes half the step, and its coefficient sum runs at 63\% of the FP64 rate.
Two of the method's hard parts have exact remedies, the seam kick on the circle and, with a plane rule of degree 14, constants
through the mortar; the drift at the shaped mortar converges with resolution.  Two remain open: the shaped mortar's seam with
adiabatic electrons, and the wall term at the edge.  The most urgent next steps are the $\rho^*$ ladder on the shaped collisional model, refining the plane with the velocity grid and the
time step fixed, with the distributed field multigrid in place of the replicated factorization, and the gyroaverage's test on the zonal-flow residual; then repetitions of every timed point.

Four extensions follow from the design.  The edge, with an X-point, open field lines and sheaths, needs the wall's term implicit, for
which Section~\ref{sec:hard} gives a Newton design.  Electromagnetic fluctuations add the parallel vector potential to the field's
unknowns and the shear Alfv\'en wave to the stiff modes; slow tearing and island evolution, over many Alfv\'en times, are where an
implicit step gains most.  Fusion alphas, one helium species slowing down through the same Landau operator on its own graded velocity
grid, would let the code compute the transport inputs that burn-accessibility models assume: the ash's confinement time relative to the
energy's, the energy confinement's dependence on temperature and heating power, and the tungsten concentration~\cite{DelgadoAparicio2026}.
A stellarator removes the toroidal symmetry that the fluid-electron block uses: its operators repeat each field period, not each
wedge, so the transform over the periods would leave one system per harmonic that couples a period's wedges, for the plane multigrid
with the toroidal coupling in its smoother.  The sweep, the plane multigrid and the collision solve do not use the symmetry, and we keep
them free of it.

\appendices

\begin{table}[t]
\centering
\caption{Toroidal Landau damping: $\gamma$ over $t \in [2.5, 9]$ against the root $-0.068384$.  Model: a one-dimensional model of the
toroidal scheme.  The 8-wedge 5D run predates the implicit-explicit split of Section~\ref{sec:step} and takes the toroidal streaming explicitly; at
16 wedges the two splits differ by 1\% in $\gamma$.  Level 3 runs with the equilibrium
inflow condition $f_0(H_0)$ at the wall; with the Boltzmann inflow condition it needs $\Delta t = 0.025$ (Appendix~\ref{app:wall}) and
gives $-0.06895$.}
\label{tab:landau}
\footnotesize
\begin{tabular}{@{}llrr@{}}
\toprule
level & wedges & 5D $\gamma$ & model $\gamma$ \\
\midrule
1 & 8 & $-0.0792$ & $-0.0817$ \\
1 & 16 & $-0.0648$ & $-0.0691$ \\
1 & 32 & $-0.0639$ & $-0.06842$ \\
2 & 16 & $-0.0680$ & \\
3 & 16 & $-0.06900$ & \\
\bottomrule
\end{tabular}
\end{table}

\begin{table}[t]
\centering
\caption{The slab ITG mode: $\gamma$ against the radial eigenproblem's $0.085561$ at the code's velocity quadrature, the fit's rms log
residual, and the difference of its two halves' rates.}
\label{tab:itg}
\footnotesize
\begin{tabular}{@{}lrrrr@{}}
\toprule
level & $\gamma$ & difference & residual & halves \\
\midrule
2 & & $-10.6\%$ & & \\
3 & 0.084798 & $-0.89\%$ & $1.9 \times 10^{-3}$ & 0.61\% \\
4 & 0.084715 & $-0.99\%$ & $2.0 \times 10^{-3}$ & 0.61\% \\
\bottomrule
\end{tabular}
\end{table}

\begin{table}[t]
\centering
\caption{The slab ITG mode with and without the gyroaverage at level 4: growth rate and frequency against the radial eigenproblem's
roots at the code's velocity quadrature, fit over $t \in [60, 100]$.}
\label{tab:itg-gk}
\footnotesize
\begin{tabular}{@{}lrrrr@{}}
\toprule
 & \multicolumn{2}{c}{gyrokinetic} & \multicolumn{2}{c}{drift-kinetic} \\
\cmidrule(lr){2-3}\cmidrule(l){4-5}
 & $\gamma$ & $\omega$ & $\gamma$ & $\omega$ \\
\midrule
reference & 0.079880 & $-0.389428$ & 0.085566 & $-0.232666$ \\
level 4 & 0.079904 & $-0.390793$ & 0.085420 & $-0.232480$ \\
difference & $+0.03\%$ & $+0.35\%$ & $-0.17\%$ & $-0.08\%$ \\
\bottomrule
\end{tabular}
\end{table}

\begin{table*}[t]
\centering
\caption{Conservation and null tests of the 5D wedge.  Defects are relative: the tiling to each cell's moments, constants and the
equilibrium to the state's maximum, the torque to the summed magnitudes of the momentum residual's terms, the budgets per step.}
\label{tab:null}
\footnotesize
\begin{tabular}{@{}>{\raggedright}p{0.19\textwidth}>{\raggedright}p{0.36\textwidth}>{\raggedright\arraybackslash}p{0.38\textwidth}@{}}
\toprule
test & configuration & measured \\
\midrule
supermesh tiling & concentric and Solov'ev, levels 2--6, 8--32 wedges & $10^{-15}$--$3 \times 10^{-14}$ \\
constants through the mortar & the same, both shapes & $2 \times 10^{-15}$--$5 \times 10^{-14}$ \\
well-balanced equilibrium & circle level 1, 8 wedges, 1000 steps of $\Delta t = 0.1$ & $1.5 \times 10^{-15}$; mass $2 \times 10^{-16}$, energy $3 \times 10^{-16}$ per step \\
numerical torque, $n = 0$ & zonal seed, 8 wedges, levels 1 / 2 & $1.2 \times 10^{-4}$ / $7.9 \times 10^{-6}$, order 3.9 \\
collisional null, unperturbed & circle level 2, 8 wedges, field, collisions every 2 steps, 2000 steps & no change; mass $1.3 \times 10^{-16}$, $K + U$ $1.1 \times 10^{-16}$ per step \\
collisional null, seeded $10^{-5}$ & the same & $n \ne 0$: $10^{-5} \to 1.3 \times 10^{-6}$; mass $5.5 \times 10^{-16}$, $K + U$ $5.3 \times 10^{-16}$ per step \\
anisotropic relaxation & uniform slab, every node one batch & each node = the lone batch to $10^{-15}$; mass $2 \times 10^{-14}$, energy $3 \times 10^{-14}$ \\
\bottomrule
\end{tabular}
\end{table*}

\section{Toroidal Landau damping}
\label{app:landau}
This test checks the streaming along $\varphi$ and the field's toroidal coupling against an analytic damping rate.  The wedges carry a
slab: a uniform field and a uniform $k_\parallel$ with $R_0 = 3$, $\tau = 8$, a toroidal mode $n = 2$, a velocity grid of 16
segments of degree 3, the adiabatic field without polarization, and $\Delta t = 0.1$.  The damping rate is fit to the potential inside
$r < a/2$ over $t \in [2.5, 9]$ and compared with the root of the dispersion relation~\cite{LandauDamping1946,FriedConte1961}, $\gamma = -0.068384$ (Table~\ref{tab:landau}).

Two resolutions control the rate.  Along $\varphi$, a one-dimensional model of the same toroidal scheme converges to the root from
8 to 32 wedges, and the 5D period follows the model's.  In the plane, the wall's slivers and inflow condition make a boundary layer in the
plane's charge, which on level 1's two rings of cells reaches $r < a/2$; the 5D rate at level 1 therefore stops short of the root even
at 32 wedges, and it converges as the plane refines, to 0.9\% of the root at level 3.  The level-3 rate is the same to six digits on one
MI250X GCD and on one A100.  With the electrons kinetic at the mass ratio 3670.5, on their own 64 graded segments
and the ions on 16 uniform ones, level 2 damps at $-0.0893$, 2.5\% from the two-species root $-0.0871$
(Table~\ref{tab:verification}); collisionless, the electrons need the 64 segments: a grid of 48 does not reproduce the damping, in the
code or in a model of the same discretization.

\section{A slab ITG mode}
\label{app:itg}
This test checks a growing mode against an exact reference.  The wedges carry a slab with a rigid $q = 1.4$, a uniform field
($B_0 = 13.333$, so the magnetic drift and the mirror force vanish), $\tau = 1$, and the $(m, n) = (2, 2)$ ITG mode~\cite{Coppi1967} driven by a
temperature profile of width $0.1\,a$ with peak gradient $R/L_T = 20$; 8 wedges, a velocity grid of 12 segments of degree 3 by 18
$\mu$ rows, and $\Delta t = 0.2$.  In this configuration the linear dynamics is local in $r$ except for the polarization term of the
field equation, so the growing mode is the ground state of an exact one-dimensional radial eigenproblem.  Solved with the code's own
velocity quadrature, that eigenproblem gives $\gamma = 0.085561$, 1.2\% above the root with continuous velocities~\cite{FriedConte1961}.  The growth rate is
fit to the envelope of the field's $n \ne 0$ part over $t \in [40, 65]$ at mid-radius (Table~\ref{tab:itg}).

Level 2 misses by 10.6\% because the plane does not resolve the eigenfunction's radial width; levels 3 and 4 agree with the reference
to 1\%, converged in $h$, and the remaining 1\% belongs to the fixed parts, the 8 wedges, the time step and the $\mu$ grid.  Level 3 on
16 GCDs reproduces the A100's values digit for digit.

\textbf{With the gyroaverage.}  The same mode with gyrokinetic ions is the ground state of the gyrokinetic radial eigenproblem, with the
polar ring average of each $\mu$ row~\cite{Steiner2015} and the code's Pad\'e polarization, solved at the code's velocity quadrature:
finite Larmor radius raises the frequency 1.67$\times$ and lowers the rate 6.6\% (Table~\ref{tab:itg-gk}).  Fit over $t \in [60, 100]$,
past the gyrokinetic run's start-up transient, the level-4 rate and frequency are within 0.03\% and 0.35\% of the reference, and the
finite-Larmor-radius shift of the rate, the gyrokinetic rate less the drift-kinetic one, is 97\% of the reference's.  At level 3 the
shift is 83\%: there the largest rings, $\rho = 0.225$, span four or five cells of the mesh, and the gyrokinetic rate is 1.1\% high.

\section{Conservation and null tests}
\label{app:null}
The gates of Section~\ref{sec:discipline} run a few steps.  Table~\ref{tab:null} collects the tests that hold the discrete
structure over a whole run or over every mesh level.  The supermesh tiles every cell's moments to rounding on both mesh shapes, and
the streaming preserves constants to rounding (Section~\ref{sec:hard}).  The well-balanced equilibrium stays put to rounding over
1000 steps.  The toroidal angular momentum's budget closes by construction, and its numerical torque, zero in the continuum for an
axisymmetric state~\cite{ScottSmirnov2010,BrizardTronko2011}, falls at fourth order in $h$.  The collisional null test runs the field, the well-balanced advection and the
collision stage together for 2000 steps: unperturbed, no collision stage changes the state and the budgets hold mass and $K + U$ to
rounding; seeded with a non-axisymmetric perturbation, the perturbation decays and does not grow.  The anisotropic relaxation of a
spatially uniform state checks the collision stage inside the wedge: every node evolves as one batch stepped alone, and the stage
conserves mass and energy to its Newton tolerance.

\section{The wall}
\label{app:wall}
This appendix gives the wall's Lyapunov functional, the term that sets its time-step limit, and the slivers' integral
(Section~\ref{sec:hard}).

\textbf{The Lyapunov functional.}  With the Boltzmann inflow condition, the relative entropy to the equilibrium plus the field energy,
$\Lambda = (f, \ln(f/f_0) - 1) + U$, changes only through the wall.  Each wall face contributes $-|\mathbf{F} \cdot \mathbf{n}|\,\rho$,
where $\mathbf{F}$ is the guiding-centre flux, $\mathbf{n}$ the face's outward normal and $\rho(f, g) = f \ln(f/g) - f + g \ge 0$,
taken on an outflow face between the trace and $f_0(H_0 + \phi)$ and on an inflow face between the inflow state and $f_0(H_0 + \phi)$.
The Boltzmann inflow condition makes the inflow state equal to $f_0(H_0 + \phi)$, so the inflow faces contribute zero, the outflow faces
contribute no more than zero, and $\Lambda$ cannot grow.

\textbf{The time-step limit.}  To find the wall term that limits the explicit step, we computed the leading eigenvalue of the explicit
terms' Jacobian by Arnoldi iterations~\cite{Arnoldi1951}, with one part of the wall masked out at a time.  Masking the slivers leaves the
eigenvalue unchanged; masking the lateral wall's faces divides it by nine.  The limit is therefore the lateral wall's.  There the inflow
condition's upwind term is quadratic in the perturbation, and the term's Jacobian about a wall potential $\phi_w$ is a tangential
derivative of size $\pi \tau \phi_w / (B_0 h^2)$, so the term's explicit limit on $\Delta t$ falls as $h^2$ and grows as $B_0$.

\textbf{The slivers' integral.}  Only a wall that is a flux surface of the discrete map would remove the slivers.  The integral treats
each sliver as its own region: a closed chain of the mortar's own curve segments, joined bitwise at the pieces' ends, integrated by a fan
rule with about 165 points per sliver.  The alternative, integration by subtraction (the whole face by the cell's rule less the face's
supermesh pieces by theirs), needs up to 2400 quadrature points per wall cell, each with the equilibrium evaluated from the invariants by
a Newton solve, and with gradient profiles it is inexact, since it integrates over a cell hundreds of times the sliver's size across a
square-root singularity of $f_0$.  On the CBC at level 4 on 16 GCDs, on the code of the comparison, the sliver integral took 5.9\% of the step against subtraction's
29\%, and the three ion verification tests do not change between the two; on this paper's code the slivers take 5.1\% of the CBC
step and 3.0\% of the shaped GAM's.

\section{The step of the adiabatic configurations}
\label{app:util}

\begin{table}[t]
\centering
\caption{The step of the two ion configurations, drift-kinetic ions with adiabatic electrons: seconds per step after
one warm-up step, uninstrumented, and GMRES iterations per stage.  GCD/A100 is the ratio of the two steps.}
\label{tab:steps}
\footnotesize
\setlength{\tabcolsep}{4pt}
\begin{tabular}{@{}lrrrrr@{}}
\toprule
configuration & devices & GCD & A100 & GCD/A100 & GMRES \\
\midrule
CBC, circle L4 & 16 & 5.30 & 3.72 & 1.43 & 5.0 \\
shaped GAM, Solov'ev L4 & 16 & 8.24 & 8.69 & 0.95 & 8.0 \\
\bottomrule
\end{tabular}
\end{table}

\textbf{The step.}  On 16 GCDs at level 4, optimizations of the residual's kernels took the shaped GAM configuration from 22.5 to 10.7\,s per step,
every change bitwise, and later ones, the wall slivers' integral of Section~\ref{sec:hard} among them, to 8.19\,s.  Table~\ref{tab:steps}
gives the step of the two ion configurations.  The A100 is 1.43$\times$ faster on the circle, and the GCD 5\% faster on the shaped mesh.

\section{The field multigrids' design}
\label{app:fieldmg}
\begin{table}[t]
\centering
\caption{The field multigrids' design, measured on the code's own operators: the circle at $B_0 = 70$, $\Delta t = 0.25$, $\tau = 1$,
gyrokinetic ions.  Top: errors relative to the block system's Cholesky solve at level 2.  Bottom: two-grid convergence factors from level 3
to level 2 per symmetric V(1,1) cycle, Galerkin, with the exact coarse solve, the asymptotic factor over 40 cycles.  Jacobi--Chebyshev:
point Jacobi with Chebyshev(3) on $[0.1, 1.1]\,\lambda_{\max}$; annulus: the one-annulus block Gauss--Seidel of
Figure~\ref{fig:annuli}.  The row $S_\parallel \times 37.7$ scales the parallel-polarization term by $(430/70)^2$, its growth against
$K_{\mathrm{pol}}$ from $B_0 = 70$ to $B_0 = 430$ ($\rho^* = 1/430$).}
\label{tab:fieldmg}
\footnotesize
\begin{tabular}{@{}lr@{}}
\toprule
check, level 2 & error \\
\midrule
(\ref{eq:identity}), adiabatic fluctuations, $c = 1$ & $5.2 \times 10^{-16}$ \\
(\ref{eq:identity}), kinetic electrons, $c = 0$, gauged & $5.5 \times 10^{-15}$ \\
zonal response, inner error $3.4 \times 10^{-5}$: & \\
\quad $G$, $C$ and the correction from the solve & $6.4 \times 10^{-4}$ \\
\quad $G$, $C$ at set-up, $G C^{-1} G^T l$ per solve & $8.3 \times 10^{-8}$ \\
\end{tabular}

\smallskip
\begin{tabular}{@{}lrr@{}}
\toprule
operator, level 3 & Jacobi--Chebyshev & annulus \\
\midrule
$H$, adiabatic fluctuations & 0.090 & 0.022 \\
$K_{\mathrm{pol}}$, kinetic-electron fluctuations & 0.443 & 0.122 \\
mean, adiabatic & 0.089 & 0.109 \\
mean, adiabatic, $S_\parallel \times 37.7$ & 0.847 & 0.076 \\
mean, kinetic electrons & 0.663 & 0.116 \\
\bottomrule
\end{tabular}
\end{table}

Table~\ref{tab:fieldmg} gives the design's tests of Section~\ref{sec:fieldmg}, on the code's own operators.  The
identity~(\ref{eq:identity}) matches the block system's Cholesky solve to rounding on the level-2 fluctuation operator, with adiabatic
and with kinetic electrons.  Formed from a solve with a relative error of $3.4 \times 10^{-5}$, the Woodbury correction's $G$ and $C$
carry an error of $6.4 \times 10^{-4}$ into the zonal response; formed once at set-up, with the cycle iterated to $10^{-12}$, they
leave $8.3 \times 10^{-8}$.  In the two-grid test the annulus smoother holds every operator at 0.02 to 0.12 per cycle, also with the
parallel-polarization term scaled 37.7-fold, its growth against $K_{\mathrm{pol}}$ from $B_0 = 70$ to $\rho^* = 1/430$.  Point Jacobi
with Chebyshev acceleration matches it on the Helmholtz operator and the adiabatic mean, and has two-grid factors of 0.85 on the scaled
mean and 0.66 on the kinetic-electron mean, which lacks the adiabatic mass.  On a turbulence run at level 3 with adiabatic electrons,
as few as two annulus cycles, a relative error of at most $10^{-4}$, leave the factorization's Newton and GMRES counts and its heat flux unchanged to the
printed digits over 40 steps.

\section{The inflow transfer}
\label{app:transfer}
\begin{table}[t]
\centering
\caption{The inflow transfer on point A's grid of the weak series (level 2, 8 wedges, one A100) at the base series' step there,
$\Delta t = 0.5$, the step fixed and the toroidal blocks thinned.  GMRES iterations per Newton iteration of the kinetic-electron stage; $\xi$: the blocks an electron at the
thermal speed crosses in a stage; $J$: the lean passes, about $2\xi$ in these runs.}
\label{tab:transfer}
\footnotesize
\begin{tabular}{@{}lrrrr@{}}
\toprule
wedges per block & 8 & 4 & 2 & 1 \\
blocks crossed in a stage, $\xi$ & 0.76 & 1.51 & 3.03 & 6.05 \\
\midrule
two sweeps & 14.7 & 18.0 & 29.2 & 47.4 \\
lean passes, $J$ & 2 & 3 & 6 & 12 \\
the transfer (three sweeps) & 12.7 & 12.2 & 12.3 & 12.8 \\
\bottomrule
\end{tabular}
\end{table}

The sweep of Section~\ref{sec:advection} solves $T_s y = x$, where $T_s$ is the stage operator $T$ without the couplings the sweep
omits, among them the couplings $C_b$ across the toroidal block boundaries.  The inflows $c$ at those boundaries satisfy $c = r + Gc$, with $r = P_{\mathrm{out}}\,T_s^{-1} x$ the outflow the
sweep delivers to each boundary and $G = P_{\mathrm{out}}\,T_s^{-1} C_b$: $C_b$ injects an inflow across the mortar at a block's upstream boundary, $T_s^{-1}$
carries it through the block's wedges, and $P_{\mathrm{out}}$ restricts the result to the downstream boundary.  $G$ has the same form in every block,
but its values differ, since each block's inverses depend on the potential.
A lean pass applies $G$ with the sweep's stored $72 \times 72$ inverses, over the electrons' rows alone.  It needs no right side; no
in-plane face, since the drift moves an electron a fraction of a cell while it crosses a block; and no $\vpar$ face, since the mirror
force inside a segment stays in the block's inverse.  On one A100 at point A a lean pass takes 3.9 to 5.8\,ms with its exchange,
against 17 to 26\,ms for a sweep.

The default number of passes is $J = \lceil \xi \rceil$, with $\xi = N_B\,\sigma_e / (2\pi R_0\, s)$, $N_B$ the toroidal blocks and
$\sigma_e/s = \sigma_e \gamma \Delta t$ the distance an electron at the thermal speed travels in a stage (\texttt{-pc\_kinsweep\_series
auto}; an integer sets $J$).  Table~\ref{tab:transfer} thins the blocks at a fixed step: the iterations of two sweeps triple, and with
$J \approx 2\xi$ the transfer holds them between 12 and 13.  At point B, at $\xi = 3.03$, $J = 3$, 4 and 7 give 15.3, 15.2 and 15.1
iterations per Newton iteration against two sweeps' 22.6, and at most 22 in a solve against 30; $J = 3$ has the shortest step, 20.7\,s against two sweeps' 21.8\,s (with collisions every fourth step), so the transfer pays for
itself at that step, and more passes add time but not iterations.  The weak series runs $J = 3$ at every point.

The integral transport matrix method iterates on the same system of interface fluxes, with the subdomain operators formed and
stored~\cite{AnistratovAzmy2015}; the lean pass applies $G$ matrix-free, from the inverses the sweep already holds.

\ifanon\else
\section{Artifact description}
\label{app:ad}
\newcommand{\adsection}{\subsection}
\input{kintora-sc-ad-body}
\clearpage   %
\fi

\section*{Acknowledgments}

\textbf{Use of AI.}  KINTORA's source code, its tests, the analysis scripts and the figure generators were written, and much of the
design of Sections~\ref{sec:background}--\ref{sec:discipline} developed, with Anthropic's Claude~\cite{claude2026}
(Fable~5, 5.1 and~5.2, Opus~5 and~5.5, Sonnet~5), an AI assistant working under \ifanon the authors' \else the author's \fi direction.  The
text of every section and appendix, Sections~\ref{sec:intro}--\ref{sec:conclusion} and Appendices~\ref{app:landau}--\ifanon\ref{app:transfer}\else\ref{app:ad}\fi, was
drafted with the same assistant and reviewed and revised by \ifanon the authors\else the author\fi, who \ifanon are \else is \fi
accountable for its content.  The references it located were checked against their DOIs, and each result's data file names its runs and the
script that generates it.  This paragraph is the disclosure of AI-generated text that SC and IEEE policy
require.
\ifanon

Further acknowledgments, among them funding and facilities, are withheld for double-anonymous review.
\else

This material is based upon work supported by the U.S. Department of Energy, Office of Science, Office of Advanced Scientific Computing
Research, Scientific Discovery through Advanced Computing (SciDAC) Program through the FASTMath Institute, under contract number
DE-AC02-05CH11231 at Lawrence Berkeley National Laboratory.  This research used resources of the National Energy Research Scientific
Computing Center, a DOE Office of Science User Facility supported by the Office of Science of the U.S. Department of Energy under Contract
No.\ DE-AC02-05CH11231 using NERSC award ASCR-ERCAP0039362.
This research used resources of the Oak Ridge Leadership Computing Facility at the Oak Ridge National Laboratory, which is supported by
the Advanced Scientific Computing Research programs in the Office of Science of the U.S. Department of Energy under Contract
No.\ DE-AC05-00OR22725.
\fi

\bibliographystyle{IEEEtran}
\bibliography{kintora-sc}

\end{document}

%% file: sec-turbulence.tex
\subsection{Turbulence}
\label{sec:turb}
The solvers' purpose is the developed turbulence, so this section runs two nonlinear cases past saturation: the smallest system of the
global ion-temperature-gradient scan of~\cite{McMillan2010}, a published benchmark, and a shaped collisional model with kinetic electrons at
deuterium's mass ratio, for which no published benchmark exists; it demonstrates the solvers in the regime they are built for.  The first runs a circular equilibrium in straight-field-line coordinates at
$1/\rho^* = 70$, with gyrokinetic ions, adiabatic electrons and no collisions.  Its temperature profile is a flat top with
$R_0/L_T = 7.1$ in the gradient region, held by a Krook heat source that acts on the energy alone, at the rate of the GENE--ORB5 benchmark of~\cite{Lapillonne2010},
0.0125, which is 3.6 times the ORB5 rate of the benchmark we compare with~\cite{McMillan2010}, and its
initial state is a local Maxwellian.  The grid is the level-5 isotropic-triangle plane, 6,435 cells, with 48 wedges, 12 $\vpar$
segments of degree 3 and 24 $\mu$ rows, at $\Delta t = 0.25$ on six Frontier nodes (48 GCDs).  We measured the resolution in the
nonlinear phase: 64 wedges move $\chi_i$ by 4\% and the level-6 plane by 11--17\%, against 62--72\% from level 4 to level 5.

The benchmark defines the match: $\chi_i$ over $r/a \in [0.4, 0.6]$ averaged in windows of 500 $a/c_s$ after a transient of 600, against
the gradient the run realizes, for which ORB5 and GENE give 0.41 and 0.39 in units of the gyro-Bohm diffusivity $\chi_{\mathrm{GB}}$ at $R_0/L_T = 6.96$~\cite{McMillan2010}.

\begin{figure}[t]
\centering
\includegraphics[width=\columnwidth]{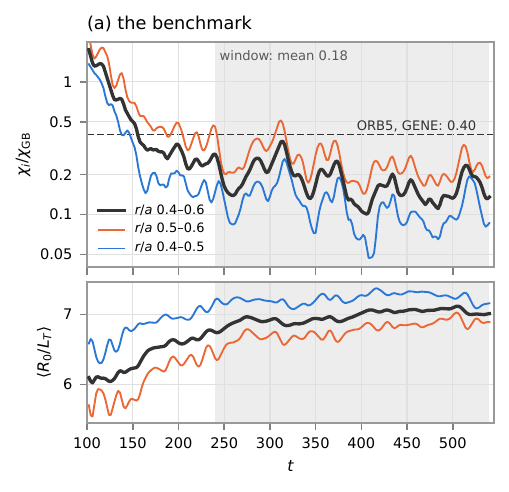}
\includegraphics[width=\columnwidth]{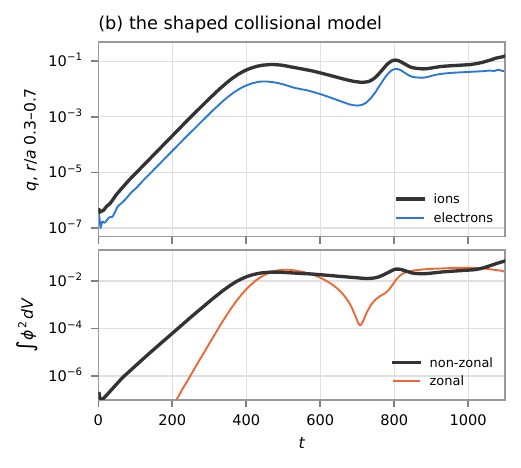}
\caption{Saturated turbulence.  (a) The benchmark's ion heat diffusivity over $r/a \in [0.4, 0.6]$ and over its halves, and their gradients; shaded, $t \in [240, 540]$.  (b) The shaped collisional model's ion and
electron heat fluxes over $r/a \in [0.3, 0.7]$, and its potential's energy, zonal and not.}
\label{fig:turb}
\end{figure}

The turbulence saturates after its first burst, at $t \approx 70$, and the comparison needs more: a quasi-steady state, in which
$\chi_i$, the zonal flow and the gradient fluctuate about steady means, tested by the agreement of the window's halves.  Over $t \in
[240, 540]$, inside the benchmark's transient, our $\chi_i$ is 0.18 $\chi_{\mathrm{GB}}$ at the realized $R_0/L_T = 6.97$, about half the
references' 0.40.  The run shows where the flux is lost: the burst builds a zonal $E \times B$ shear layer at $r/a \approx 0.4$ whose shearing
rate keeps growing after saturation, to 2.0--2.4 times the linear growth rate, and inside it, at $r/a = 0.38$, $\chi_i$ falls to 0.03--0.09 while the
gradient stays at its steepest; the outer half of the window carries the flux (Figure~\ref{fig:turb}a).  Whether the layer erodes over McMillan's window, $t \in
[600, 1100]$, decides the comparison.

The shaped collisional model runs a Solov'ev plane with SPARC's elongation, triangularity and aspect ratio~\cite{Creely2020}, 234
cells in six flux rings, with 16 wedges, gyrokinetic ions and kinetic electrons at deuterium's mass ratio, 3670.5, both species on 25
$\vpar$ segments, Landau collisions between every pair of species every step at the collision frequency of SPARC's reference plasma, and
the flat-top profile of the benchmark at $R_0/L_T = 7.1$, at $\Delta t = 1$ on one Frontier node (eight GCDs) to $t = 1100$.  The turbulence saturates in a first burst at $t \approx 480$, falls quiet and bursts again at $t \approx 800$, and its
ion heat flux is still rising at $t = 1100$ (Figure~\ref{fig:turb}b): the run reaches no stationary level, and the case shows the
solvers' behaviour through saturation, not a transport result.

\textbf{The solvers in saturated turbulence.}  The counts hold through saturation: over the benchmark's 1,760 steps from $t = 100$ to 540 the
advection solve takes at most 7 GMRES iterations and Newton at most 2, and a step costs 12.8\,s on six
Frontier nodes.  Over the
shaped model's 1,100 steps the kinetic-electron stage's 6,004 solves take at most 21 GMRES iterations, 10.2 on average, and Newton at most 2 a stage.  The collision stage's Newton takes at most 7 iterations a batch, summed over its two substeps.  A step of $\Delta t = 1$ costs 27 to 41\,s on eight GCDs, rising with the turbulence, 47 to 51\% of it the collision stage.  Over four
steps from $t = 1000$, the budget, the change less the wall's flux, closes the mass to $1.2 \times 10^{-7}$ and $K + U$ to $7 \times 10^{-8}$ of their totals a step at the solves' tolerances; with the
solves 100$\times$ tighter it closes the mass to $2 \times 10^{-11}$ and $K + U$ to $1.2 \times 10^{-8}$, the time integrator's $O(\Delta t^4)$ part
(Section~\ref{sec:conservation}).

%% file: fig/tab_kernels_turb.tex
\begin{tabular}{@{}lrrrrrrrrrrrrr@{}}
\toprule
kernel & \% step & ms & GB/s & DRAM & L2 & L1 & SM & occ. & regs & local & lanes & FP64 & FP32 \\
\midrule
\multicolumn{14}{@{}l}{\emph{the collisionless step, 16.25\,s a step, 47\% of the step (on the GCD 19.80\,s, 48\%)}} \\
sweep, one launch per level & 10.4 & 0.137 & 567 & 28 & 13 & 19 & 15 & 15 & 129 & 0.0 & 26.2 & 3.7 & 0.0 \\
Jacobian, wall slivers & 9.3 & 23.7 & 172 & 8 & 23 & 37 & \textbf{71} & 17 & 185 & 10.3 & 32.0 & 51.3 & 0.2 \\
Jacobian, per-point terms & 4.5 & 11.4 & 107 & 5 & 14 & 59 & \textbf{64} & 17 & 168 & 6.5 & 25.4 & 36.7 & 0.3 \\
sweep, the blocks' inverses & 3.6 & 3.20 & 71 & 4 & \textbf{63} & 58 & 38 & 44 & 81 & 3.2 & 24.9 & 12.5 & 0.0 \\
Jacobian, lateral faces & 2.3 & 5.81 & 253 & 12 & 17 & 37 & \textbf{62} & 12 & 255 & 0.6 & 32.0 & 37.1 & 0.0 \\
Jacobian, mortar faces & 1.9 & 4.72 & 336 & 16 & 29 & \textbf{74} & 21 & 25 & 128 & 1.6 & 32.0 & 19.3 & 0.0 \\
Jacobian, seam kick & 1.8 & 4.60 & 177 & 9 & 17 & 42 & 22 & 12 & 253 & 3.6 & 25.0 & 15.1 & 0.0 \\
\midrule
\multicolumn{14}{@{}l}{\emph{the collision stage, 18.06\,s a step, 53\% of the step (on the GCD 21.05\,s, 52\%)}} \\
coefficient sum & 39.4 & 190 & 4 & 0 & 6 & 15 & \textbf{80} & 23 & 128 & 0.7 & 31.9 & 62.9 & 0.2 \\
Gauss--Jordan, the blocks' factors & 7.7 & 0.995 & 48 & 2 & 2 & 31 & 55 & 47 & 64 & 0.0 & 22.6 & 0.0 & 6.5 \\
Gauss--Jordan, the blocks' assembly & 2.7 & 0.353 & 233 & 11 & 31 & 37 & 18 & 12 & 212 & 0.0 & 26.6 & 5.9 & 0.1 \\
\bottomrule
\end{tabular}

%% file: kintora-sc-ad-body.tex
\adsection{Artifact identification}

\textbf{Contributions.}
\begin{itemize}
\item[C1] Five solvers published as verified examples, four with porting prompts for an LLM code assistant, listed in PETSc issue
\#2015 (\url{https://gitlab.com/petsc/petsc/-/issues/2015}); one, K4, ported into a clone of PETSc's main branch (the paper's Table~I, Section~X).
\item[C2] The advection solve (K1): GMRES preconditioned by one upwind sweep, on the GPU as dependence-ordered launches.
\item[C3] The kinetic-electron preconditioner (K1, K5): the sweep, an inflow transfer across the toroidal blocks, and a fluid-electron block solved by toroidal modes.
\item[C4] The plane multigrid (K2), smoothed over flux-surface annuli, for the fluid-electron block and the Pad\'e field solves.
\item[C5] The collision solve (K3): a batched block-tridiagonal direct solver for the Newton steps of Landau collisions.
\item[C6] A conservative Eulerian DG discretization: exact mass, and energy except for $O(\Delta t^4)$, by a kick at the mortar.
\item[C7] A gyroaverage on the unstructured P2 plane: a Galerkin ring form by exact clipping, its exact adjoint the deposit.
\item[C8] A 4D decomposition of the 5D phase space, measured in weak scaling to 256 devices and strong scaling to 192.
\item[C9] The hardware utilization on the A100 of a shaped model with gyrokinetic ions, kinetic electrons at deuterium's mass ratio and Landau collisions.
\end{itemize}
The results that support them without being one (V): the verification tests, the saturated turbulence, the gates of the check suite.

\textbf{Computational artifacts.}  All are public in \url{https://gitlab.com/markadams4/petsc-claude-papers}, directory
\texttt{kintora-eulerian/} (A1, A3, A4) and \texttt{petsc/} and \texttt{kintora/} (A2, A5); the paper's source and PDF are in
\texttt{kintora-eulerian/paper/kintora-sc/}.
\begin{itemize}
\item[A1] The solver, frozen at each commit a result ran on (this description's Table~\ref{tab:versions}), with the library \texttt{src/kin}, the
driver \texttt{src/sldg/ex12.cxx} and the meshes \texttt{src/sldg/meshes/}: \texttt{code/} holds commit 99c011d, \texttt{code-v0.2/}
tag \texttt{v0.2} (commit 2f145d2), \texttt{code-d44b12e/} commit d44b12e and \texttt{code-2de209d/} the weak series' base point A.  Every other
commit of Table~\ref{tab:versions} is a patch, \path{code-<commit>.patch}, the difference of its tree from the frozen tree its row names
(\texttt{git apply} in that directory); \path{PATCHES.txt} lists each patch's base.  The 4ca57b7 patch is v0's code.
\item[A2] PETSc from our fork \texttt{petsc-claude}, whose current tree is the \texttt{petsc/} directory of
\url{https://gitlab.com/markadams4/petsc-claude-papers}; its \texttt{petsc/FORK-COMMITS.md} lists the commits named here.  On Frontier, commit
\texttt{eda0b4c370a} of its \texttt{main} branch (it adds the supernodal triangular solve, \texttt{-mat\_kokkos\_sptrsv supernodal}); for v0.2,
commit \texttt{70fd866} and, on Perlmutter, its child \texttt{ba193c0} (64-bit indices in the cuDSS interface); every v0 run on Perlmutter (the weak
series, both its points and its base, Table~XI and the runs of Section~V-A's inflow transfer) ran on \texttt{5ce415651a7}, a descendant of
\texttt{eda0b4c370a} that syncs the cuDSS interface with its upstream merge request, as their logs print.  Configured with 64-bit indices,
Kokkos and Kokkos Kernels (Table~\ref{tab:sw}).
\item[A3] The run scripts, in the working repository's layout: \texttt{scripts/v0/} for the v0 campaign (its \texttt{README.md} maps each
script to its data file), \texttt{scripts/s4/} (the weak series), \texttt{scripts/turb/} (the McMillan benchmark), \texttt{scripts/c1d/}
(the shaped collisional model, its \texttt{README.md} the order of the runs), \texttt{scripts/turbutil/} (Table~XI), \texttt{scripts/sc27/}
(v0.2's runs) and \texttt{scripts/frontier/} (Frontier's environment and gates); the check suite \path{scripts/check_ex12_eul.sh} and its
parallel driver \path{scripts/check_par.*}; the fitters and figure generators the data generators import, \texttt{scripts/*.py}.
\item[A4] The data and the generators: \texttt{paper/kintora-sc/data/*.csv}, one row per run or per number with its job id and log path, each
file's header saying how it was measured, which claim of the paper it serves (by the paper's words and its \LaTeX\ labels, not by number)
and the code it ran on; the runs' logs that the generators read, \texttt{paper/kintora-sc/data/hist/} (gzipped where large); the generators,
\texttt{paper/kintora-sc/*.py} (Table~\ref{tab:map}); meshes the runs read that are not in the frozen trees, \texttt{kintora-eulerian/meshes/}.  Some
numbers rest on the working repository's design ledger, which is not public; \path{data/wedge-gates.csv} transcribes each with its run.
Superseded and kept for the record: \path{collide.csv} (the collision stage over $\mu$ blocks, cut from v0), \path{s4-b0scan.csv},
\path{kinetic-itg-L4.csv}, \path{pathII-counts.csv}, \path{v0-pointA-cbc.csv}, \path{v0-weak16.csv}, \path{v0-weak16-parts.csv},
\path{v0-weak16-breakdown.csv} and \path{v0-gateD-compare.csv} (the first weak series and its gate, replaced by \path{s4-weak.csv}).
\item[A5] The solvers' current source, public in the \texttt{kintora/} directory of the same repository and refreshed from our working
repository: \texttt{kintora/PORTING.md} lists the solvers of the paper's Table~I with their files and the prompts that port them,
\texttt{kintora/README.md} maps the rest of the code, and \texttt{kintora/SNAPSHOT.txt} names the commit, 4ca57b7 (v0's code) at this writing.
\end{itemize}

\begin{table}[t]
\centering
\caption{The commits the paper's results ran on, of the working repository \texttt{plasma-kinetics}, with where their source is in
\texttt{kintora-eulerian/}: a frozen tree, or a patch to the frozen tree named.  Every data file names its code; a run on a clone with
uncommitted work names the commit that later carried it.}
\label{tab:versions}
\footnotesize
\setlength{\tabcolsep}{3pt}
\begin{tabular}{@{}ll>{\raggedright\arraybackslash}p{0.27\textwidth}@{}}
\toprule
commit & date & results (source) \\
\midrule
e0e4c7c & 10-14 & Tables XII (level 1, 8 wedges) and XV (tiling, constants, the well-balanced equilibrium, the torque) (patch to \texttt{code-v0.2/}) \\
69cfdb6 & 10-14 & Table XII (level 1, 16 and 32 wedges; level 2) (patch to \texttt{code-v0.2/}) \\
3d1fd82 & 10-15 & Table XIII (patch to \texttt{code-v0.2/}) \\
a01b79e & 10-15 & Tables XII (level 3) and XIII (level 3 on the GCDs) (patch to \texttt{code-v0.2/}) \\
dd9a5eb & 10-15 & Table XV (the anisotropic relaxation) (patch to \texttt{code-v0.2/}) \\
8639050 & 10-16 & Section III-B's schedule ablation, the base (patch to \texttt{code-v0.2/}) \\
aeb4d45 & 10-18 & Table VII's drift-kinetic level-4 row and the $\nu^* = 0.063$ decay (patch to \texttt{code-v0.2/}) \\
f5bde6e & 10-18 & Fig.~\ref{fig:hr} and the $\nu^* = 0.25$ decay, Table XV (the collisional nulls), Section VIII's entropy (patch to \texttt{code-v0.2/}) \\
4af462a & 10-21 & Sections II-A's and III-B's decomposition and segment-item ablations (patch to \texttt{code-v0.2/}) \\
975b423 & 10-22 & Fig.~\ref{fig:scaling}(a)'s level-4 to level-6 rows on the MI250X, the advection counts, Section X's decomposition spread (patch to \texttt{code-v0.2/}) \\
2f145d2 & 10-22 & tag \texttt{v0.2}: Fig.~\ref{fig:scaling}(a), Appendix E's optimization steps (\texttt{code-v0.2/}) \\
99c011d & 10-28 & Table VII's verification scripts, Table XVI (\texttt{code/}) \\
ed99aff & 10-29 & the A100's circle rows of Table XVI, the seam kick's cost (patch to \texttt{code/}) \\
761bf27 & 10-31 & Table VII's e--D relaxation (patch to \texttt{code/}) \\
4fe03a0 & 10-31 & Table XIV (patch to \texttt{code-d44b12e/}) \\
d12de2a & 11-01 & Table VII's gyrokinetic CBC rows at $B_0 = 50$ (patch to \texttt{code-d44b12e/}) \\
4baea16 & 11-01 & the CBC rows at $B_0 = 100$ and at 3670.5 (patch to \texttt{code-d44b12e/}) \\
d44b12e & 11-02 & the base of the two patches above (\texttt{code-d44b12e/}) \\
c88b383 & 11-09 & Fig.~\ref{fig:turb}(a)'s start (E8) and the benchmark's resolution pairs (patch to \texttt{code-d44b12e/}) \\
62d87d8 & 11-14 & the fluid-electron block's three terms, Section V-B (patch to \texttt{code-2de209d/}) \\
5688691 & 11-16 & the ions' pole in the fluid-electron block (patch to \texttt{code-2de209d/}) \\
174f8a9 & 11-19 & Fig.~\ref{fig:turb}(a), the McMillan benchmark (patch to \texttt{code-2de209d/}) \\
e766e43 & 11-21 & the shaped collisional model at $\Delta t = 0.25$, Table XI's checkpoint (patch to \texttt{code-2de209d/}) \\
2de209d & 11-22 & the weak series' base (Table IX's gain), point A on the A100 (\texttt{code-2de209d/}) \\
5d942b5 & 11-22 & the base: B and C on the A100; f11b442's \texttt{s4opts.sh} (patch to \texttt{code-2de209d/}) \\
6a11dcd & 11-23 & Table XI's base ($\Delta t = 0.25$) on the A100; the weak series' base on the MI250X (patch to \texttt{code-2de209d/}) \\
375872b & 11-23 & Table XI's base on the MI250X (patch to \texttt{code-2de209d/}) \\
4d17abf & 11-24 & the inflow transfer at point A (branch \texttt{xport}, Section V-A) (patch to \texttt{code-2de209d/}) \\
7afd303 & 11-24 & the inflow transfer at point B; the step's CBC rows; the check suite's decomposition legs (patch to \texttt{code-2de209d/}) \\
4ca57b7 & 11-24 & v0's code: Tables VIII--XI, Fig.~\ref{fig:turb}(b); edfdb1c's \texttt{s4opts.sh} (patch to \texttt{code-2de209d/}) \\
4e2d724 & 11-25 & Table XI's Nsight Compute passes; its changes are defaults that case does not use (patch to \texttt{code-2de209d/}) \\
\bottomrule
\end{tabular}
\end{table}

\begin{table*}[t]
\centering
\caption{Each result of the paper, its data, its runs and its generator.  Data files are under \texttt{paper/kintora-sc/data/} and their
logs under its \texttt{hist/}, run scripts under \texttt{scripts/}, generators in \texttt{paper/kintora-sc/} unless a path is given.
Contributions C1--C9 and V as in Section~I.}
\label{tab:map}
\footnotesize
\setlength{\tabcolsep}{4pt}
\begin{tabular}{@{}>{\raggedright}p{0.2\textwidth}>{\raggedright}p{0.06\textwidth}>{\raggedright}p{0.25\textwidth}>{\raggedright}p{0.27\textwidth}>{\raggedright\arraybackslash}p{0.13\textwidth}@{}}
\toprule
result & contrib. & data & runs & generator \\
\midrule
Figs.~\ref{fig:decomp}--\ref{fig:annuli} (the method) & C2--C8 & Fig.~2(b): \path{sweep-order-solovev_L3.txt} and \path{.faces} (row 0) & \texttt{ex12 -eul\_wedge\_pc\_dump\_order} & \path{scripts/gen_sc_*_figure.py} \\
Table I (the solvers as components) & C1 & the solvers' files and prompts, \path{kintora/PORTING.md} (A5) & -- & -- \\
Table II (the grid), Section III-A (the integrator's memory) & C1, C8 & \path{k4-memory.csv}, \path{s4-weak.csv} & the weak series' memory runs (\path{hist/s4w/*mem*}) & -- \\
Section II-A, III-B (the decomposition order, the sweep on the device, segment items) & C2, C8 & \path{sweep-device.csv}; \path{k1-advection-counts.csv} (the counts) & Frontier, \path{hist/sweep/}, \path{hist/scaling/} & \path{sweep_device.py} \\
Section IV (constants, the shaped drift, the seam kick, the null test) & C6 & \path{g1b-quadrature.csv}, \path{seam-kick-cost.csv}, \path{wedge-gates.csv} & Mac (G1b); PM 60487993 (\path{scripts/v0/kcabr.sbatch}) & \path{seam_kick_cost.py}, \path{wedge_gates.py} \\
Section V-A (the inflow transfer, the stage's tolerance) & C3 & \path{inflow-transfer.csv}; \path{fluid-block.csv} (katol) & PM 61925505 (point A), 61948135, 61942359 (point B) & \path{inflow_transfer.py} \\
Section V-B (the fluid-electron block's terms, the ions' pole) & C3 & \path{fluid-block.csv} & PM 61277523, 61291733, 61346599, 61348937, 61350182 & \path{fluid_block.py} \\
Table III, Section VI-B (the plane multigrid as built) & C4 & \path{k2-multigrid.csv} & Frontier 5769019, 5773874, 5901621, 5901722, 5913510; PM 61091137, 61106090, 61127814 & -- \\
Table XVII, Appendix F (the field multigrids' design) & C4 & \path{fieldmg-design.csv} (the design harness, ledger D1350--D1364) & -- & -- \\
Section VII (the gyroaverage's gates) & C7 & \path{wedge-gates.csv} & the check suite (G-K0, G-K1) & \path{wedge_gates.py} \\
Table IV (the collision solve) & C5 & \path{k3-collision-solve.csv} & PM 60555267, 60555268, 60557628, 60553319; the proxy of \texttt{-coll\_check\_aniso2} & -- \\
Section VIII (the collision stage's checks) & C5, V & \path{coll-gates-3670.csv}, \path{landau-relax-3670.csv}, \path{wedge-gates.csv} (entropy, gates) & the check suite's multi-species leg; \texttt{-coll\_check\_aniso2}, PM 60617351; PM 59966181, 59980711 & \path{coll_gates.py}, \path{relax_fit.py}, \path{wedge_gates.py} \\
Section VIII (the collision stage's costs) & C5 & \path{k3-collision-solve.csv}, its costs rows (ledger D1057, D1061, D1062, D1094) & PM 60642585, 60642738 & -- \\
Table V, Section IX (the energy budget) & C6, C7 & \path{wedge-gates.csv} (the ledger rows with their runs); \path{c1d-solvers.csv} (the shaped model's rows) & the check suite; \path{scripts/sc27/pmseam.sbatch} (v0.2); FT 6011620, 6011813 & \path{wedge_gates.py}, \path{c1d_data.py} \\
Section X (verification discipline) & C1--C8 & \path{decomp-states.csv} (several processes against one), \path{wedge-gates.csv} (the kernels, the sweep, the gates) & PM CPU 61941973 (the full check on v0's tree, \path{hist/checks/}); Frontier, \path{hist/scaling/} & \path{decomp_states.py}, \path{wedge_gates.py} \\
Table VII (verification) & C2, C4, C6, V & \path{v0-verification.csv} & \path{scripts/v0/pmverif16.sbatch}, \path{fverif.sbatch} & -- \\
\quad its CBC rows and the text's diagnostics & C2, C7, V & \path{cbc-B.csv} (gyrokinetic, the rho$^*$ point, the step), \path{cbc-diag.csv} (drift-kinetic level 4, the end spectrum, A100 against MI250X, the equilibrium), \path{external-cbc.csv} & \path{scripts/s4/} (\texttt{RUN=fit}); Frontier 5709502, 5709918, 5733888, 5988414, 5988415; PM 60140345 & \path{cbc_fit.py}, \path{cbc_diag.py} \\
\quad its Landau damping with kinetic electrons & V & \path{landau-3670.csv} & PM 60539036 & \path{landau_e_fit.py} \\
\quad its level-4 shaped GAM & V & \path{shaped-gam-L4.csv} & \path{pmgamL4.sbatch}, \path{pmgamL4_2d.sbatch} & \path{gam_fit.py} \\
Fig.~\ref{fig:hr}, Table VII (the zonal-flow decay) & V & \path{hr_rt.csv} ($\nu^* = 0.25$), \path{hr_rt_nu063.csv} ($\nu^* = 0.063$), \path{hr-xiao.csv} & PM 59966181, 59966183, 59966208; 59995753 & \path{fig_hr.py}, \path{hr_xiao.py} \\
Fig.~\ref{fig:turb}, Section XI-C (saturated turbulence) & C2, C3, C4, V & \path{r5-chi.csv}, \path{r5-chi_in.csv}, \path{r5-chi_out.csv}, \path{r5-chi-e8.csv}, \path{r5-rings.csv}, \path{r5-resolution.csv}, \path{r5-solvers.csv} (the benchmark); \path{c1d-hist.csv}, \path{c1d-hist-dt025.csv}, \path{c1d-solvers.csv} (the shaped model) & \path{scripts/turb/fte8.sbatch} (the benchmark; Frontier 5891858 (E8), 5950128, 5950207, 5955410, 5957580; the resolution pairs PM 61018731, Frontier 5891859, 5893987, 5895463); \path{scripts/c1d/c1dt0.sbatch} (the shaped model at $\Delta t = 1$; Frontier 5998185, 5998186, 6001492, 6001493, 6006700--6006705; the ledger runs 6011620, 6011812, 6011813); \path{scripts/c1d/} (its run at $\Delta t = 0.25$, in the order of its \path{README.md}; Perlmutter 61764251, 61791500, 61797518, 61797631, 61803286, 61812781, 61819618, 61852038) & \path{r5_data.py}, \path{scripts/turb_chi.py}, \path{scripts/turbana.py}, \path{c1d_data.py}, \path{fig_turb.py} \\
Fig.~\ref{fig:scaling}(a), Section XI-D (strong scaling) & C2, C8 & \path{scaling.csv}, \path{scaling-reps.csv} (v0.2) & \path{scripts/sc27/s43series.sh}, \path{pms43series.sh} & \path{fig_scaling.py} \\
Fig.~\ref{fig:scaling}(b, c), Tables VIII and IX (weak scaling; by part) & C2--C5, C7, C8 & \path{s4-weak.csv}; the base, \path{s4-weak-base.csv} & \path{scripts/s4/pms4.sbatch}, \path{fts4.sbatch}; PM 61955851 (A, B), 61956368 (C); Frontier 5989337, 5989338, 5989432 (A, B, C); the base PM 61789994, 61817699, 61817698 (61789995, B's memory), Frontier 5978370, 5978371, 5981409 & \path{weak_s4.py}, \path{fig_scaling.py} \\
Table X (the gyroaverage's cost) & C7 & \path{s4-weak.csv} & Table VIII's & \path{weak_s4.py} \\
Table XI, Section XI-E (the current model's kernels) & C2--C5, C9 & \path{turb-step-a100.csv}, \path{turb-util-a100.csv}, \path{turb-step-mi250x.csv}; the base \path{turb-step-a100-dt025.csv}, \path{turb-step-mi250x-dt025.csv} & \path{scripts/turbutil/pmutil_c1d.sbatch}, \path{ftab_c1d.sbatch}; PM 61971838, 62014201, 62025986; Frontier 5992585; the base PM 61879444, Frontier 5979406 (logs \path{hist/turbstep/}) & \path{scripts/turbutil/gen_step.py}, \path{ncutime.py}, \path{turbtab.py}, \path{tab_turb.py} \\
Tables XII, XIII, XV, Appendices A--C (toroidal Landau damping, the slab ITG, conservation and null tests) & C6, V & \path{wedge-gates.csv} (the ledger rows with their runs; the nulls and the entropy from \path{hist/null/}); \path{landau-3670.csv} & the check suite; the runs \path{wedge-gates.csv} names, at the commits of Table~\ref{tab:versions} & \path{wedge_gates.py} \\
Table XIV (the slab ITG with the gyroaverage) & C7 & \path{gk-slab-itg.csv} & PM 60637637 & \path{gk_slab_fit.py}; \path{scripts/ex11_radial_itg.py --flr pade} (the reference) \\
Appendix D, Section XI-D (the wall, the slivers' share of the step) & C6 & \path{wedge-gates.csv}; \path{v0-kernels-mi250x.csv} & \path{scripts/v0/futil.sbatch} & \path{wedge_gates.py} \\
Table XVI (the steps, Appendix E) & V & \path{v0-steps.csv} (the A100's circle row on ed99aff) & \path{fgam.sbatch}, \path{scripts/sc27/fcost43.sbatch}, \path{pmutil.sbatch} & -- \\
\bottomrule
\end{tabular}
\end{table*}

\adsection{Dependencies}

\textbf{Hardware.}  Frontier (OLCF): nodes of four AMD MI250X (eight GCDs of 64\,GB), one process per GCD.  Perlmutter (NERSC): nodes of
four NVIDIA A100s, 40\,GB (the \texttt{hbm40g} constraint) for the weak series and 80\,GB (\texttt{hbm80g}) for Table~XI and the other
timed runs, one process per GPU, every GPU visible to every process
(\texttt{--gpus-per-node=4 --gpu-bind=none}; with one visible device per process the device halos' IPC handles cannot be opened).  Both use
GPU-aware Cray MPICH over Slingshot 11 (\texttt{MPICH\_GPU\_SUPPORT\_ENABLED=1}).

\begin{table}[h]
\centering
\caption{Software of the v0 runs, read from the run logs and the builds' configure logs (the arches are \texttt{arch-frontier-opt-gcc-kokkos-hip-r64-i64} and \texttt{arch-perlmutter-opt-gcc-kokkos-cuda-i64}).  PETSc's \texttt{-log\_view} header prints the commit it was configured at: 7ee3f992905 on Frontier, where the
library was rebuilt at eda0b4c370a without reconfiguring, and 5ce415651a7 on Perlmutter (A2).}
\label{tab:sw}
\footnotesize
\begin{tabular}{@{}lll@{}}
\toprule
 & Frontier & Perlmutter \\
\midrule
PETSc arch & \texttt{hip-r64-i64} & \texttt{cuda-i64} \\
PETSc (fork) & eda0b4c370a & 5ce415651a7 \\
compiler & GCC 14.2 (CPE 25.09) & GCC 14.3 \\
GPU stack & ROCm 6.4.2 & CUDA 13.2 \\
MPI & Cray MPICH 9.0.1 & Cray MPICH 9.1.0 \\
Kokkos & 5.2.2 & 5.2.2 \\
field solve & Kokkos Cholesky & cuDSS 0.8.0 \\
\bottomrule
\end{tabular}
\end{table}

\textbf{Inputs.}  The meshes of A1 and \texttt{kintora-eulerian/meshes/}; no other data.  Every run starts from the analytic equilibrium and a seeded
mode, except restarts from a run's own checkpoint: the McMillan benchmark (from E8's $t = 100$ dump), the shaped collisional model's
restart chains, and Table~XI, which restarts from the $\Delta t = 0.25$ run's checkpoint at $t = 811$ (PM 61812781).  The checkpoints are
not in the artifact (several GB a run); a chain is reproduced from its first segment.

\adsection{Installation}

PETSc is configured with \texttt{--with-64-bit-indices}, Kokkos, Kokkos Kernels and the device's backend by the configure scripts
in \texttt{configure/}, \path{arch-frontier-opt-gcc-kokkos-hip-r64-i64.py} and \path{arch-perlmutter-opt-gcc-kokkos-cuda-i64.py}, run
from the fork's root.  On Frontier, \texttt{code/scripts/frontier/env.sh} loads the toolchain.  Then
\texttt{make libkin.a} in \texttt{src/kin} and \texttt{make ex12} in \texttt{src/sldg} build the solver, with no compiler warnings on
either machine, in under 10 minutes on a login node.  The run trees of the campaign were \texttt{git archive}s of 99c011d, the same files
as \texttt{code/}, and of the later commits of Table~\ref{tab:versions}.  The generators need Python 3 with NumPy and SciPy.

\adsection{Experiments}

Every experiment is one batch script of A3 with its options in the script; the solver prints the quantities the paper reports.  The step
time is PETSc's \texttt{-log\_view}: the \texttt{Solve} stage over its steps, the collision stage included, after one warm-up step in the weak
series and four in Table~XI, and the \texttt{TSStep} event over the timed steps in the scaling series and the ablations.  Each v0 point is one run.
\begin{itemize}
\item \textbf{Verification} (Table VII): \texttt{pmverif16.sbatch} (16 A100s) and \texttt{fverif.sbatch} (16 GCDs) with \texttt{CASE=itg},
\texttt{hr} or \texttt{gam} run the level-3 cases at their published layouts: the CBC linear ITG, 500 steps without and with the
flux-surface-averaged electrons, about 9 minutes; the zonal-flow decay pair at $\nu^* = 0.25$, 500 steps each, 6 to 9 minutes; the shaped
GAM to $t = 25$, 6 minutes.  The CBC growth rates reproduce v0.2's to 0.06\% in every radial bin, the shaped GAM's zonal potential v0.2's to
$9 \times 10^{-12}$ of its maximum; the zonal-flow decay moves by $1.1 \times 10^{-3}$, the seam kick's change.  The fits are
\texttt{ex11\_rh\_fit.py} (GAM) and \texttt{eul\_n1\_fit.py} (decay).
\item \textbf{Weak scaling} (Fig.~\ref{fig:scaling}(b, c), Tables VIII--X): \texttt{scripts/s4/pms4.sbatch} (A100 40\,GB) and \texttt{fts4.sbatch}
(MI250X GCD) run a point with the options of \texttt{s4opts.sh}, gyrokinetic and drift-kinetic ions and a memory run, 8 steps with a
collision stage every step.  Every point ran on code 4ca57b7 on both machines with the \texttt{s4opts.sh} of edfdb1c (case \texttt{e}:
the time steps of Table VIII, the cut series, \texttt{-pc\_kinsweep\_series 3}, and the predictor guard).  The base series of Table IX's
gain ran with a quarter of those steps and a collision stage every 4: on the A100 point A on 2de209d and points B and C on 5d942b5, with
the \texttt{s4opts.sh} of f11b442 (B's memory run is 2de209d's), and on the MI250X on 6a11dcd.  \texttt{weak\_s4.py} reads their
\texttt{-log\_view} logs, kept in \texttt{data/hist/s4w/}, into \path{s4-weak.csv} and \path{s4-weak-base.csv}.  Point C takes 64
Perlmutter nodes or 32 Frontier nodes.
\item \textbf{The inflow transfer} (Section V-A): point A of the weak series, 2 processes on one A100, 4 steps, with the toroidal blocks
thinned by \texttt{-pc\_kinsweep\_lpr} 8, 4, 2 and 1 at a fixed step (PM 61925505, branch \texttt{xport} at 4d17abf), and point B, 16 A100s,
$\Delta t = 1$, two sweeps against the transfer with 3, 4 and the automatic number of lean passes (PM 61948135, 61942359, 7afd303's code);
the runs' options are in the logs' option tables, \texttt{inflow\_transfer.py} reads them into \path{inflow-transfer.csv}.
\item \textbf{The wall slivers' share} (Appendix D): \texttt{futil.sbatch} with \texttt{CFG=CBC} or \texttt{GAM} profiles rank 0
under rocprofv3 (seven counter passes and a kernel trace), and \texttt{kwutil2.py} reduces them into \path{v0-kernels-mi250x.csv}.
\item \textbf{The current model's kernels} (Table XI): \texttt{scripts/turbutil/pmutil\_c1d.sbatch} restarts the shaped
collisional model, its options those of \texttt{scripts/c1d/} with \texttt{XO="-dt 1 -coll\_every 1"}, on 8 A100s 80\,GB from its
checkpoint at $t = 811$ and runs 4 warm-up and 8 timed steps per pass: uninstrumented (the step by part, \texttt{gen\_step.py} into
\path{turb-step-a100.csv}, the slowest and the fastest process's time of each part), under Nsight Compute on one process (\texttt{PR}), one pass per kernel group, and once more under Nsight
Compute timing every launch of that process (\texttt{ncutime.py} with \texttt{WSTAGE=4}, the shares of the step; jobs 61971838, 62014201
and 62025986); \texttt{turbtab.py} merges them into
\path{turb-util-a100.csv} and \texttt{tab\_turb.py} writes the table.  The MI250X's column, the step by part, is \texttt{ftab\_c1d.sbatch}'s
uninstrumented pass on 8 GCDs (job 5992585; \texttt{XO="-dt 1 -coll\_every 1 -eul\_wedge\_fluid\_emap\_project"}), into \path{turb-step-mi250x.csv}.  The base, at $\Delta t = 0.25$ with collisions every
fourth step, is the same scripts' plain passes without \texttt{XO} (PM 61879444, Frontier 5979406), in the \path{-dt025} files.
\item \textbf{The steps} (Table XVI): the uninstrumented runs of \texttt{fgam.sbatch} (shaped GAM), \texttt{scripts/sc27/fcost43.sbatch}
(CBC, MI250X) and \texttt{pmutil.sbatch}'s plain leg (A100).
\item \textbf{Strong scaling} (Fig.~\ref{fig:scaling}(a); v0.2): \texttt{s43series.sh} and \texttt{pms43series.sh} submit
the strong-scaling points, five runs a point on the MI250X.  The level-4 to level-6 rows are \texttt{fcost41.sbatch}'s runs on 975b423
(\path{hist/scaling/}), whose states across the layouts are \path{decomp-states.csv}'s second part.
\item \textbf{The seam kick's cost} (Section~IV): \texttt{kcabr.sbatch} runs the CBC at level 3 on 2 A100s with
\texttt{-eul\_wedge\_seam\_kick 1} and \texttt{0}, 4 timed steps each, then profiles the level-4 proxy's mortar kernel under Nsight
Compute; \texttt{seam\_kick\_cost.py} reads them into \path{seam-kick-cost.csv}.
\item \textbf{The check suite} (Section X): \path{scripts/check_par.sbatch} runs \path{check_ex12_eul.sh}'s legs in parallel on a
Perlmutter CPU node (job 61941973: 354 assertions over 206 legs, 0 FAIL); its output, \path{hist/checks/}, holds the gates' numbers that
\path{wedge-gates.csv} and \path{decomp-states.csv} read.
\end{itemize}

%% file: kintora-sc.bbl
\begin{thebibliography}{100}
\providecommand{\url}[1]{#1}
\csname url@samestyle\endcsname
\providecommand{\newblock}{\relax}
\providecommand{\bibinfo}[2]{#2}
\providecommand{\BIBentrySTDinterwordspacing}{\spaceskip=0pt\relax}
\providecommand{\BIBentryALTinterwordstretchfactor}{4}
\providecommand{\BIBentryALTinterwordspacing}{\spaceskip=\fontdimen2\font plus
\BIBentryALTinterwordstretchfactor\fontdimen3\font minus
  \fontdimen4\font\relax}
\providecommand{\BIBforeignlanguage}[2]{{%
\expandafter\ifx\csname l@#1\endcsname\relax
\typeout{** WARNING: IEEEtran.bst: No hyphenation pattern has been}%
\typeout{** loaded for the language `#1'. Using the pattern for}%
\typeout{** the default language instead.}%
\else
\language=\csname l@#1\endcsname
\fi
#2}}
\providecommand{\BIBdecl}{\relax}
\BIBdecl

\bibitem{Germaschewski2021}
K.~Germaschewski, B.~Allen, T.~Dannert, M.~Hrywniak, J.~Donaghy, G.~Merlo,
  S.~Ethier, E.~D'Azevedo, F.~Jenko, and A.~Bhattacharjee, ``Toward exascale
  whole-device modeling of fusion devices: Porting the {GENE} gyrokinetic
  microturbulence code to {GPU},'' \emph{Physics of Plasmas}, vol.~28, no.~6,
  p. 062501, 2021.

\bibitem{Mandell2024GX}
N.~R. Mandell, W.~Dorland, I.~Abel, R.~Gaur, P.~Kim, M.~Martin, and T.~Qian,
  ``{GX}: a {GPU}-native gyrokinetic turbulence code for tokamak and
  stellarator design,'' \emph{Journal of Plasma Physics}, vol.~90, no.~4, p.
  905900402, 2024.

\bibitem{Malaboeuf2025}
E.~Malaboeuf, K.~Obrejan, M.~Peybernes, J.~Bigot, E.~Bourne, and
  V.~Grandgirard, ``Porting a {F}ortran plasma simulation to exascale on {AMD}
  {GPUs} using both {OpenMP} and {Kokkos},'' in \emph{Proceedings of the SC '25
  Workshops of the International Conference for High Performance Computing,
  Networking, Storage and Analysis}.\hskip 1em plus 0.5em minus 0.4em\relax
  ACM, 2025, pp. 2051--2067.

\bibitem{XGCgpu2021}
S.~M. Mniszewski, J.~Belak, J.-L. Fattebert, C.~F.~A. Negre, S.~R. Slattery,
  A.~A. Adedoyin, R.~F. Bird, C.~Chang, G.~Chen, S.~Ethier, S.~Fogerty,
  S.~Habib, C.~Junghans, D.~Lebrun-Grandi{\'e}, J.~Mohd-Yusof, S.~G. Moore,
  D.~Osei-Kuffuor, S.~J. Plimpton, A.~Pope, S.~T. Reeve, L.~Ricketson,
  A.~Scheinberg, A.~Y. Sharma, and M.~E. Wall, ``Enabling particle applications
  for exascale computing platforms,'' \emph{The International Journal of High
  Performance Computing Applications}, vol.~35, no.~6, pp. 572--597, 2021.

\bibitem{CockburnShu2001}
B.~Cockburn and C.-W. Shu, ``{Runge--Kutta} discontinuous {Galerkin} methods
  for convection-dominated problems,'' \emph{Journal of Scientific Computing},
  vol.~16, no.~3, pp. 173--261, 2001.

\bibitem{Landau1936}
L.~D. Landau, ``Kinetic equation for the {C}oulomb effect,''
  \emph{Physikalische Zeitschrift der Sowjetunion}, vol.~10, p. 154, 1936.

\bibitem{Choquette2021}
J.~Choquette, W.~Gandhi, O.~Giroux, N.~Stam, and R.~Krashinsky, ``{NVIDIA}
  {A100} {Tensor} {Core} {GPU}: Performance and innovation,'' \emph{IEEE
  Micro}, vol.~41, no.~2, pp. 29--35, 2021.

\bibitem{Atchley2023}
S.~Atchley, C.~Zimmer, J.~Lange, D.~Bernholdt, V.~Melesse~Vergara, T.~Beck,
  M.~Brim, R.~Budiardja, S.~Chandrasekaran, M.~Eisenbach, T.~Evans, M.~Ezell,
  N.~Frontiere, A.~Georgiadou, J.~Glenski, P.~Grete, S.~Hamilton, J.~Holmen,
  A.~Huebl, D.~Jacobson, W.~Joubert, K.~Mcmahon, E.~Merzari, S.~Moore,
  A.~Myers, S.~Nichols, S.~Oral, T.~Papatheodore, D.~Perez, D.~M. Rogers,
  E.~Schneider, J.-L. Vay, and P.~K. Yeung, ``{Frontier}: Exploring exascale,''
  in \emph{Proceedings of the International Conference for High Performance
  Computing, Networking, Storage and Analysis}, 2023, pp. 1--16.

\bibitem{TrottKokkos2022}
C.~R. Trott, D.~Lebrun-Grandi{\'e}, D.~Arndt, J.~Ciesko, V.~Dang,
  N.~Ellingwood, R.~Gayatri, E.~Harvey, D.~S. Hollman, D.~Ibanez, N.~Liber,
  J.~Madsen, J.~Miles, D.~Poliakoff, A.~Powell, S.~Rajamanickam, M.~Simberg,
  D.~Sunderland, B.~Turcksin, and J.~Wilke, ``Kokkos 3: Programming model
  extensions for the exascale era,'' \emph{IEEE Transactions on Parallel and
  Distributed Systems}, vol.~33, no.~4, pp. 805--817, 2022.

\bibitem{PETSc2024}
S.~Balay, S.~Abhyankar, M.~F. Adams, S.~Benson, J.~Brown, P.~Brune,
  K.~Buschelman, L.~Dalcin, A.~Dener, V.~Eijkhout \emph{et~al.},
  ``{PETS}c/{TAO} users manual,'' 2024,
  \url{https://petsc.org/release/docs/manual/}.

\bibitem{Kotschenreuther1995}
M.~Kotschenreuther, G.~Rewoldt, and W.~M. Tang, ``Comparison of initial value
  and eigenvalue codes for kinetic toroidal plasma instabilities,''
  \emph{Computer Physics Communications}, vol.~88, no.~2, pp. 128--140, 1995,
  the GS2 code.

\bibitem{Barnes2019}
M.~Barnes, F.~I. Parra, and M.~Landreman, ``stella: An operator-split,
  implicit--explicit $\delta f$-gyrokinetic code for general magnetic field
  configurations,'' \emph{Journal of Computational Physics}, vol. 391, pp.
  365--380, 2019.

\bibitem{Lee1987}
W.~W. Lee, ``Gyrokinetic particle simulation model,'' \emph{Journal of
  Computational Physics}, vol.~72, no.~1, pp. 243--269, 1987.

\bibitem{BelliHammett2005}
E.~A. Belli and G.~W. Hammett, ``A numerical instability in an {ADI} algorithm
  for gyrokinetics,'' \emph{Computer Physics Communications}, vol. 172, no.~2,
  pp. 119--132, 2005.

\bibitem{Jenko2000}
F.~Jenko, W.~Dorland, M.~Kotschenreuther, and B.~N. Rogers, ``Electron
  temperature gradient driven turbulence,'' \emph{Physics of Plasmas}, vol.~7,
  no.~5, pp. 1904--1910, 2000, the GENE code.

\bibitem{SaadSchultz1986}
Y.~Saad and M.~H. Schultz, ``{GMRES}: A generalized minimal residual algorithm
  for solving nonsymmetric linear systems,'' \emph{SIAM Journal on Scientific
  and Statistical Computing}, vol.~7, no.~3, pp. 856--869, 1986.

\bibitem{BrownComposable2012}
J.~Brown, M.~G. Knepley, D.~A. May, L.~C. McInnes, and B.~Smith, ``Composable
  linear solvers for multiphysics,'' in \emph{2012 11th International Symposium
  on Parallel and Distributed Computing}, 2012, pp. 55--62.

\bibitem{KintoraCode2026}
M.~F. Adams, ``{KINTORA}: the solver's source,''
  \url{https://gitlab.com/markadams4/petsc-claude-papers/-/tree/main/kintora},
  2026, a snapshot of the private repository; \texttt{kintora/PORTING.md} lists
  the solvers, their files and the prompts that port them, and the PETSc fork
  the code builds on is \texttt{petsc/} of the same repository.

\bibitem{Littlejohn1983}
R.~G. Littlejohn, ``Variational principles of guiding centre motion,''
  \emph{Journal of Plasma Physics}, vol.~29, no.~1, pp. 111--125, 1983.

\bibitem{CaryBrizard2009}
J.~R. Cary and A.~J. Brizard, ``Hamiltonian theory of guiding-center motion,''
  \emph{Reviews of Modern Physics}, vol.~81, no.~2, pp. 693--738, 2009.

\bibitem{Dubin1983}
D.~H.~E. Dubin, J.~A. Krommes, C.~Oberman, and W.~W. Lee, ``Nonlinear
  gyrokinetic equations,'' \emph{Physics of Fluids}, vol.~26, no.~12, pp.
  3524--3535, 1983.

\bibitem{BrizardHahm2007}
A.~J. Brizard and T.~S. Hahm, ``Foundations of nonlinear gyrokinetic theory,''
  \emph{Reviews of Modern Physics}, vol.~79, no.~2, pp. 421--468, 2007.

\bibitem{Lee1983}
W.~W. Lee, ``Gyrokinetic approach in particle simulation,'' \emph{Physics of
  Fluids}, vol.~26, no.~2, pp. 556--562, 1983.

\bibitem{Dimits2000}
A.~M. Dimits \emph{et~al.}, ``Comparisons and physics basis of tokamak
  transport models and turbulence simulations,'' \emph{Physics of Plasmas},
  vol.~7, no.~3, pp. 969--983, 2000.

\bibitem{Solovev1968}
L.~S. Solov'ev, ``The theory of hydromagnetic stability of toroidal plasma
  configurations,'' \emph{Soviet Physics JETP}, vol.~26, p. 400, 1968.

\bibitem{CerfonFreidberg2010}
A.~J. Cerfon and J.~P. Freidberg, ```{One} size fits all' analytic solutions to
  the {Grad--Shafranov} equation,'' \emph{Physics of Plasmas}, vol.~17, no.~3,
  p. 032502, 2010.

\bibitem{FarrellMaddison2011}
P.~E. Farrell and J.~R. Maddison, ``Conservative interpolation between volume
  meshes by local {G}alerkin projection,'' \emph{Computer Methods in Applied
  Mechanics and Engineering}, vol. 200, no. 1--4, pp. 89--100, 2011.

\bibitem{KennedyCarpenter2003}
C.~A. Kennedy and M.~H. Carpenter, ``Additive {Runge--Kutta} schemes for
  convection--diffusion--reaction equations,'' \emph{Applied Numerical
  Mathematics}, vol.~44, no. 1-2, pp. 139--181, 2003.

\bibitem{Anzt2019}
H.~Anzt, J.~Dongarra, G.~Flegar, N.~J. Higham, and E.~S. Quintana-Ort{\'i},
  ``Adaptive precision in block-{Jacobi} preconditioning for iterative sparse
  linear system solvers,'' \emph{Concurrency and Computation: Practice and
  Experience}, vol.~31, no.~6, p. e4460, 2019.

\bibitem{HackbuschProbst1997}
W.~Hackbusch and T.~Probst, ``Downwind {Gau{\ss}--Seidel} smoothing for
  convection dominated problems,'' \emph{Numerical Linear Algebra with
  Applications}, vol.~4, no.~2, pp. 85--102, 1997.

\bibitem{BeyWittum1997}
J.~Bey and G.~Wittum, ``Downwind numbering: robust multigrid for
  convection-diffusion problems,'' \emph{Applied Numerical Mathematics},
  vol.~23, no.~1, pp. 177--192, 1997.

\bibitem{AndersonSaad1989}
E.~Anderson and Y.~Saad, ``Solving sparse triangular linear systems on parallel
  computers,'' \emph{International Journal of High Speed Computing}, vol.~1,
  no.~1, pp. 73--95, 1989.

\bibitem{ThomasLombard1979}
P.~D. Thomas and C.~K. Lombard, ``Geometric conservation law and its
  application to flow computations on moving grids,'' \emph{AIAA Journal},
  vol.~17, no.~10, pp. 1030--1037, 1979.

\bibitem{Kahan1965}
W.~Kahan, ``Pracniques: further remarks on reducing truncation errors,''
  \emph{Communications of the ACM}, vol.~8, no.~1, p.~40, 1965.

\bibitem{Winsor1968}
N.~Winsor, J.~L. Johnson, and J.~M. Dawson, ``Geodesic acoustic waves in
  hydromagnetic systems,'' \emph{Physics of Fluids}, vol.~11, no.~11, pp.
  2448--2450, 1968.

\bibitem{DalMasoLeFlochMurat1995}
G.~Dal~Maso, P.~G. LeFloch, and F.~Murat, ``Definition and weak stability of
  nonconservative products,'' \emph{Journal de Math{\'e}matiques Pures et
  Appliqu{\'e}es}, vol.~74, pp. 483--548, 1995.

\bibitem{Angelino2006}
P.~Angelino, A.~Bottino, R.~Hatzky, S.~Jolliet, O.~Sauter, T.~M. Tran, and
  L.~Villard, ``On the definition of a kinetic equilibrium in global
  gyrokinetic simulations,'' \emph{Physics of Plasmas}, vol.~13, no.~5, p.
  052304, 2006.

\bibitem{Berberich2021}
J.~P. Berberich, P.~Chandrashekar, and C.~Klingenberg, ``High order
  well-balanced finite volume methods for multi-dimensional systems of
  hyperbolic balance laws,'' \emph{Computers \& Fluids}, vol. 219, p. 104858,
  2021.

\bibitem{HairerWanner1996}
E.~Hairer and G.~Wanner, \emph{Solving Ordinary Differential Equations
  {II}}.\hskip 1em plus 0.5em minus 0.4em\relax Springer Berlin Heidelberg,
  1996.

\bibitem{KnollKeyes2004}
D.~A. Knoll and D.~E. Keyes, ``{Jacobian}-free {Newton--Krylov} methods: a
  survey of approaches and applications,'' \emph{Journal of Computational
  Physics}, vol. 193, no.~2, pp. 357--397, 2004.

\bibitem{MurphyGolubWathen2000}
M.~F. Murphy, G.~H. Golub, and A.~J. Wathen, ``A note on preconditioning for
  indefinite linear systems,'' \emph{SIAM Journal on Scientific Computing},
  vol.~21, no.~6, pp. 1969--1972, 2000.

\bibitem{BenziGolubLiesen2005}
M.~Benzi, G.~H. Golub, and J.~Liesen, ``Numerical solution of saddle point
  problems,'' \emph{Acta Numerica}, vol.~14, pp. 1--137, 2005.

\bibitem{CurtisPowellReid1974}
A.~R. Curtis, M.~J.~D. Powell, and J.~K. Reid, ``On the estimation of sparse
  {Jacobian} matrices,'' \emph{IMA Journal of Applied Mathematics}, vol.~13,
  no.~1, pp. 117--119, 1974.

\bibitem{YavuzLarsen1992}
M.~Yavuz and E.~W. Larsen, ``Iterative methods for solving x-y geometry {$S_N$}
  problems on parallel architecture computers,'' \emph{Nuclear Science and
  Engineering}, vol. 112, no.~1, pp. 32--42, 1992.

\bibitem{AnistratovAzmy2015}
D.~Y. Anistratov and Y.~Y. Azmy, ``Iterative stability analysis of spatial
  domain decomposition based on block {J}acobi algorithm for the
  diamond-difference scheme,'' \emph{Journal of Computational Physics}, vol.
  297, pp. 462--479, 2015.

\bibitem{ChenChacon2014}
G.~Chen, L.~Chac{\'o}n, C.~A. Leibs, D.~A. Knoll, and W.~Taitano, ``Fluid
  preconditioning for {Newton--Krylov}-based, fully implicit, electrostatic
  particle-in-cell simulations,'' \emph{Journal of Computational Physics}, vol.
  258, pp. 555--567, 2014.

\bibitem{HammettPerkins1990}
G.~W. Hammett and F.~W. Perkins, ``Fluid moment models for {Landau} damping
  with application to the ion-temperature-gradient instability,''
  \emph{Physical Review Letters}, vol.~64, no.~25, pp. 3019--3022, 1990.

\bibitem{FriedConte1961}
B.~D. Fried and S.~D. Conte, \emph{The Plasma Dispersion Function}.\hskip 1em
  plus 0.5em minus 0.4em\relax New York: Academic Press, 1961.

\bibitem{Davis1979}
P.~J. Davis, \emph{Circulant Matrices}.\hskip 1em plus 0.5em minus 0.4em\relax
  New York: Wiley, 1979.

\bibitem{DayHeroux2001}
D.~Day and M.~A. Heroux, ``Solving complex-valued linear systems via equivalent
  real formulations,'' \emph{SIAM Journal on Scientific Computing}, vol.~23,
  no.~2, pp. 480--498, 2001.

\bibitem{cuDSS}
{NVIDIA}, ``{cuDSS}: the {CUDA} direct sparse solver library,''
  \url{https://docs.nvidia.com/cuda/cudss/}, 2025.

\bibitem{Brandt1977}
A.~Brandt, ``Multi-level adaptive solutions to boundary-value problems,''
  \emph{Mathematics of Computation}, vol.~31, no. 138, pp. 333--390, 1977.

\bibitem{George1973}
A.~George, ``Nested dissection of a regular finite element mesh,'' \emph{SIAM
  Journal on Numerical Analysis}, vol.~10, no.~2, pp. 345--363, 1973.

\bibitem{Hammett1993}
G.~W. Hammett, M.~A. Beer, W.~Dorland, S.~C. Cowley, and S.~A. Smith,
  ``Developments in the gyrofluid approach to tokamak turbulence simulations,''
  \emph{Plasma Physics and Controlled Fusion}, vol.~35, no.~8, pp. 973--985,
  1993.

\bibitem{DorlandHammett1993}
W.~Dorland and G.~W. Hammett, ``Gyrofluid turbulence models with kinetic
  effects,'' \emph{Physics of Fluids B: Plasma Physics}, vol.~5, no.~3, pp.
  812--835, 1993.

\bibitem{AdamsHPC2026}
M.~F. Adams, ``Parallel design of a {GPU}-resident semi-{L}agrangian
  drift-kinetic solver on unstructured extruded tokamak meshes,'' 2026,
  arXiv:2610.12585.

\bibitem{Yamazaki2020}
I.~Yamazaki, S.~Rajamanickam, and N.~Ellingwood, ``Performance portable
  supernode-based sparse triangular solver for manycore architectures,'' in
  \emph{49th International Conference on Parallel Processing - ICPP}, 2020, pp.
  1--11.

\bibitem{KokkosKernels2021}
S.~Rajamanickam, S.~Acer, L.~Berger-Vergiat, V.~Dang, N.~Ellingwood, E.~Harvey,
  B.~Kelley, C.~R. Trott, J.~Wilke, and I.~Yamazaki, ``{Kokkos} kernels:
  Performance portable {Sparse/Dense} linear algebra and graph kernels,''
  arXiv:2103.11991, 2021.

\bibitem{Jolliet2007}
S.~Jolliet, A.~Bottino, P.~Angelino, R.~Hatzky, T.~M. Tran, B.~F. McMillan,
  O.~Sauter, K.~Appert, Y.~Idomura, and L.~Villard, ``A global collisionless
  {PIC} code in magnetic coordinates,'' \emph{Computer Physics Communications},
  vol. 177, no.~5, pp. 409--425, 2007.

\bibitem{Grandgirard2016}
V.~Grandgirard, J.~Abiteboul, J.~Bigot, T.~Cartier-Michaud, N.~Crouseilles,
  G.~Dif-Pradalier, C.~Ehrlacher, D.~Esteve, X.~Garbet, P.~Ghendrih, G.~Latu,
  M.~Mehrenberger, C.~Norscini, C.~Passeron, F.~Rozar, Y.~Sarazin,
  E.~Sonnendr{\"u}cker, A.~Strugarek, and D.~Zarzoso, ``A 5{D} gyrokinetic
  full-$f$ global semi-{L}agrangian code for flux-driven ion turbulence
  simulations,'' \emph{Computer Physics Communications}, vol. 207, pp. 35--68,
  2016.

\bibitem{Adams2003}
M.~Adams, M.~Brezina, J.~Hu, and R.~Tuminaro, ``Parallel multigrid smoothing:
  polynomial versus {G}auss--{S}eidel,'' \emph{Journal of Computational
  Physics}, vol. 188, no.~2, pp. 593--610, 2003.

\bibitem{Wathen1987}
A.~J. Wathen, ``Realistic eigenvalue bounds for the {G}alerkin mass matrix,''
  \emph{IMA Journal of Numerical Analysis}, vol.~7, no.~4, pp. 449--457, 1987.

\bibitem{Hatzky2002}
R.~Hatzky, T.~M. Tran, A.~K{\"o}nies, R.~Kleiber, and S.~J. Allfrey, ``Energy
  conservation in a nonlinear gyrokinetic particle-in-cell code for
  ion-temperature-gradient-driven modes in $\theta$-pinch geometry,''
  \emph{Physics of Plasmas}, vol.~9, no.~3, pp. 898--912, 2002.

\bibitem{ScottSmirnov2010}
B.~Scott and J.~Smirnov, ``Energetic consistency and momentum conservation in
  the gyrokinetic description of tokamak plasmas,'' \emph{Physics of Plasmas},
  vol.~17, no.~11, p. 112302, 2010.

\bibitem{Gorler2011}
T.~G{\"o}rler, X.~Lapillonne, S.~Brunner, T.~Dannert, F.~Jenko, F.~Merz, and
  D.~Told, ``The global version of the gyrokinetic turbulence code {GENE},''
  \emph{Journal of Computational Physics}, vol. 230, no.~18, pp. 7053--7071,
  2011.

\bibitem{Ku2009}
S.~Ku, C.~S. Chang, and P.~H. Diamond, ``Full-$f$ gyrokinetic particle
  simulation of centrally heated global {ITG} turbulence from magnetic axis to
  edge pedestal top in a realistic tokamak geometry,'' \emph{Nuclear Fusion},
  vol.~49, no.~11, p. 115021, 2009, the XGC1 code.

\bibitem{SutherlandHodgman1974}
I.~E. Sutherland and G.~W. Hodgman, ``Reentrant polygon clipping,''
  \emph{Communications of the ACM}, vol.~17, no.~1, pp. 32--42, 1974.

\bibitem{Huba2019}
J.~D. Huba, ``{NRL Plasma Formulary},'' Naval Research Laboratory, Washington,
  DC, Tech. Rep. NRL/PU/6790--19-640, 2019.

\bibitem{HirvijokiAdams2017}
E.~Hirvijoki and M.~F. Adams, ``Conservative discretization of the {L}andau
  collision integral,'' \emph{Physics of Plasmas}, vol.~24, p. 032121, 2017.

\bibitem{AdamsLandau2025}
M.~F. Adams, P.~Wang, J.~Merson, K.~Huck, and M.~G. Knepley, ``A performance
  portable, fully implicit {L}andau collision operator with batched linear
  solvers,'' \emph{SIAM Journal on Scientific Computing}, vol.~47, no.~2, pp.
  B360--B381, 2025.

\bibitem{KrausHirvijoki2017}
M.~Kraus and E.~Hirvijoki, ``Metriplectic integrators for the {Landau}
  collision operator,'' \emph{Physics of Plasmas}, vol.~24, no.~10, p. 102311,
  2017.

\bibitem{Speth2013}
R.~L. Speth, W.~H. Green, S.~MacNamara, and G.~Strang, ``Balanced splitting and
  rebalanced splitting,'' \emph{SIAM Journal on Numerical Analysis}, vol.~51,
  no.~6, pp. 3084--3105, 2013.

\bibitem{Freund1993}
R.~W. Freund, ``A transpose-free quasi-minimal residual algorithm for
  {Non-Hermitian} linear systems,'' \emph{SIAM Journal on Scientific
  Computing}, vol.~14, no.~2, pp. 470--482, 1993.

\bibitem{Cooper1987}
G.~J. Cooper, ``Stability of {Runge--Kutta} methods for trajectory problems,''
  \emph{IMA Journal of Numerical Analysis}, vol.~7, no.~1, pp. 1--13, 1987.

\bibitem{AdamsMethods2026}
M.~F. Adams, ``{KINTORA}: a semi-{L}agrangian discontinuous {G}alerkin
  drift-kinetic, unstructured mesh tokamak solver,'' 2026, arXiv:2610.17428.

\bibitem{Gorler2016}
T.~G{\"o}rler, N.~Tronko, W.~A. Hornsby, A.~Bottino, R.~Kleiber, C.~Norscini,
  V.~Grandgirard, F.~Jenko, and E.~Sonnendr{\"u}cker, ``Intercode comparison of
  gyrokinetic global electromagnetic modes,'' \emph{Physics of Plasmas},
  vol.~23, p. 072503, 2016.

\bibitem{Gao2010}
Z.~Gao, ``Plasma shaping effects on the geodesic acoustic mode in the large
  orbit drift width limit,'' \emph{Physics of Plasmas}, vol.~17, no.~9, p.
  092503, 2010.

\bibitem{HintonRosenbluth1999}
F.~L. Hinton and M.~N. Rosenbluth, ``Dynamics of axisymmetric {$E\times B$} and
  poloidal flows in tokamaks,'' \emph{Plasma Physics and Controlled Fusion},
  vol.~41, no.~3A, pp. A653--A662, 1999.

\bibitem{Xiao2007}
Y.~Xiao, P.~J. Catto, and K.~Molvig, ``Collisional damping for ion temperature
  gradient mode driven zonal flow,'' \emph{Physics of Plasmas}, vol.~14, p.
  032302, 2007.

\bibitem{Frei2021}
B.~J. Frei, J.~Ball, A.~C.~D. Hoffmann, R.~Jorge, P.~Ricci, and L.~Stenger,
  ``Development of advanced linearized gyrokinetic collision operators using a
  moment approach,'' \emph{Journal of Plasma Physics}, vol.~87, no.~5, p.
  905870501, 2021.

\bibitem{HelanderSigmar2002}
P.~Helander and D.~J. Sigmar, \emph{Collisional Transport in Magnetized
  Plasmas}.\hskip 1em plus 0.5em minus 0.4em\relax Cambridge: Cambridge
  University Press, 2002.

\bibitem{RosenbluthHinton1998}
M.~N. Rosenbluth and F.~L. Hinton, ``Poloidal flow driven by
  ion-temperature-gradient turbulence in tokamaks,'' \emph{Physical Review
  Letters}, vol.~80, no.~4, pp. 724--727, 1998.

\bibitem{Romanelli1989}
F.~Romanelli, ``Ion temperature-gradient-driven modes and anomalous ion
  transport in tokamaks,'' \emph{Physics of Fluids B: Plasma Physics}, vol.~1,
  no.~5, pp. 1018--1025, 1989.

\bibitem{DongHortonKim1992}
J.~Q. Dong, W.~Horton, and J.~Y. Kim, ``Toroidal kinetic $\eta_i$-mode study in
  high-temperature plasmas,'' \emph{Physics of Fluids B: Plasma Physics},
  vol.~4, no.~7, pp. 1867--1876, 1992.

\bibitem{XiaoCatto2006shaping}
Y.~Xiao and P.~J. Catto, ``Plasma shaping effects on the collisionless residual
  zonal flow level,'' \emph{Physics of Plasmas}, vol.~13, no.~8, p. 082307,
  2006.

\bibitem{McMillan2010}
B.~F. McMillan, X.~Lapillonne, S.~Brunner, L.~Villard, S.~Jolliet, A.~Bottino,
  T.~G{\"o}rler, and F.~Jenko, ``System size effects on gyrokinetic
  turbulence,'' \emph{Physical Review Letters}, vol. 105, p. 155001, 2010.

\bibitem{Lapillonne2010}
X.~Lapillonne, B.~F. McMillan, T.~G{\"o}rler, S.~Brunner, T.~Dannert, F.~Jenko,
  F.~Merz, and L.~Villard, ``Nonlinear quasisteady state benchmark of global
  gyrokinetic codes,'' \emph{Physics of Plasmas}, vol.~17, p. 112321, 2010.

\bibitem{Creely2020}
A.~J. Creely, M.~J. Greenwald, S.~B. Ballinger, D.~Brunner, J.~Canik, J.~Doody
  \emph{et~al.}, ``Overview of the {SPARC} tokamak,'' \emph{Journal of Plasma
  Physics}, vol.~86, no.~5, p. 865860502, 2020.

\bibitem{YangKurthWilliams2020}
C.~Yang, T.~Kurth, and S.~Williams, ``Hierarchical roofline analysis for
  {GPUs}: Accelerating performance optimization for the {NERSC-9} {Perlmutter}
  system,'' \emph{Concurrency and Computation: Practice and Experience},
  vol.~32, no.~20, p. e5547, 2020.

\bibitem{Candy2016}
J.~Candy, E.~A. Belli, and R.~V. Bravenec, ``A high-accuracy {E}ulerian
  gyrokinetic solver for collisional plasmas,'' \emph{Journal of Computational
  Physics}, vol. 324, pp. 73--93, 2016, the CGYRO code.

\bibitem{CGYROgpu2018}
I.~Sfiligoi, J.~Candy, and M.~Kostuk, ``{CGYRO} performance on {Power9} {CPUs}
  and {Volta} {GPUs},'' in \emph{High Performance Computing: ISC High
  Performance 2018 International Workshops, Revised Selected Papers}, ser.
  Lecture Notes in Computer Science, R.~Yokota, M.~Weiland, J.~Shalf, and
  S.~Alam, Eds., vol. 11203.\hskip 1em plus 0.5em minus 0.4em\relax Springer,
  2018, pp. 365--372.

\bibitem{Idomura2008}
Y.~Idomura, M.~Ida, T.~Kano, N.~Aiba, and S.~Tokuda, ``Conservative global
  gyrokinetic toroidal full-f five-dimensional {V}lasov simulation,''
  \emph{Computer Physics Communications}, vol. 179, no.~6, pp. 391--403, 2008.

\bibitem{Dorf2013}
M.~A. Dorf, R.~H. Cohen, M.~Dorr, T.~Rognlien, J.~Hittinger, J.~Compton,
  P.~Colella, D.~Martin, and P.~McCorquodale, ``Simulation of neoclassical
  transport with the continuum gyrokinetic code {COGENT},'' \emph{Physics of
  Plasmas}, vol.~20, no.~1, p. 012513, 2013.

\bibitem{Dorf2020}
M.~Dorf and M.~Dorr, ``Progress with the {5D} full-{F} continuum gyrokinetic
  code {COGENT},'' \emph{Contributions to Plasma Physics}, vol.~60, no. 5-6, p.
  e201900113, 2020.

\bibitem{Mandell2020}
N.~R. Mandell, A.~Hakim, G.~W. Hammett, and M.~Francisquez, ``Electromagnetic
  full-\emph{f} gyrokinetics in the tokamak edge with discontinuous {G}alerkin
  methods,'' \emph{Journal of Plasma Physics}, vol.~86, no.~1, p. 905860109,
  2020.

\bibitem{Jo2025}
G.~Jo, J.~Seo, J.-M. Kwon, and E.~Yoon, ``A field-aligned gyrokinetic solver
  based on discontinuous {G}alerkin in tokamak geometry,'' \emph{Computer
  Physics Communications}, vol. 316, p. 109769, 2025.

\bibitem{Grandgirard2006}
V.~Grandgirard, M.~Brunetti, P.~Bertrand, N.~Besse, X.~Garbet, P.~Ghendrih
  \emph{et~al.}, ``A drift-kinetic semi-{L}agrangian 4{D} code for ion
  turbulence simulation,'' \emph{Journal of Computational Physics}, vol. 217,
  no.~2, pp. 395--423, 2006.

\bibitem{Bourne2025}
E.~Bourne, V.~Grandgirard, Y.~Asahi, J.~Bigot, P.~Donnel, A.~Hoffmann, A.~Kara,
  P.~Krah, B.~Legouix, E.~Malaboeuf, Y.~Munschy, K.~Obrejan, T.~Padioleau,
  M.~Protais, and P.~Vidal, ``Gyselalib++: A portable {C++} library for
  semi-{L}agrangian kinetic and gyrokinetic simulations,'' \emph{Journal of
  Open Source Software}, vol.~10, no. 113, p. 8582, 2025.

\bibitem{XGCgpu2017}
E.~D'Azevedo, S.~Abbott, T.~Koskela, P.~Worley, S.-H. Ku, S.~Ethier, E.~Yoon,
  M.~S. Shephard, R.~Hager, J.~Lang, J.~Choi, N.~Podhorszki, S.~Klasky,
  M.~Parashar, and C.-S. Chang, ``The fusion code {XGC},'' in \emph{Exascale
  Scientific Applications: Scalability and Performance Portability}, T.~P.
  Straatsma, K.~B. Antypas, and T.~J. Williams, Eds.\hskip 1em plus 0.5em minus
  0.4em\relax Chapman and Hall/CRC, 2017, pp. 529--552.

\bibitem{Lanti2020}
E.~Lanti, N.~Ohana, N.~Tronko, T.~Hayward-Schneider, A.~Bottino, B.~F.
  McMillan, A.~Mishchenko, A.~Scheinberg, A.~Biancalani, P.~Angelino,
  S.~Brunner, J.~Dominski, P.~Donnel, C.~Gheller, R.~Hatzky, A.~Jocksch,
  S.~Jolliet, Z.~X. Lu, J.~P. Martin~Collar, I.~Novikau, E.~Sonnendr{\"u}cker,
  T.~Vernay, and L.~Villard, ``{ORB5}: A global electromagnetic gyrokinetic
  code using the {PIC} approach in toroidal geometry,'' \emph{Computer Physics
  Communications}, vol. 251, p. 107072, 2020.

\bibitem{Idomura2020}
Y.~Idomura, T.~Ina, Y.~Ali, and T.~Imamura, ``Acceleration of fusion plasma
  turbulence simulations using the mixed-precision communication-avoiding
  {K}rylov method,'' in \emph{SC20: International Conference for High
  Performance Computing, Networking, Storage and Analysis}.\hskip 1em plus
  0.5em minus 0.4em\relax IEEE, 2020, pp. 1--13.

\bibitem{Ghosh2018}
D.~Ghosh, M.~A. Dorf, M.~R. Dorr, and J.~A.~F. Hittinger, ``Kinetic simulation
  of collisional magnetized plasmas with semi-implicit time integration,''
  \emph{Journal of Scientific Computing}, vol.~77, no.~2, pp. 819--849, 2018.

\bibitem{Manteuffel2018}
T.~A. Manteuffel, J.~Ruge, and B.~S. Southworth, ``Nonsymmetric algebraic
  multigrid based on local approximate ideal restriction ($\ell${AIR}),''
  \emph{SIAM Journal on Scientific Computing}, vol.~40, no.~6, pp.
  A4105--A4130, 2018.

\bibitem{Dorf2025}
M.~Dorf, M.~Dorr, D.~Ghosh, M.~Umansky, and V.~Soukhanovskii, ``Implicit
  full-{F} simulations of neoclassical ion transport,'' \emph{Physics of
  Plasmas}, vol.~32, no.~8, p. 083904, 2025.

\bibitem{Evans2026}
T.~M. Evans, R.~Glasby, C.~Hauck, S.~R. Schnake, K.~J. Schwiebert, L.~Shoemake,
  and S.~Slattery, ``Sweep-based, implicit solutions of the multidimensional
  {BGK} equation on unstructured grids,'' 2026, arXiv:2608.19150.

\bibitem{HiguerasRoldan2016}
I.~Higueras and T.~Rold{\'a}n, ``Construction of additive semi-implicit
  {R}unge--{K}utta methods with low-storage requirements,'' \emph{Journal of
  Scientific Computing}, vol.~67, no.~3, pp. 1019--1042, 2016.

\bibitem{CavaglieriBewley2015}
D.~Cavaglieri and T.~Bewley, ``Low-storage implicit/explicit {R}unge--{K}utta
  schemes for the simulation of stiff high-dimensional {ODE} systems,''
  \emph{Journal of Computational Physics}, vol. 286, pp. 172--193, 2015.

\bibitem{Pautz2002}
S.~D. Pautz, ``An algorithm for parallel {$S_n$} sweeps on unstructured
  meshes,'' \emph{Nuclear Science and Engineering}, vol. 140, no.~2, pp.
  111--136, 2002.

\bibitem{Plimpton2005}
S.~J. Plimpton, B.~Hendrickson, S.~P. Burns, W.~McLendon, III, and
  L.~Rauchwerger, ``Parallel {$S_n$} sweeps on unstructured grids: Algorithms
  for prioritization, grid partitioning, and cycle detection,'' \emph{Nuclear
  Science and Engineering}, vol. 150, no.~3, pp. 267--283, 2005.

\bibitem{BakerKoch1998}
R.~S. Baker and K.~R. Koch, ``An {$S_n$} algorithm for the massively parallel
  {CM-200} computer,'' \emph{Nuclear Science and Engineering}, vol. 128, no.~3,
  pp. 312--320, 1998.

\bibitem{Vermaak2021}
J.~I.~C. Vermaak, J.~C. Ragusa, M.~L. Adams, and J.~E. Morel, ``Massively
  parallel transport sweeps on meshes with cyclic dependencies,'' \emph{Journal
  of Computational Physics}, vol. 425, p. 109892, 2021.

\bibitem{Haut2019}
T.~S. Haut, P.~G. Maginot, V.~Z. Tomov, B.~S. Southworth, T.~A. Brunner, and
  T.~S. Bailey, ``An efficient sweep-based solver for the {$S_N$} equations on
  high-order meshes,'' \emph{Nuclear Science and Engineering}, vol. 193, no.~7,
  pp. 746--759, 2019.

\bibitem{Suau2025}
G.~Suau, T.~Gautier, A.~Calloo, R.~Baron, and R.~Le~Tellier, ``Performance
  portable batched linear algebra kernels for transport sweeps using
  {K}okkos,'' in \emph{Proceedings of the SC '25 Workshops of the International
  Conference for High Performance Computing, Networking, Storage and
  Analysis}.\hskip 1em plus 0.5em minus 0.4em\relax ACM, 2025, pp. 1147--1158.

\bibitem{Choi2026}
N.~Choi, J.~Hanophy, and Y.~Wang, ``Development of a {GPU}-based discrete
  ordinates transport sweeper in the {G}riffin reactor physics application,''
  \emph{Nuclear Science and Engineering}, pp. 1--28, 2026, published online 3
  August 2026.

\bibitem{Warsa2004}
J.~S. Warsa, T.~A. Wareing, and J.~E. Morel, ``{K}rylov iterative methods and
  the degraded effectiveness of diffusion synthetic acceleration for
  multidimensional {$S_N$} calculations in problems with material
  discontinuities,'' \emph{Nuclear Science and Engineering}, vol. 147, no.~3,
  pp. 218--248, 2004.

\bibitem{AdamsLarsen2002}
M.~L. Adams and E.~W. Larsen, ``Fast iterative methods for discrete-ordinates
  particle transport calculations,'' \emph{Progress in Nuclear Energy},
  vol.~40, no.~1, pp. 3--159, 2002.

\bibitem{Hanophy2020}
J.~Hanophy, B.~S. Southworth, R.~Li, T.~Manteuffel, and J.~Morel, ``Parallel
  approximate ideal restriction multigrid for solving the {$S_N$} transport
  equations,'' \emph{Nuclear Science and Engineering}, vol. 194, no.~11, pp.
  989--1008, 2020.

\bibitem{Kanschat2008}
G.~Kanschat, ``Robust smoothers for high-order discontinuous {G}alerkin
  discretizations of advection--diffusion problems,'' \emph{Journal of
  Computational and Applied Mathematics}, vol. 218, no.~1, pp. 53--60, 2008.

\bibitem{Krzysik2026}
O.~A. Krzysik, B.~S. Southworth, and B.~Philip, ``Overlapping {S}chwarz methods
  are not anisotropy-robust multigrid smoothers,'' \emph{Numerical Linear
  Algebra with Applications}, vol.~33, no.~2, p. e70070, 2026.

\bibitem{Ottaviani2011}
M.~Ottaviani, ``An alternative approach to field-aligned coordinates for plasma
  turbulence simulations,'' \emph{Physics Letters A}, vol. 375, no.~15, pp.
  1677--1685, 2011.

\bibitem{Hariri2013}
F.~Hariri and M.~Ottaviani, ``A flux-coordinate independent field-aligned
  approach to plasma turbulence simulations,'' \emph{Computer Physics
  Communications}, vol. 184, no.~11, pp. 2419--2429, 2013.

\bibitem{Stegmeir2016}
A.~Stegmeir, D.~Coster, O.~Maj, K.~Hallatschek, and K.~Lackner, ``The field
  line map approach for simulations of magnetically confined plasmas,''
  \emph{Computer Physics Communications}, vol. 198, pp. 139--153, 2016.

\bibitem{Michels2021}
D.~Michels, A.~Stegmeir, P.~Ulbl, D.~Jarema, and F.~Jenko, ``{GENE-X}: A
  full-$f$ gyrokinetic turbulence code based on the flux-coordinate independent
  approach,'' \emph{Computer Physics Communications}, vol. 264, p. 107986,
  2021.

\bibitem{Latu2017}
G.~Latu, M.~Mehrenberger, Y.~G{\"u}{\c{c}}l{\"u}, M.~Ottaviani, and
  E.~Sonnendr{\"u}cker, ``Field-aligned interpolation for semi-{L}agrangian
  gyrokinetic simulations,'' \emph{Journal of Scientific Computing}, vol.~74,
  no.~3, pp. 1601--1650, 2017.

\bibitem{Dingfelder2020}
B.~Dingfelder and F.~J. Hindenlang, ``A locally field-aligned discontinuous
  {G}alerkin method for anisotropic wave equations,'' \emph{Journal of
  Computational Physics}, vol. 408, p. 109273, 2020.

\bibitem{Held2016}
M.~Held, M.~Wiesenberger, and A.~Stegmeir, ``Three discontinuous {G}alerkin
  schemes for the anisotropic heat conduction equation on non-aligned grids,''
  \emph{Computer Physics Communications}, vol. 199, pp. 29--39, 2016.

\bibitem{Wiesenberger2023}
M.~Wiesenberger and M.~Held, ``A finite volume flux coordinate independent
  approach,'' \emph{Computer Physics Communications}, vol. 291, p. 108838,
  2023.

\bibitem{Kopriva1996}
D.~A. Kopriva, ``A conservative staggered-grid {Chebyshev} multidomain method
  for compressible flows. {II}. {A} semi-structured method,'' \emph{Journal of
  Computational Physics}, vol. 128, no.~2, pp. 475--488, 1996.

\bibitem{Laughton2021}
E.~Laughton, G.~Tabor, and D.~Moxey, ``A comparison of interpolation techniques
  for non-conformal high-order discontinuous {G}alerkin methods,''
  \emph{Computer Methods in Applied Mechanics and Engineering}, vol. 381, p.
  113820, 2021.

\bibitem{Francisquez2024}
M.~Francisquez, N.~R. Mandell, A.~Hakim, and G.~W. Hammett, ``Conservative
  discontinuous {G}alerkin interpolation: Sheared boundary conditions,''
  \emph{Computer Physics Communications}, vol. 298, p. 109109, 2024.

\bibitem{LinLee1995}
Z.~Lin and W.~W. Lee, ``Method for solving the gyrokinetic {P}oisson equation
  in general geometry,'' \emph{Physical Review E}, vol.~52, no.~5, pp.
  5646--5652, 1995.

\bibitem{Steiner2015}
C.~Steiner, M.~Mehrenberger, N.~Crouseilles, V.~Grandgirard, G.~Latu, and
  F.~Rozar, ``Gyroaverage operator for a polar mesh,'' \emph{The European
  Physical Journal D}, vol.~69, no.~1, p.~18, 2015.

\bibitem{CrouseillesMehrenbergerSellama2010}
N.~Crouseilles, M.~Mehrenberger, and H.~Sellama, ``Numerical solution of the
  gyroaverage operator for the finite gyroradius guiding-center model,''
  \emph{Communications in Computational Physics}, vol.~8, no.~3, pp. 484--510,
  2010.

\bibitem{Hager2016}
R.~Hager, E.~S. Yoon, S.~Ku, E.~F. D'Azevedo, P.~H. Worley, and C.~S. Chang,
  ``A fully non-linear multi-species {F}okker--{P}lanck--{L}andau collision
  operator for simulation of fusion plasma,'' \emph{Journal of Computational
  Physics}, vol. 315, pp. 644--660, 2016.

\bibitem{Abdelfattah2023}
A.~Abdelfattah, S.~Tomov, P.~Luszczek, H.~Anzt, and J.~Dongarra, ``{GPU}-based
  {LU} factorization and solve on batches of matrices with band structure,'' in
  \emph{Proceedings of the SC '23 Workshops of the International Conference on
  High Performance Computing, Network, Storage, and Analysis}, 2023, pp.
  1672--1679.

\bibitem{Nayak2025}
P.~Nayak, I.~Aggarwal, and H.~Anzt, ``Efficient solution of batched band linear
  systems on {GPUs},'' \emph{The International Journal of High Performance
  Computing Applications}, vol.~39, no.~5, pp. 615--630, 2025.

\bibitem{Liegeois2023}
K.~Liegeois, S.~Rajamanickam, and L.~Berger-Vergiat, ``Performance portable
  batched sparse linear solvers,'' \emph{IEEE Transactions on Parallel and
  Distributed Systems}, vol.~34, no.~5, pp. 1524--1535, 2023.

\bibitem{KleinStrzodka2023}
C.~Klein and R.~Strzodka, ``Tridigpu: a {GPU} library for block tridiagonal and
  banded linear equation systems,'' \emph{ACM Transactions on Parallel
  Computing}, vol.~10, no.~1, pp. 1--33, 2023.

\bibitem{Jin2025}
D.~Jin, A.~Montoison, and S.~Shin, ``Harnessing batched {BLAS}/{LAPACK} kernels
  on {GPUs} for parallel solutions of block tridiagonal systems,''
  arXiv:2509.03015, 2025.

\bibitem{Sadr2026}
M.~Sadr, E.~Lanti, A.~Mishchenko, X.~Wang, and L.~Villard, ``{ORB5X}: a
  performance-portable global electromagnetic gyrokinetic {PIC} code in
  {C++}/{K}okkos built using agentic {AI},'' arXiv:2607.17363, 2026.

\bibitem{Koldunov2026}
N.~V. Koldunov, S.~K. Cheedela, S.~Danilov, D.~Sidorenko, S.~Beyer, and
  T.~Jung, ``An ocean model ported by a large language model: experience and
  lessons from {FESOM2} ({F}ortran to {C} to {C++}/{K}okkos),''
  arXiv:2606.11356, 2026.

\bibitem{Gupta2026}
S.~Gupta, K.~Kamalakkannan, M.~Moraru, G.~Shipman, and P.~Diehl, ``From legacy
  {F}ortran to portable {K}okkos: an autonomous agentic {AI} workflow,''
  \emph{IEEE Transactions on Software Engineering}, 2026, early access.

\bibitem{Godoy2024}
W.~F. Godoy, P.~Valero-Lara, K.~Teranishi, P.~Balaprakash, and J.~S. Vetter,
  ``Large language model evaluation for high-performance computing software
  development,'' \emph{Concurrency and Computation: Practice and Experience},
  vol.~36, no.~26, p. e8269, 2024.

\bibitem{Ludwig2026}
O.~Ludwig, N.~Anklesaria, Z.~Jin, S.~Pophale, K.~Moshood, C.~J. DeVore
  \emph{et~al.}, ``Kaizen: metamorphic fuzzing and differential testing for
  {LLM}-translated {HPC} applications,'' arXiv:2607.04058, 2026.

\bibitem{XiaoCatto2006}
Y.~Xiao and P.~J. Catto, ``Short wavelength effects on the collisionless
  neoclassical polarization and residual zonal flow level,'' \emph{Physics of
  Plasmas}, vol.~13, no.~10, p. 102311, 2006.

\bibitem{PETScIssue2015}
M.~F. Adams, ``{GPU} features of the petsc-claude fork and {KINTORA}'s solvers,
  for {LLM}-assisted ports,'' PETSc issue \#2015,
  \url{https://gitlab.com/petsc/petsc/-/issues/2015}, 2026, the solvers' tables
  and the prompts that port them into an application or into PETSc.

\bibitem{DelgadoAparicio2026}
L.~F. Delgado-Aparicio, M.~Ono, and J.~E. Menard, ``Generalized
  {L}awson--{C}ordey--{M}ills accessibility of fusion ignition,''
  \emph{Physical Review Letters}, vol. 137, no.~11, p. 115102, 2026.

\bibitem{LandauDamping1946}
L.~D. Landau, ``On the vibrations of the electronic plasma,'' \emph{Journal of
  Physics (USSR)}, vol.~10, pp. 25--34, 1946.

\bibitem{Coppi1967}
B.~Coppi, M.~N. Rosenbluth, and R.~Z. Sagdeev, ``Instabilities due to
  temperature gradients in complex magnetic field configurations,''
  \emph{Physics of Fluids}, vol.~10, no.~3, pp. 582--587, 1967.

\bibitem{BrizardTronko2011}
A.~J. Brizard and N.~Tronko, ``Exact momentum conservation laws for the
  gyrokinetic {Vlasov--Poisson} equations,'' \emph{Physics of Plasmas},
  vol.~18, no.~8, p. 082307, 2011.

\bibitem{Arnoldi1951}
W.~E. Arnoldi, ``The principle of minimized iterations in the solution of the
  matrix eigenvalue problem,'' \emph{Quarterly of Applied Mathematics}, vol.~9,
  no.~1, pp. 17--29, 1951.

\bibitem{claude2026}
{Anthropic}, ``Claude ({Fable}~5, {Fable}~5.1, {Fable}~5.2, {Opus}~5,
  {Opus}~5.5 and {Sonnet}~5 models),'' \url{https://claude.ai}, 2026, large
  language model, used as an AI coding and writing assistant.

\end{thebibliography}
